\documentclass[11pt, amsmath, a4paper]{article}

\usepackage{amsfonts}
\usepackage{amssymb}
\usepackage{mathrsfs}
\usepackage{graphicx}

\newcommand   \Integers {\mathbb Z}
\newcommand   \PositiveIntegers {{\mathbb N}}

\newcommand   \finiteIntegerRing[1] {\basel{\Integers}{{#1}}}
\newcommand   \invertibleRingElements[1] {\basel{\Integers^{\ast}}{{#1}}}
\newcommand  \cnt      {\mathfrak n}
\newcommand  \primeInt  {\mathtt p}
\newcommand  \coprimeInt   {\mathtt q}

\newcommand   \finiteIntegerField {\basel{\mathbb Z}{\primeInt}} 
 
\newcommand   \scalars  {\mathbb F}

\def  \mod  {\small{\textsf{ mod }}}

\def  \gcd  {\small \textsf{gcd}}

\newcommand   \powerset {\mathfrak P}

\newcommand  \HashTable {\textsf{H}}
\newcommand  \LookupTable {\textsf{T}}
\newcommand  \SignatureTable {\textsf{S}}
\newcommand  \BackSubstitutionTable {\textsf{B}}
\newcommand  \SignatureVerificationTable {\textsf{V}}
\newcommand  \SignatureAuthenticationTable {\textsf{A}}

\newcommand  \mygrp    {\mathsf G}
\newcommand  \eltSet    {\mathsf E}
\newcommand  \orderofthegroup  {\mathsf n}
\newcommand  \nonzeroscalars {{\mathbb F}^{\ast}}

\newcommand  \false {\mathtt{false}}
\newcommand  \true {\mathtt{true}}

\newcommand  \xx   {\mathbf x}
\newcommand  \yy   {\mathbf y}
\newcommand  \zz   {\mathbf z}

\newcommand \oomega {\boldmath \omega}

\newcommand  \bglb {\big (}
\newcommand  \bgrb {\big )}
\newcommand  \bglc {\big \{}
\newcommand  \bgrc {\big \}}
\newcommand  \bgls {\big [}
\newcommand  \bgrs {\big ]}

\newcommand  \PSPACE {\mathsf {PSPACE}}

\newcommand   \boolean  {{\scriptstyle{\mathcal B}}\displaystyle{}}
\newcommand   \QBF  {{\scriptstyle{\basel{\boolean}{\scriptscriptstyle{\mathcal Q}}}}\displaystyle{}}
\newcommand   \QSAT  {{\scriptstyle{\basel{\boolean}{\scriptscriptstyle{{\mathcal Q}\rm{-SAT}}}}}\displaystyle{}}
\newcommand   \ArithmeticExpressions  {{\scriptstyle{{\mathcal ARITH}\textrm{-}{\mathcal EXP}}}\displaystyle{}}
 
\newcommand   \QuantifiedArithmeticExpressions  {{\scriptstyle{\basel{\ArithmeticExpressions}{\scriptscriptstyle{\mathcal Q}}}}\displaystyle{}}
\newcommand   \SatisfiableQuantifiedArithmeticExpressions  {{\scriptstyle{\basel{\ArithmeticExpressions}{\scriptscriptstyle{\mathcal Q}\scriptscriptstyle{\rm{-SAT}}}}}\displaystyle{}}

\newcommand \leftinverse {\textrm{\tiny{(-L)}}} 
\newcommand \rightinverse {\textrm{\tiny{(-R)}}}

\newtheorem{theorem}{Theorem}

\newtheorem{corollary}{Corollary}[theorem]

\newcommand \proof {\rm{\textbf{Proof.}}~~}

\newcommand {\basel}[2]{#1_{_{#2}}}

\newcommand  \polynomials[3] { #1{\mathbf{[}}\basel{#2}{\mathrm{1}},\, \ldots,\, \basel{#2}{\mathit{#3}}{\mathbf{]}} }

\newcommand  \Expressions[3] { {\mathcal {EXP}} {\mathbf{\bglb}} {#1\, ; \, \mathbf{[}}\basel{#2}{\mathrm{1}},\, \ldots,\, \basel{#2}{\mathit{#3}}{\mathbf{]}} {\mathbf{\bgrb}}} 

\newcommand  \multipolynomials[5] { #1{\mathbf{[}}\basel{#2}{\mathrm{1}},\, \ldots,\, \basel{#2}{\mathit{#4}},\, \basel{#3}{\mathrm{1}},\, \ldots,\, \basel{#3}{\mathit{#5}} {\mathbf{]}} }

\newcommand  \multiexpressions[5] {{\mathcal {EXP}} {\mathbf{\bglb}} {#1\,;\, \mathbf{[}}\basel{#2}{\mathrm{1}},\, \ldots,\, \basel{#2}{\mathit{#4}},\, \basel{#3}{\mathrm{1}},\, \ldots,\, \basel{#3}{\mathit{#5}} {\mathbf{]}} {\mathbf{\bgrb}}}

\newcommand  \singlevariablepolynomials[2] {#1{\mathbf{[}#2\mathbf{]}}}

\newcommand  \singlevariableexpressions[2] {{\mathcal {EXP}} {\mathbf{\bglb}} {#1\,;\,{\mathbf{[}#2\mathbf{]}}}{\mathbf{\bgrb}}}

\newcommand \qed  {\hfill{} $\blacksquare$}

\def \lspace  {\hspace*{-0.1cm}}

\def \tab  {\hspace*{0.5cm}}
\def \ltab  {\hspace*{-0.5cm}}

\def \shiftright  {\hspace*{2.0cm}}
\def \shiftleft  {\hspace*{-2.0cm}}

\title{Multivariate Cryptography with Mappings of Discrete Logarithms and Polynomials}

\author{Duggirala Meher Krishna\thanks{Department of Electronics and Communication Engineering (ECE), Gayatri Vidya Parishad College of Engineering (Autonomous), Madhurawada, VISAKHAPATNAM -- 530 048, Andhra Pradesh, India. E-mail~: ~~ \underline{duggiralameherkrishna@gmail.com}}  \tab \tab Duggirala Ravi\thanks{
Department of Computer Science and Engineering (CSE), Gayatri Vidya Parishad College of Engineering (Autonomous), Madhurawada, VISAKHAPATNAM -- 530 048, Andhra Pradesh, India. E-mail ~: ~ \underline{ravi@gvpce.ac.in}; ~ \underline{duggirala.ravi@yahoo.com}; ~
~ \underline{duggirala.ravi@rediffmail.com}; ~ and ~ \underline{drdravi2000@yahoo.com}}}

\date{}

\begin{document}

\maketitle
\tableofcontents
\begin{abstract}
{In this paper, algorithms for multivariate public key cryptography and digital signature are described. Plain messages and encrypted messages are arrays, consisting of elements from a fixed finite ring or field. The encryption and decryption algorithms are based on multivariate mappings. The security of the private key depends on the difficulty of solving a system of parametric simultaneous multivariate equations involving polynomial or exponential mappings. The method is a general purpose utility for most data encryption, digital certificate or digital signature applications. }
\end{abstract}
\section{Introduction}

\subsection{Preliminary Discussion}
The role of cryptographic algorithms is to provide information security  \cite{Buchmann:2004, Koblitz:1994, Schneier:1996, StallingsCNS:2011, StallingsNSE:2011, Stinson:2005}. In general, proper data encryption and authentication mechanisms with access control are the preferred means for a trusted secure system \cite{StallingsCNS:2011, StallingsNSE:2011}. The most popular public key cryptosystems are the RSA \cite{RSA:1978}, NTRU Encrypt algorithm \cite{HLS:1999, HPS:1998, HPS:2001, HS:2001}, elliptic curve cryptography (ECC)  \cite{Koblitz:1987, Miller:1985, Smart:1999, Washington:2008}, the algorithms based on diophantine equations and discrete logarithms \cite{LCL:1995, ElGamal:1985}, and those based on multivariate quadratic polynomials \cite{BBD:2009, Koblitz:1999}. The RSA, the NTRU and the ECC are assumed to be secure algorithms unless there are new breakthroughs in integer factoring (for RSA), or in lattice reduction (for NTRU), or in elliptic curve discrete logarithm techniques (for ECC)  \cite{CS:1997, Gentry:2001}.

 In this paper, algorithms for public key cryptography as well as digital signature based on multivariate mappings are described, with plain and encrypted message arrays consisting of elements from a fixed commutative and finite ring or field. The keys can be built up starting from independently chosen small degree polynomial or easy exponential mappings, resulting in fast key generation and facilitating easy changes of keys as often as required. The security depends on the difficulty of solving parametric simultaneous multivariate equations involving polynomial or exponential mappings \cite{Buchberger:1965, CGHMP:2003, Faugere:1999, Faugere:2002, Marker:2000, MMP:1996, vanDalen:1994, vandenDries:2000} in the case of straightforward attacks, and on the difficulty of finding the private keys in the case of key recovery attacks.

\subsection{\label{Sec-notation}Notation}

 In the sequel, let $\Integers$ be the set of integers, and let
 $\PositiveIntegers$ be the set of positive integers.  For a positive integer $\cnt \geq 2$, let $\finiteIntegerRing{\cnt}$ be the ring of integers with addition and multiplication $\mod \cnt$, and  $\invertibleRingElements{\cnt}$ be the commutative group  of invertible elements in $\finiteIntegerRing{\cnt}$, with respect to multiplication operation in $\finiteIntegerRing{\cnt}$. Let $\scalars$ be a finite field, consisting of $\primeInt^{n}$ elements for some positive integer $n$ and prime number $\primeInt$, and let $\nonzeroscalars$ be the multiplicative group of nonzero elements in $\scalars$. Let $\mygrp$ be a finite cyclic group of order $\cnt \geq 2$.  Let $\eltSet$ be either $\scalars$ or $\finiteIntegerRing{\cnt}$ or $\mygrp$. If $\eltSet = \mygrp$, where $\mygrp$ is equipped with only the group operation, then $\mygrp$ is isomorphic to $\finiteIntegerRing{\cnt}$, where the group operation in   $\mygrp$ is identified with the addition operation of $\finiteIntegerRing{\cnt}$. The addition operation of $\Integers$ is a primary operation, and the multiplication operation, that can be treated as a secondary operation \cite{Manin:2010} over the additive group $\Integers$, is defined uniquely such that the distribution laws hold true, with $1$ as the multiplicative identity, rendering $\Integers$ as the commutative ring, and the same holds for $\finiteIntegerRing{\cnt}$. Let $\polynomials{\eltSet}{x}{m}$, for $m \in \PositiveIntegers$, be the algebra of multivariate polynomials in $m$ formal variables $\basel{x}{1},\ldots,\basel{x}{m}$ with coefficients in $\eltSet$.  Now, if $\mygrp = \nonzeroscalars$, for a finite field $\scalars$, then the group operation in $\mygrp$ coincides with the multiplication operation in $\scalars$ and $\polynomials{\mygrp}{x}{m} = \polynomials{\scalars}{x}{m}$. If $m = 1$, then $\polynomials{\eltSet}{x}{m}$ is denoted by $\singlevariablepolynomials{\eltSet}{x}$, with $x = \basel{x}{1}$.  A variable with its name expressed in bold face assumes values from a product space, which is a product of finitely many copies of the same set, and each component of the variable, expressed in the corresponding case without boldness and a positive integer subscript, assumes values from the constituent component space, succinctly as, for example, $\xx = (\basel{x}{1},\, \ldots,\, \basel{x}{m}) \in \eltSet^{m}$,  for some $m \in \PositiveIntegers$.

\subsection{\label{Sec-polynomials-over-Zn}Polynomials over $\finiteIntegerRing{\cnt}$}
  Let $\cnt = \prod_{i = 1}^{r} \basel{\primeInt^{\basel{l}{i}}}{i}$, where $r$ and $\basel{l}{i}$ are positive integers, and $\basel{\primeInt}{i}$ are distinct prime numbers, for $1 \leq i \leq r$. Let $\basel{\coprimeInt}{i} = \basel{\primeInt^{-\basel{l}{i}}}{i}\cnt =
 \prod_{\tiny{\begin{array}{c}
 j = 1\\
 j \neq i
\end{array} } }^{r} \basel{\primeInt^{\basel{l}{j}}}{j} $, 
and let $\basel{m}{i} \in \PositiveIntegers$ be such that 
$\basel{m}{i}\basel{\coprimeInt}{i} \equiv 1 \mod \basel{\primeInt^{\basel{l}{i}}}{i}$, for $1 \leq i \leq r$. Then, $\finiteIntegerRing{\cnt} = \oplus_{i = 1}^{r} \basel{m}{i}\basel{\coprimeInt}{i}\finiteIntegerRing{\basel{\primeInt^{\basel{l}{i}}}{i}}$. 

Now, a polynomial $f(x) \in \singlevariablepolynomials{\basel{\Integers}{\cnt}}{x}$ can be expressed as $\sum_{i = 1}^{r} \basel{m}{i} \basel{\coprimeInt}{i} \basel{f}{i}(x)$, for some unique polynomials $\basel{f}{i}(x) \in \singlevariablepolynomials{\basel{\Integers}{\basel{\primeInt^{\basel{l}{i}}}{i}}}{x}$, for $1 \leq i \leq r$.  For some $x \in \Integers$ and index $i$, where $1 \leq i \leq r$, if $\basel{\primeInt}{i} \mid  f(x)$, then
$\gcd \bglb f(x) \mod \basel{\primeInt^{\basel{l}{i}}}{i}\, , \, \basel{\primeInt}{i}\bgrb$
$=$
$\gcd \bglb \basel{f}{i}(x) \, , \, \basel{\primeInt}{i}\bgrb$
$=$
$\basel{\primeInt}{i} \neq 1$. Thus, $\gcd(f(x), \, \cnt) = 1$, for every $x \in \basel{\Integers}{\cnt}$, if and only if $\gcd(\basel{f}{i}(x),\, \basel{\primeInt}{i}) = 1$, for every $x \in \basel{\Integers}{\basel{\primeInt^{\basel{l}{i}}}{i}}$, for every index $i$, where $1 \leq i \leq r$. Similarly, $f$ is a surjective (hence bijective) mapping from $\basel{\Integers}{\cnt}$ onto $\basel{\Integers}{\cnt}$, if and only if $\basel{f}{i}$ is a surjective (hence bijective) mapping from $\basel{\Integers}{\basel{\primeInt^{\basel{l}{i}}}{i}}$ onto $\basel{\Integers}{\basel{\primeInt^{\basel{l}{i}}}{i}}$,  or equivalently, $\basel{f}{i}(x) \mod \basel{\primeInt}{i}$ is a bijective mapping from $\basel{\Integers}{\basel{\primeInt}{i}}$ into itself and, when $\basel{l}{i} \geq 2$,  $\basel{f'}{i}(x)\not \equiv 0 \mod \basel{\primeInt}{i}$, for all $x \in  \basel{\Integers}{\basel{\primeInt^{\basel{l}{i}}}{i}}$, where $\basel{f'}{i}$ is the formal algebraic derivative of $\basel{f}{i}$, for every index $i$, where $1 \leq i \leq r$ \cite{LMT:1993}. Now, if $g(x) \in \singlevariablepolynomials{\basel{\Integers}{\cnt}}{x}$, where $g(x) = \sum_{i = 1}^{r} \basel{m}{i} \basel{\coprimeInt}{i} \basel{g}{i}(x)$, for some $\basel{g}{i}(x) \in \singlevariablepolynomials{\basel{\Integers}{\basel{\primeInt^{\basel{l}{i}}}{i}}}{x}$, for $1 \leq i \leq r$, then $f(x)g(x) = \sum_{i = 1}^{r} \basel{m}{i} \basel{\coprimeInt}{i} \basel{f}{i}(x)\basel{g}{i}(x)$. Thus, (A) $f(x)$ is a unit in $\singlevariablepolynomials{\basel{\Integers}{\cnt}}{x}$, if and only if  $\basel{f}{i}(x)$ is a unit, {\em i.e.},  $\basel{f}{i}(x) \mod \basel{\primeInt}{i} \in  \basel{\Integers^{\ast}}{\basel{\primeInt}{i}}$, for every index $i$, where $1 \leq i \leq r$, (B)  $f(x)$ is reducible in $\singlevariablepolynomials{\basel{\Integers}{\cnt}}{x}$, if and only if $\basel{f}{i}(x)$ is reducible in $\singlevariablepolynomials{\basel{\Integers}{\basel{\primeInt^{\basel{l}{i}}}{i}}}{x}$, for some index $i$, where $1 \leq i \leq r$, and (C) $f(x)$ is irreducible in $\singlevariablepolynomials{\basel{\Integers}{\cnt}}{x}$, if and only if  $\basel{f}{i}(x)$ is irreducible in $\singlevariablepolynomials{\basel{\Integers}{\basel{\primeInt^{\basel{l}{i}}}{i}}}{x}$, or equivalently, $\basel{f}{i}(x) \mod \basel{\primeInt}{i}$ is irreducible in $\singlevariablepolynomials{\basel{\Integers}{\basel{\primeInt}{i}}}{x}$, for every index $i$, where $1 \leq i \leq r$.  Thus, for any positive integer $k$,  $\polynomials{\finiteIntegerRing{\cnt}}{x}{k}$ can be expressed as $\oplus_{i = 1}^{r}\basel{m}{i} \basel{\coprimeInt}{i} \polynomials{\finiteIntegerRing{\basel{\primeInt^{\basel{l}{i}}}{i}}}{x}{k}$.

\subsection{\label{Sec-modular-exponentiation-over-Zn}Modular Exponentiation over $\finiteIntegerRing{\cnt}$}

The modular exponentiation operation is extensively studied in connection with the RSA cryptosystem  \cite{Buchmann:2004, Koblitz:1994, RSA:1978, Schneier:1996, StallingsCNS:2011, StallingsNSE:2011, Stinson:2005}. In this section, the modular exponentiation is extended to the situation, wherein the exponents are functions. The security of the RSA system depends on the difficulty of factorization of a positive integer into its prime factors. However, simplification of computations as well as porting of variables from base level to exponentiation level by a homomorphism requires availability of prime factors in advance for both encryption and decryption, while working with multivariate mappings involving functions as exponents. In the sequel, let  $\varphi$ be the Euler phi function \cite{Buchmann:2004, Koblitz:1994, Schneier:1996, Stinson:2005}. Let $\cnt = \prod_{i = 1}^{r} \basel{\primeInt^{\basel{l}{i}}}{i}$, where $r \in \PositiveIntegers$, $\basel{l}{i} \in \PositiveIntegers \backslash \{1\}$ and $\basel{\primeInt}{i}$ are distinct prime numbers, for $1 \leq i \leq r$. Let  $\Expressions{\finiteIntegerRing{\cnt}}{x}{m}$ be the smallest set of  expressions,  closed with respect to addition and multiplication, and containing expressions of the form  $a(\basel{x}{1},\, \ldots,\, \basel{x}{m})^{ b(\basel{x}{1},\, \ldots,\, \basel{x}{m}) }$,  where $a(\basel{x}{1},\, \ldots,\, \basel{x}{m}) $
 $\in$
 $ \polynomials{\finiteIntegerRing{\cnt}}{x}{m}$, and either 
\begin{enumerate}
\item {\label{condition-1-in-modular-exponentiation-over-Zn}}as a formal expression, $b(\basel{x}{1},\, \ldots,\, \basel{x}{m})$ does not depend on $(\basel{x}{1},\, \ldots,\, \basel{x}{m})$ and evaluates to any fixed positive integer, or

\item   $a(\basel{x}{1},\, \ldots,\, \basel{x}{m})$ evaluates to elements in $\invertibleRingElements{\cnt}$, for all values of $(\basel{x}{1},\, \ldots,\, \basel{x}{m})$ in some domain of interest, which is a subset of $\finiteIntegerRing{\cnt}^{m}$,  and  $b(\basel{x}{1},\, \ldots,\, \basel{x}{m})$ is of the form $c(h(\basel{x}{1}),\, \ldots,\, h(\basel{x}{m}))$, for some expression $c(\basel{z}{1},\, \ldots,\, \basel{z}{m}) \in \Expressions{\finiteIntegerRing{\varphi(\cnt)}}{z}{m}$ and ring homomorphism $h$ from $\finiteIntegerRing{\cnt}$ into $\finiteIntegerRing{\varphi(\cnt)}$.  

\end{enumerate}
\noindent The condition in (\ref{condition-1-in-modular-exponentiation-over-Zn}) above implies that  {\small{$\polynomials{\finiteIntegerRing{\cnt}}{x}{m} \subseteq  \Expressions{\finiteIntegerRing{\cnt}}{x}{m}$}}. Thus, the integers in $\Integers$ and those in $\finiteIntegerRing{\cnt}$, for various modulus positive integers $\cnt \geq 2$, need to be distinguished clearly as separate elements. The expressions in $\Expressions{\finiteIntegerRing{\cnt}}{x}{m}$ are turned into mappings, by identifying appropriate domains of values and interpretation for variables and operations in the respective domains \cite{vanDalen:1994, vandenDries:2000, Manin:2010, Marker:2000}. For $\xx \in \finiteIntegerRing{\cnt}^{m}$ and $s \in \PositiveIntegers \backslash \{1\}$, such that $s \mid \cnt$, let $\xx \mod s = \bglb \basel{x}{1} \mod s,\, \ldots,\, \basel{x}{m} \mod s \bgrb$.  Let  $f(\xx) \in \polynomials{\basel{\Integers}{\cnt}}{x}{m}$ be  such that $f(\xx)$ evaluates to elements in $\invertibleRingElements{\cnt}$, for $\xx \in  X$, for some  $X \subseteq \finiteIntegerRing{\cnt}^{m}$, and let $\basel{f}{i}(\xx) \in \polynomials{\basel{\Integers}{\basel{\primeInt^{\basel{l}{i}}}{i}}}{x}{m}$, for $1 \leq i \leq r$, be such that $f(\xx) = \sum_{i = 1}^{r} \basel{m}{i} \basel{\coprimeInt}{i} \basel{f}{i}(\xx  \mod \basel{\primeInt^{\basel{l}{i}}}{i})$. Now, for $\xx \in X$ and  $k \in \Integers$, the following holds: $\bglb f(\xx)\bgrb^{k}$
 $ =$
$\bglb f(\xx)\bgrb^{k \mod \varphi(\cnt)}$
 $=$
$  \sum_{i = 1}^{r} \basel{m}{i} \basel{\coprimeInt}{i} \bglb\basel{f}{i}(\xx \mod \basel{\primeInt^{\basel{l}{i}}}{i})\bgrb^{k \mod \varphi(\cnt)}$
 $=$
$  \sum_{i = 1}^{r} \basel{m}{i} \basel{\coprimeInt}{i} \bglb\basel{f}{i}(\xx \mod \basel{\primeInt^{\basel{l}{i}}}{i})\bgrb^{k \mod \varphi(\basel{\primeInt^{\basel{l}{i}}}{i})}$. Let $g(\yy) \in \polynomials{\basel{\Integers}{\varphi(\basel{\Integers}{\cnt})}}{y}{n}$ and $\basel{g}{i}(\zz) \in \polynomials{\basel{\Integers}{\varphi(\basel{\primeInt^{\basel{l}{i}}}{i})}}{z}{n}$ be such that the following holds: $ \basel{g}{i}\bglb\yy \mod \varphi(\basel{\primeInt^{\basel{l}{i}}}{i})\bgrb = g(\yy) \mod \varphi\bglb \basel{\primeInt^{\basel{l}{i}}}{i} \bgrb $, for $1 \leq i \leq r$. 
Thus, $f^{g(\yy)}(\xx) = \sum_{i = 1}^{r}  \basel{m}{i} \basel{\coprimeInt}{i} \basel{f^{g(\yy )}}{i}(\xx) = \sum_{i = 1}^{r}  \basel{m}{i} \basel{\coprimeInt}{i} \basel{f^{\basel{g}{i}(\yy \mod \varphi(\basel{\primeInt^{\basel{l}{i}}}{i}))}}{i}(\xx\mod \basel{\primeInt^{\basel{l}{i}}}{i})$, for independent vectors $\xx \in X$ and $\yy \in \finiteIntegerRing{\varphi(\cnt)}^{n}$. Now, $\varphi(\basel{\primeInt^{\basel{l}{i}}}{i}) = (\basel{\primeInt}{i}-1)\basel{\primeInt^{\basel{l}{i}-1}}{i}$, where $\basel{l}{i} \geq 2$,  for $1 \leq i \leq r$. Let $\basel{w}{i} = (\basel{\primeInt}{i}-1)^{-1} \mod \basel{\primeInt^{\basel{l}{i}-1}}{i}$, and let $\basel{h}{i}\,:\, \finiteIntegerRing{\basel{\primeInt^{\basel{l}{i}}}{i}} \rightarrow  \finiteIntegerRing{\varphi(\basel{\primeInt^{\basel{l}{i}}}{i})}$ be the map defined by $\basel{h}{i}(x) = (\basel{\primeInt}{i}-1)(\basel{w}{i} x \mod \basel{\primeInt^{\basel{l}{i}-1}}{i})$, for $1 \leq i \leq r$. Then, $\basel{h}{i}$  is a ring homomorphism, for $1 \leq i \leq r$. Now,  let $h\bglb\sum_{i =1}^{r}\basel{m}{i}\basel{\coprimeInt}{i}\basel{z}{i}\bgrb = \bglb  \basel{h}{1}(\basel{z}{1}),\, \ldots, \,  \basel{h}{r}(\basel{z}{r})\bgrb$, for $\basel{z}{i} \in \finiteIntegerRing{\basel{\primeInt^{\basel{l}{i}}}{i}}$ and $1 \leq i \leq r$. Then, the map $h$ is a ring homomorphism from the ring $\oplus_{i = 1}^{r} \basel{m}{i} \basel{\coprimeInt}{i}\finiteIntegerRing{\basel{\primeInt^{\basel{l}{i}}}{i}}$ into the ring of direct product $\prod_{i = 1}^{r} \finiteIntegerRing{\varphi(\basel{\primeInt^{\basel{l}{i}}}{i})}$. If the base level and exponentiation level interpretation maps are $\basel{\mathcal I}{\textrm{base}}$ and  $\basel{\mathcal I}{\mathrm{exponent}}$, respectively, then $\basel{\mathcal I}{\mathrm{exponent}}$ can be chosen to be $h \circ \basel{\mathcal I}{\mathrm{base}}$, applied from right to left in the written order, preserving the respective ring operations in the base level and exponentiation level subexpressions. If $\basel{l}{i} = 1$, for some index $i$, where $1\leq i \leq r$, then exponentiation along $i$th component can be carried by interpreting $\finiteIntegerRing{\basel{\primeInt}{i}}$ to be a finite field, and porting values of base level  expressions to exponentiation level expressions by discrete logarithm mapping, as discussed in section \ref{Sec-modular-exponentiation-over-Finite-Fields}.

\subsection{\label{Sec-modular-exponentiation-over-Finite-Fields}Modular Exponentiation over $\scalars$}

  Let $\scalars$ be a finite field containing $\primeInt^{n}$ elements and $\cnt = \primeInt^{n}-1$, for some prime number $\primeInt$ and positive integer $n$. Let  $\Expressions{\scalars}{x}{m}$ be the smallest set of  expressions,  closed with respect to addition and multiplication, and containing expressions of the form  $a(\basel{x}{1},\, \ldots,\, \basel{x}{m})^{ b(\basel{x}{1},\, \ldots,\, \basel{x}{m}) }$,  where $a(\basel{x}{1},\, \ldots,\, \basel{x}{m}) $
 $\in$
 $ \polynomials{\scalars}{x}{m}$, and either  
\begin{enumerate}
\item \label{item-1-in-modular-exponentiation-over-Finite-Fields}as a formal expression, $b(\basel{x}{1},\, \ldots,\, \basel{x}{m})$ does not depend on $(\basel{x}{1},\, \ldots,\, \basel{x}{m})$ and evaluates to any fixed positive integer, or 
\item  \label{item-2-in-modular-exponentiation-over-Finite-Fields} $a(\basel{x}{1},\, \ldots,\, \basel{x}{m})$ evaluates to elements in $\nonzeroscalars$, for all values of $(\basel{x}{1},\, \ldots,\, \basel{x}{m})$ in some domain of interest, which is a subset of $\mygrp^{m}$, where $\mygrp = \nonzeroscalars$, and  $b(\basel{x}{1},\, \ldots,\, \basel{x}{m})$ is of the form $c(h(\basel{x}{1}),\, \ldots,\, h(\basel{x}{m}))$, for some expression $c(\basel{z}{1},\, \ldots,\, \basel{z}{m}) \in \Expressions{\finiteIntegerRing{\cnt}}{z}{m}$ and group isomorphism $h$ from $\mygrp$ into $\finiteIntegerRing{\cnt}$.    
\end{enumerate}
The condition in (\ref{item-1-in-modular-exponentiation-over-Finite-Fields}) above implies that {\small{$\polynomials{\scalars}{x}{m} \subseteq \Expressions{\scalars}{x}{m}$}}. For a primitive element $a \in \nonzeroscalars$, let $\basel{\log}{a} : \nonzeroscalars \rightarrow \finiteIntegerRing{\cnt}$ be the discrete logarithm function defined by $\basel{\log}{a}(g) = x$, exactly when $a^{x} = g$, for $g \in \nonzeroscalars$ and $x \in \finiteIntegerRing{\cnt}$. Thus, the group homomorphism $h$ can be taken to be $\basel{\log}{a}$. If the base level and exponentiation level interpretation maps are $\basel{\mathcal I}{\textrm{base}}$ and  $\basel{\mathcal I}{\textrm{exponent}}$, respectively, then  $\basel{\mathcal I}{\textrm{exponent}}$ can be chosen to be $\basel{\log}{a} \circ ~ \basel{\mathcal I}{\textrm{base}}$, applied from right to left in the written order. For porting a subexpression involving addition operation in $\scalars$, such as, for example,  $f(\xx) \in \polynomials{\scalars}{x}{m}$, where $f(\xx) \neq 0$, for $\xx \in \mygrp^{m}$, where $\mygrp = \nonzeroscalars$, occurring in a base level expression to an exponentiation level,  the base level subexpression is replaced by a supplementary variable $z$, which is ported to first exponentiation level by the discrete logarithm mapping. In the subsequent levels of exponentiation, the interpretation is performed by applying ring homomorphisms, as  discussed in section \ref{Sec-modular-exponentiation-over-Zn}.

\section{Main Results}

\subsection{\label{Sec-parametric-injective-mappings}Parametric Injective Mappings}

 Let $\eltSet$ be either $\scalars$ or $\finiteIntegerRing{\cnt}$.  Let $\mygrp \subseteq \eltSet$ be the domain of interpretation for the variables occurring in the mappings.   For  $l \in \{0\} \cup \PositiveIntegers$ and  $m \in \PositiveIntegers$. a parametric multivariate injective mapping $\eta\bglb \basel{z}{1}, \, \ldots,\, \basel{z}{l};\, (\basel{x}{1},\,\ldots,\, \basel{x}{m})\bgrb$ from $\mygrp^{m}$ into $\eltSet^{m}$ is a multivariate injective mapping, which is an expression from either $\multipolynomials{\eltSet}{x}{z}{m}{l}$ or $\multiexpressions{\eltSet}{x}{z}{m}{l}$ with interpretation conventions as discussed in sections \ref{Sec-modular-exponentiation-over-Zn}-\ref{Sec-modular-exponentiation-over-Finite-Fields}, as appropriate, for $(\basel{x}{1},\, \ldots,\, \basel{x}{m}) \in \mygrp^{m}$ and  $(\basel{z}{1}, \, \ldots,\, \basel{z}{l}) \in {\mathcal Z} \subseteq \eltSet^{l}$,  and its parametric inverse $\eta^{-1}\bglb \basel{z}{1}, \, \ldots,\, \basel{z}{l};\, (\basel{y}{1},\,\ldots,\, \basel{y}{m})\bgrb$ is such that, for every fixed $(\basel{z}{1}, \, \ldots,\, \basel{z}{l}) \in {\mathcal Z} $, the following holds:  if $\eta\bglb\basel{z}{1}, \, \ldots,\, \basel{z}{l};\, (\basel{x}{1},\,\ldots,\, \basel{x}{m})\bgrb$
 $ = $
 $(\basel{y}{1},\, \ldots,\, \basel{y}{m})$, then $(\basel{x}{1},\,\ldots,\, \basel{x}{m}) =  \eta^{-1}\bglb\basel{z}{1}, \, \ldots,\, \basel{z}{l};\, (\basel{y}{1},\, \ldots,\, \basel{y}{m})\bgrb$, for every $(\basel{x}{1},\, \ldots,\, \basel{x}{m}) \in \mygrp^{m}$ and  $(\basel{y}{1},\, \ldots,\, \basel{y}{m}) \in \eltSet^{m}$. For example, let $\cnt$ be the set cardinality of $\mygrp = \nonzeroscalars$, $a \in \nonzeroscalars$ be a fixed primitive element, which is made known in the public key, and $\eta\bglb\basel{z}{1}, \, \ldots,\, \basel{z}{l};\, x \bgrb = f(\basel{z}{1},\, \ldots,\, \basel{z}{l})x^{g(\basel{\log}{a}(\basel{z}{1}),\, \ldots,\, \basel{\log}{a}(\basel{z}{l}))}$, where {\small{$f(\basel{z}{1},\, \ldots,\, \basel{z}{l}) \in \Expressions{\scalars}{z}{l}$}} and {\small{$g(\basel{t}{1},\, \ldots,\, \basel{t}{l})\in \Expressions{\basel{\Integers}{\orderofthegroup}}{t}{l}$}} are such that {\small{$f(\basel{z}{1},\, \ldots,\, \basel{z}{l}) \neq 0$}}, for {\small{$\basel{z}{1},\, \ldots,\, \basel{z}{l} \in \nonzeroscalars$}}, and {\small{$\gcd\bglb g(\basel{t}{1},\, \ldots,\, \basel{t}{l}),\, \cnt) = 1$}}, for {\small{$\basel{t}{1},\, \ldots,\, \basel{t}{l} \in \basel{\Integers}{\cnt}$}}. Then, {\small{$\eta\bglb\basel{z}{1}, \, \ldots,\, \basel{z}{l};\, x\bgrb$}} is a parametric bijective mapping from $\nonzeroscalars$ into $\nonzeroscalars$, with $\basel{z}{1},\, \ldots,\, \basel{z}{l} \in \nonzeroscalars$ as parameters, and {\small{$\eta^{-1}\bglb\basel{z}{1}, \, \ldots,\, \basel{z}{l};\, x\bgrb $
 $ = $
 $[~[f(\basel{z}{1},\, \ldots,\, \basel{z}{l})]^{-1}x~]^{[~[g(\basel{\log}{a}(\basel{z}{1}),\, \ldots,\, \basel{\log}{a}(\basel{z}{l}))]^{-1}\mod \cnt~]}$}}.

\subsubsection{\label{sec-parametrization-of-permutation-polynomials}Parametrization Methods}

Let, for some positive integers $k$, $l$ and $m$, 
$\basel{g}{i}\bglb\basel{z}{1},\, \ldots,\, \basel{z}{l}\bgrb$, $1 \leq i \leq k$,
be a partition of unity of $\eltSet^{l}$, {\em i.e.},
$\sum_{i = 1}^{k} \basel{g}{i}\bglb\basel{z}{1},\, \ldots,\, \basel{z}{l}\bgrb = 1$
and $ \basel{g}{i}\bglb\basel{z}{1},\, \ldots,\, \basel{z}{l}\bgrb
\cdot \basel{g}{j}\bglb\basel{z}{1},\, \ldots,\, \basel{z}{l}\bgrb$
$  = 0$,
$i \neq j$, $1 \leq i,\, j \leq k$, for every
$\bglb\basel{z}{1},\, \ldots,\, \basel{z}{l}\bgrb \in \eltSet^{l}$.
Let $\basel{\zeta}{i}\bglb \basel{z}{1},\, \ldots,\, \basel{z}{l};\, \xx \bgrb$,
$1 \leq i \leq k$, $\xx = (\basel{x}{1}, \ldots,\, \basel{x}{m})$,
be parametric multivariate injective mappings from $\mygrp^{m}$ into $\eltSet^{m}$,
that may or may not depend on the parameters $ \basel{z}{1},\, \ldots,\, \basel{z}{l}$. Let
 $\basel{\phi}{i}\bglb \basel{z}{1},\, \ldots,\, \basel{z}{l}\bgrb$ and
$\, \basel{\chi}{i}\bglb \basel{z}{1},\, \ldots,\, \basel{z}{l}\bgrb $ 
 be expressions such that 
$\basel{\phi}{i}\bglb \basel{z}{1},\, \ldots,\, \basel{z}{l}\bgrb$ evaluates to invertible elements in $\eltSet$, for all $\bglb\basel{z}{1},\, \ldots,\, \basel{z}{l}\bgrb \in \eltSet^{l}$, $1 \leq i \leq k$. Then, the expression
{\small{$\eta (\basel{z}{1},\, \ldots,\, \basel{z}{l};\, \xx)$}}
$ =$
{\small{$\sum_{i = 1}^{k} \basel{g}{i}(\basel{z}{1},\, \ldots,\, \basel{z}{l}) \cdot
\basel{\phi}{i}(\basel{z}{1},\, \ldots,\, \basel{z}{l}) \cdot
[\basel{\zeta}{i}(\basel{z}{1},\, \ldots,\, \basel{z}{l};\, \xx) +
\basel{\chi}{i}(\basel{z}{1},\, \ldots,\, \basel{z}{l})]$}}
is a parametric multivariate injective mapping, with its
parametric inverse
{\small{$\eta^{-1}(\basel{z}{1},\, \ldots,\, \basel{z}{l};\, \xx)$}}
$ =$
{\small{$\sum_{i = 1}^{k} \basel{g}{i}(\basel{z}{1},\, \ldots,\, \basel{z}{l}) \cdot
\basel{\zeta^{-1}}{i}(\basel{z}{1},\, \ldots,\, \basel{z}{l};\,  \basel{\yy}{i})$}\,,}
where 
{\small{ $\basel{y}{i,\, j} = [\basel{\phi}{i}(\basel{z}{1},\, \ldots,\, \basel{z}{l})]^{-1}
 \cdot \basel{x}{j} - \basel{\chi}{i}(\basel{z}{1},\, \ldots,\, \basel{z}{l})$,
 $1 \leq j \leq m$},} 
 {\small{ $\xx= (\basel{x}{1},\, \ldots,\, \basel{x}{m})$},} and
{\small{ $\basel{\yy}{i} = (\basel{y}{i,\, 1},\, \ldots,\, \basel{y}{i,\, m})$,
 $1 \leq i \leq k$}.} For public key cryptography hashing keys, it is possible to construct parametric multivariate injective mappings in section \ref{Sec-HashKeys} with any expressions $\basel{\phi}{i}(\basel{z}{1},\, \ldots,\, \basel{z}{l})$ that evaluate to only invertible elements, $1 \leq i \leq k$, having only a small number of terms. For digital signature hashing keys in section \ref{Sec-HashKeys}, however, the multivariate expressions $\basel{\phi}{i}(
 \basel{z}{1},\, \ldots,\, \basel{z}{l}) \neq 0$, $1 \leq i \leq k$, must be such that both $\basel{\phi}{i}(\basel{z}{1},\, \ldots,\, \basel{z}{l})$ and its multiplicative inverse $[\basel{\phi}{i}(\basel{z}{1},\, \ldots,\, \basel{z}{l})]^{-1}$ are expressible with only a small number of terms each.

\subsubsection{\label{Sec-partition-of-unity}Partition of Unity of $\scalars$}

 Let $f(z) \in \singlevariableexpressions{\scalars}{z}$, which is
called a discriminating function, and let $\basel{K}{f}$ be the
codomain of $f$, {\em i.e.}, $\basel{K}{f} =
 \{f(x)\,:\, x \in \scalars\} = \{\basel{a}{i}\,:\, 1 \leq i \leq k\}$,
 for some positive integer $k$.
Let {\small{$\basel{\ell}{i}(x) = \bigg[\prod_{\tiny{
 \begin{array}{c}   j = 1\\  j \neq i \end{array} } }^{k} 
\bglb \basel{a}{i}-\basel{a}{j}\bgrb\bigg]^{-1}\cdot
  \prod_{\tiny{
 \begin{array}{c}  j = 1\\  j \neq i \end{array} } }^{k} 
\bglb f(x)-\basel{a}{j}\bgrb$},} $1 \leq i \leq k$.
Then, $\basel{\ell}{i}(x) = 1$, for $x \in \basel{E}{i} = \{z \in \scalars\,:\, f(z) - \basel{a}{i} = 0\}$,
and $\basel{\ell}{i}(x) = 0$, for $x \in \scalars \backslash \basel{E}{i}$, $1 \leq i \leq k$.
Thus, $\{\basel{E}{i}\,:\,1 \leq i \leq k\}$ is a partition of $\scalars$, and $\basel{\ell}{i}(x)$
is the characteristic function of the equivalence class $\basel{E}{i}$, $1 \leq i \leq k$.
Now, the set
 $\{ \basel{g}{i}(\basel{z}{1},\, \ldots,\, \basel{z}{l})
= \basel{\ell}{i}\bglb h(\basel{z}{1},\, \ldots,\, \basel{z}{l})\bgrb\,:\,
 1 \leq i \leq k\}$, where
 $h(\basel{z}{1},\, \ldots,\, \basel{z}{l}) \in \Expressions{\scalars}{z}{{\mathit l}}$,
is a partition of unity of $\scalars^{l}$.

\vspace*{-0.2cm}

\begin{small}
\paragraph{\small{\bf{Examples.}}}   (A) ~  Let the vector space dimension of $\scalars$ be $n$
 as an extension field of $\basel{\Integers}{\primeInt}$, and
 let $f(z) = \sum_{i = 1}^{n}\basel{a}{i}z^{\primeInt^{i-1}}$, 
 where $\basel{a}{i} \in \scalars$, $1 \leq i \leq n$,
 be a noninvertible linear operator from $\scalars$ into $\scalars$,
 with $\basel{\Integers}{\primeInt}$ as the field.
 For every linear operator $T$ from $\scalars$ into $\scalars$
 with $\basel{\Integers}{\primeInt}$ as the field, there exist
 scalars $\basel{c}{i} \in \scalars$, $1 \leq i \leq n$,
 such that $Tz = \sum_{i=1}^{n}\basel{c}{i}z^{\primeInt^{i-1}}$
 \cite{LN:1986}. Now, each equivalence class is an affine
 vector subspace of the form $\{y+x\,:\, f(x) = 0, ~x \in \scalars\}$,
 for some $y \in \scalars$. Thus, if $r$ is the rank of $f$
 as linear operator from $\scalars$ into $\scalars$ with
 $\basel{\Integers}{\primeInt}$ as the field, then the nullity
 of $f$ is $n-r$, each equivalence class has $\primeInt^{n-r}$
 elements, and there are $k = \primeInt^{r}$ equivalence classes.
 For the number of equivalence classes to be small, the rank $r$
 of $f$ must be small, such as $r = 1$ or $r = 2$.
 ~~(B) ~ Let $f(z) = z^{r}$,  where  $r$ is a large positive integer
 dividing $\primeInt^{n}-1$.  Now, the equivalence classes are $\{0\}$
 and the cosets of the congruence relation $x \sim y$ if and only if
 $( x^{-1}y )^{r} = 1$, for $x,\, y \in \scalars \backslash \{0\}$.
 Since $\basel{K}{f} = \{0\} \cup \{z^{r}\,:\, z \in \scalars \backslash \{0\}\}$, 
 there are $k = 1+(\primeInt^{n}-1)/r$ equivalence classes. 
 
\end{small}

\subsubsection{\label{partition-of-unity-of-Zn}Partition of Unity of $\basel{\Integers}{\primeInt^{l}}$}

Let $s \in \PositiveIntegers$ be a divisor of $(\primeInt-1)$  and $k = 1+\frac{(\primeInt-1)}{s}$. Now, $\primeInt^{l-1} \geq l$, for any $l \in \PositiveIntegers$ and prime number $\primeInt$. Let $h(x) = x^{s\primeInt^{l-1}}$, for $x \in \basel{\Integers}{\primeInt^{l}}$. Then, $\bglb h(x) \bgrb^{k-1} = 1$, for $x \in \basel{\Integers^{\star}}{\primeInt^{l}}$, and $h(x) = 0$, for $x \in \basel{\Integers}{\primeInt^{l}} \backslash \basel{\Integers^{\star}}{\primeInt^{l}}$.  Thus, the set $\{x^{s\primeInt^{l-1}}\,:\, x \in \basel{\Integers}{\primeInt^{l}}\}$ contains $k$ distinct elements. Let $x, \, y \in \basel{\Integers}{\primeInt^{l}}$ be such that $h(x) \neq h(y)$. If $h(x) = 0$ or $h(y) = 0$, then $(h(y)-h(x))  \in \basel{\Integers^{\star}}{\primeInt^{l}}$. Now, let $x,\, y \in \basel{\Integers^{\star}}{\primeInt^{l}}$. If $(x^{-1}y)^{s\primeInt^{l-1}} = 1+b\primeInt^{t}$, for some $b \in \basel{\Integers^{\star}}{\primeInt^{l}}$ and $t \in \PositiveIntegers$, then, since  $1 + b \primeInt^{t} \sum_{i = 1}^{k-1} \frac{(k-1)!}{i!(k-i-1)!}  b^{i-1}\primeInt^{(i-1)t} = 
(1+b\primeInt^{t})^{k-1} = \bglb(x^{-1}y)^{s\primeInt^{l-1}}\bgrb^{k-1} = 1 \mod \primeInt^{l}$, it follows that either $t \geq l$ or  $(k-1)+\sum_{i = 2}^{k-1} \frac{(k-1)!}{i!(k-i-1)!} b^{i-1}\primeInt^{(i-1)t} = 0 \mod \primeInt^{l-t}$. However, since $k = 1+\frac{\primeInt-1}{s}$, and therefore, $1 \leq k-1 \leq \primeInt-1$, it follows that $(k-1)+\sum_{i = 2}^{k-1} \frac{(k-1)!}{i!(k-i-1)!} b^{i-1}\primeInt^{(i-1)t} = k-1 \mod \primeInt$. Thus, if  $x,\, y \in \basel{\Integers^{\star}}{\primeInt^{l}}$ and 
$h(x) \neq h(y)$, then $(x^{-1}y)^{s\primeInt^{l-1}} - 1  \neq 0 \mod\primeInt$, and hence if  $x,\, y \in \basel{\Integers}{\primeInt^{l}}$ and  $h(x) \neq h(y)$, then $(h(y)-h(x)) \in \basel{\Integers^{\star}}{\primeInt^{l}}$. If $\basel{a}{j} \in \basel{\Integers}{\primeInt^{l}}$, $1 \leq j \leq k$, are such that $\{x^{s\primeInt^{l-1}}\,:\, x \in \basel{\Integers}{\primeInt^{l}}\} = \{\basel{a}{j}\,:\, 1 \leq j \leq k\}$, then $(\basel{a}{i}-\basel{a}{j}) \in \basel{\Integers^{\star}}{\primeInt^{l}}$, for $i \neq j$, $1 \leq i,\, j \leq k$, and the Lagrange interpolation polynomials  $\basel{g}{j}(x) \in \singlevariablepolynomials{\basel{\Integers}{\primeInt}}{x}$ can be obtained for the equivalence classes $\basel{E}{j} = \{x^{s\primeInt^{l-1}} = \basel{a}{j}\,:\, x \in \basel{\Integers}{\primeInt^{l}}\}$.  Thus, corresponding to every homomorphism of $\basel{\Integers^{\star}}{\primeInt}$ into $\basel{\Integers^{\star}}{\primeInt}$, a partition of unity of $\basel{\Integers}{\primeInt^{l}}$ can be obtained.

\subsubsection{\label{Sec-multivariate-polynomials-that-evaluate-to-only-invertible-elements}Multivariate Polynomials that Evaluate to only Invertible Elements}

Let $f(z) \in \singlevariablepolynomials{\scalars}{z}$ be a polynomial which is not surjective as a mapping from $\scalars$ into $\scalars$. Then, there exists an element $c \in \scalars$, such that $f(z)-c \neq 0$, for every $z \in \scalars$. For $a \in \scalars \backslash \{0\}$ and $g(\basel{z}{1},\, \ldots,\, \basel{z}{l}) \in \polynomials{\scalars}{z}{l}$,  $a\bglb f({\small{g(\basel{z}{1},\, \ldots,\, \basel{z}{l})}}) -c \bgrb \neq 0$, for every $(\basel{z}{1},\, \ldots,\, \basel{z}{l}) \in \scalars^{l}$.

\vspace*{-0.2cm}

\begin{small}
\paragraph{\small{\bf{Examples.}}} (A)~ Let $f(z)$ be a product of irreducible polynomials
in $\singlevariablepolynomials{\scalars}{z}$ of degree $2$ or more each. 
Then, $c$ can be chosen to be $0$.~~
(B)~ Let the vector space dimension of $\scalars$ be $n$
as an extension field of $\basel{\Integers}{\primeInt}$, and
let $f(z) = \sum_{i = 1}^{n}\basel{a}{i}z^{\primeInt^{i-1}}$, 
where $\basel{a}{i} \in \scalars$, $1 \leq i \leq n$,
be a noninvertible linear operator from $\scalars$ into $\scalars$,
with $\basel{\Integers}{\primeInt}$ as the field. Then,
for any basis $\{\basel{\alpha}{1},\, \ldots,\, \basel{\alpha}{n} \}$
for $\scalars$, with $\basel{\Integers}{\primeInt}$ as the field,
there exists an index $j$, $1 \leq j \leq n$, such that 
$\sum_{i = 1}^{n}\basel{a}{i}z^{\primeInt^{i-1}} - \basel{\alpha}{j}  
\neq 0$, for every $z \in \scalars$, and $c$ can be taken
to be $\basel{\alpha}{j}$.~~
(C) ~ Let $r \geq 2$ be a positive integer divisor of $\primeInt^{n}-1$,
and let $f(z) = z^{r}$. Then, there exists an element
$c \in \scalars \backslash \{0\}$, such that $c^{(\primeInt^{n}-1)/r} \neq 1$.
Now, since $c^{(\primeInt^{n}-1)/r} \neq 0$ and
$c^{(\primeInt^{n}-1)/r} \neq 1$, it follows that
 $f(z) - c \neq 0$, for every $z \in \scalars$.
\\

\end{small}

If $f(z) \in \singlevariablepolynomials{\scalars}{z}$
is such that $f(z) \neq 0$, for every $z \in \scalars$, then
$[f(z)]^{-1} = \sum_{i = 1}^{k} \basel{a^{-1}}{i}\basel{\ell}{i}(z)$, where $\{\basel{a}{i}\,:\, 1 \leq i \leq k\}  = 
 \{f(z)\,:\, z \in \scalars\} $, and $\basel{\ell}{i}(z) = 
\bigg[\prod_{\tiny{ \begin{array}{c}   j = 1\\  j \neq i \end{array} } }^{k} 
\bglb \basel{a}{i}-\basel{a}{j}\bgrb\bigg]^{-1}\cdot
  \prod_{\tiny{
 \begin{array}{c}  j = 1\\  j \neq i \end{array} } }^{k} 
\bglb f(z)-\basel{a}{j}\bgrb$, $1 \leq i \leq k$.
 Thus,  for digital signature hashing keys in section \ref{Sec-HashKeys}, the appropriate choices for a nonvanishing function $f(z) \neq 0$, $z \in \scalars$, are those similar to the choice of discriminating functions discussed at the  end of section \ref{Sec-partition-of-unity}.
 
Let $\cnt = \prod_{i = 1}^{r} \basel{\primeInt^{\basel{l}{i}}}{i}$, where $r \in \PositiveIntegers$, $\basel{l}{i} \in \PositiveIntegers$ and $\basel{\primeInt}{i}$ are distinct prime numbers, for $1 \leq i \leq r$,  and  $f(z) \in \singlevariablepolynomials{\finiteIntegerRing{\cnt}}{z}$.
From section \ref{Sec-polynomials-over-Zn}, it can be recalled that, $f(z) \in \invertibleRingElements{\cnt}$, for $z \in \finiteIntegerRing{\cnt}$, if and only if for every $i$, where $1 \leq i \leq r$, $f(z) \mod \basel{\primeInt}{i} \in \invertibleRingElements{\basel{\primeInt}{i}}$, for $z \in \finiteIntegerRing{\cnt}$.

\subsection{\label{Sec-nonparametric-univariate-bijective-mappings}Univariate Bijective Mappings without Parameters}

\subsubsection{Single Variable Permutation Polynomials without Hashing} 

{\underline{\small{\bf{Examples in} $\singlevariablepolynomials{\scalars}{x}$}}}~~
Bijective mappings in  $\singlevariablepolynomials{\scalars}{x}$, also called  permutation polynomials, are extensively studied as Dickson polynomials \cite{Dickson:1897} in the literature. A comprehensive survey on Dickson polynomials can be found in \cite{AG:1991, GM:1994,  LMT:1993,  Mullen:2000,  MN:1987}. Some recent results are
presented in \cite{AAW:2008,  AGW:2009,  AW:2005}. If $f(z) \in \singlevariablepolynomials{\scalars}{z}$
is a permutation polynomial, then, for every $a \in \scalars \backslash \{0\}$, 
$b \in \scalars$ and nonnegative integer $i$, the polynomial $af(z^{\primeInt^{i}})-b$
is a permutation polynomial. Some easy examples are described in the following.

\vspace*{-0.2cm}

\begin{small}
\paragraph{\small{\bf{Examples.}}}   (A) ~ Let $\scalars$ be a finite dimensional
extension field of $\basel{\Integers}{\primeInt}$ of vector space dimension $n$. Any polynomial  $f(z) = \sum_{i = 1}^{n}\basel{a}{i}z^{\primeInt^{i-1}}$, where $\basel{a}{i} \in \scalars$, $1 \leq i \leq n$, that is an invertible linear operator from $\scalars$ onto $\scalars$, with
$\basel{\Integers}{\primeInt}$ as the field, is a permutation polynomial.~~
(B)~ Let $r$ be a positive integer divisor of $n$,
and $f(z) = z^{^{\primeInt^{r}}} - a z$,
where $a^{^{(\sum_{i = 1}^{n/r}\primeInt^{(i-1)r})}} \neq 1$.
Then, for every  $z \in \scalars \backslash \{0\}$,
$z^{^{(\primeInt^{r}-1)}}-a \neq 0$, since 
$z^{^{\primeInt^{n}}-1}  =  z^{^{(\primeInt^{r}-1)
\sum_{i = 1}^{n/r}\primeInt^{(i-1)r}}} = 1$,  and
therefore, the null space of $f(z)$, as a linear operator
from $\scalars$ into $\scalars$ with $\basel{\Integers}{\primeInt}$
as the field, is $\{0\}$. Thus, $f(z)$ is a permutation polynomial.~~
(C) ~Let $r$ be a positive integer relatively prime to $(\primeInt^{n}-1)$. Then, the polynomial $f(z) = z^{r}$ is a permutation polynomial.
\\

\end{small}

\noindent
{\underline{\small{\bf{Examples in} $\singlevariablepolynomials{\finiteIntegerRing{\primeInt^{l}}}{x}$}}}~~ Let $l \in \PositiveIntegers$ and $\primeInt$ be a prime number. For any positive integer $n$, Dickson polynomials that are permutation polynomials, having nonvanishing derivatives over the finite field containing $\primeInt^{n}$ elements, are found in \cite{AG:1991, AAW:2008, AGW:2009,  AW:2005, GM:1994, LMT:1993, Mullen:2000, MN:1987}. For a small prime number $\primeInt$, two methods for construction of permutation polynomials $f(x) \in \singlevariablepolynomials{\finiteIntegerField}{x}$, such that $f'(x) \neq 0 \mod \primeInt$, are described below. As a set, $\finiteIntegerField$ is taken to be the set of integers $i$, where $0 \leq i \leq \primeInt-1$. For $\primeInt = 2$, the only permutation polynomials are $f(x) = x$ and $f(x) = x-1$, and in both cases, $f'(x) = 1 \mod 2$. Now, let $\primeInt \geq 3$ be a small prime number, such that the computations below are not difficult for implementation.   Let {\small{$\basel{\ell}{i}(x) = \bgls \prod_{{\tiny{\begin{array}{c} j = 0\\ j \neq i \end{array}}}}^{\primeInt-1} (i-j)\bgrs^{-1} \cdot
 \prod_{{\tiny{\begin{array}{c} j = 0\\ j \neq i \end{array}}}}^{\primeInt-1} (x-j) = 
- \prod_{{\tiny{\begin{array}{c} j = 0\\ j \neq i \end{array}}}}^{\primeInt-1} (x-j) $},} for $i \in \finiteIntegerField$. Now, {\small{$\basel{\ell'}{i}(x) = - \sum_{{\tiny{\begin{array}{c} j = 0\\ j \neq i \end{array}}}}^{\primeInt-1}   \prod_{{\tiny{\begin{array}{c} k = 0\\ k \not \in \{i,\, j\} \end{array}}}}^{\primeInt-1} (x-k)$, for $i \in \finiteIntegerField$}~,}~ which implies that {\small{ $\basel{\ell'}{i}(j) =  - \prod_{{\tiny{\begin{array}{c} k = 0\\ k \not \in \{i,\, j\} \end{array}}}}^{\primeInt-1} (j-k) ~ = ~ (j-i)^{-1}$~,~~ for $j \neq i$ and $j \in \finiteIntegerField$},} and  {\small{$\basel{\ell'}{i}(i) =   - \sum_{{\tiny{\begin{array}{c} j = 0\\ j \neq i \end{array}}}}^{\primeInt-1}   \prod_{{\tiny{\begin{array}{c} k = 0\\ k \not \in \{i,\, j\} \end{array}}}}^{\primeInt-1} (i-k) ~ = ~ \sum_{{\tiny{\begin{array}{c} j = 0\\ j \neq i \end{array}}}}^{\primeInt-1}  (i-j)^{-1} = 0~$},}  for $i \in \finiteIntegerField$, since $\primeInt \geq 3$. For a fixed permutation sequence $\{\basel{a}{i} \in \finiteIntegerField \,:\, 0 \leq i \leq \primeInt-1\}$ of  $\finiteIntegerField$, either of the two procedures described below constructs a permutation polynomial in $f(x) \in \singlevariablepolynomials{\finiteIntegerField}{x}$, such that $f(i) = \basel{a}{i}$ and $f'(i) \not \equiv 0 \mod \primeInt$, for $i \in \finiteIntegerField$.

  \noindent
{\small{\fbox{\sf{Method 1}}}}~~
Let $\sum_{i = 0}^{\primeInt-1} \basel{a}{i} \basel{\ell}{i}(x) = \basel{b}{0} + \sum_{i = 1}^{\primeInt-1} \basel{b}{i} x^{i}$, for some $\basel{b}{i} \in \finiteIntegerField$, for $0 \leq i \leq \primeInt-1$, and let $g(x) = \basel{c}{1} + \sum_{i=2}^{\primeInt-1} \basel{c}{i} x^{i-1}$, for some $\basel{c}{i} \in \finiteIntegerField$, for $1 \leq i \leq \primeInt-1$, be such that $g(x) \not \equiv 0 \mod \primeInt$, for every $x \in \finiteIntegerField$. Let $\basel{\rho}{i} = i^{-1} \basel{c}{i}$ and $\basel{\sigma}{i} = \basel{b}{i}-\basel{\rho}{i}$, for $1 \leq i \leq \primeInt-1$. Let $f(x) = \basel{b}{0} + \sum_{i = 1}^{\primeInt-1} (\basel{\rho}{i} x^{i} + \basel{\sigma}{i}x^{i\primeInt})$. Then, $f(x) \equiv \basel{b}{0} + \sum_{i = 1}^{\primeInt-1} \basel{b}{i} x^{i} \mod \primeInt$, for every $x \in \finiteIntegerField$, and $f'(x) \equiv \basel{\rho}{1} + \sum_{i = 2}^{\primeInt-1} i \basel{\rho}{i} x^{i-1} \equiv  \basel{c}{1} + \sum_{i = 2}^{\primeInt-1}  \basel{c}{i} x^{i-1} \mod \primeInt$, for every $x \in \finiteIntegerField$, satisfying the stated requirement.

  \noindent
{\small{\fbox{\sf{Method 2}}}}~~
Let $\basel{b}{i}, \, \basel{c}{i},\, \sigma \in \finiteIntegerField$, for $ 0 \leq i \leq \primeInt-1$, be such that $\basel{b}{0} = \basel{a}{0}$ and $\basel{b}{j}+\basel{c}{j} = \basel{a}{j}$, for $1 \leq j \leq \primeInt-1$, and let $f(x) =  \sum_{i = 0}^{\primeInt-1} (\basel{b}{i} + x^{\primeInt-1}\basel{c}{i} - \sigma i)\basel{\ell}{i}(x) + \sigma x^{\primeInt}$. It can be immediately verified that $f(i) \equiv \basel{a}{i} \mod \primeInt$, for $0 \leq i \leq \primeInt-1$, and $f'(x) =  \sum_{i = 0}^{\primeInt-1} (\basel{b}{i}  +  x^{\primeInt-1} \basel{c}{i} - \sigma i) \basel{\ell'}{i}(x) + \primeInt \sigma x^{\primeInt-1} + (\primeInt-1)x^{\primeInt-2} \sum_{i = 0}^{\primeInt-1} \basel{c}{i}\basel{\ell}{i}(x)$, where $\primeInt \geq 3$. Thus, the parameters $\basel{c}{0}$, $\sigma$, $\basel{b}{j}$ and $\basel{c}{j}$, for $1 \leq j \leq \primeInt-1$, need to be chosen such that $f'(x) \not \equiv 0 \mod \primeInt$, for all $x \in \finiteIntegerField$. Now, $f(x)+ \sigma x =  \sum_{i = 0}^{\primeInt-1} (\basel{b}{i} + \basel{c}{i} x^{\primeInt-1}) \basel{\ell}{i}(x) + \sigma x^{\primeInt}$, and $f'(x)+\sigma =  \sum_{i = 0}^{\primeInt-1} (\basel{b}{i} + \basel{c}{i} x^{\primeInt-1}) \basel{\ell'}{i}(x) + \primeInt \sigma x^{\primeInt-1} + (\primeInt-1)x^{\primeInt-2} \sum_{i = 0}^{\primeInt-1} \basel{c}{i}\basel{\ell}{i}(x)$.  Thus,  $f'(0)+\sigma \equiv  - \sum_{i = 1}^{\primeInt-1} i^{-1} \basel{b}{i} \mod \primeInt$ and $f'(j) +\sigma \equiv  \sum_{{\tiny{\begin{array}{c} i = 0\\ i \neq j \end{array}}}}^{\primeInt-1} \basel{a}{i} (j-i)^{-1} + \basel{c}{0} j^{-1} - j^{-1} \basel{c}{j} \mod \primeInt$, for $1 \leq j \leq \primeInt-1$, which implies that every element in the sequence of numbers $(f'(i)+\sigma) \mod \primeInt$, for $0 \leq i \leq \primeInt-1$, is independent of the choice of $\sigma$, and the condition that $f'(i) \not \equiv 0 \mod \primeInt$, for $0 \leq i \leq \primeInt-1$, is equivalent to that $\sigma \not \in \{(f'(i)+\sigma) \mod \primeInt\,:\, 0 \leq i \leq \primeInt-1\}$. For $\primeInt \geq 3$, $\sum_{i = 0}^{\primeInt-1} i \equiv \sum_{i = 0}^{\primeInt-1} 1 \equiv 0 \mod \primeInt$, and since $\finiteIntegerField$ is the splitting field of the polynomial $x^{\primeInt} - x = \prod_{i = 0}^{\primeInt-1}(x-i)$, the elementary symmetric polynomials $\basel{s}{r}(\basel{t}{1},\, \basel{t}{2},\, \ldots,\, \basel{t}{n})$, which are homogeneous of degree $r$ in $n$ variables, for the particular instances of parameters $n = \primeInt$ and $\basel{t}{i}= i-1$, for $1 \leq i \leq \primeInt$, as defined in \cite{Lang:2002}, are all congruent to $0 \mod \primeInt$, for $1 \leq r \leq \primeInt-2$. Thus, $\sum_{i = 0}^{\primeInt-1} i^{r} \equiv \sum_{i = 0}^{\primeInt-1} 1 \equiv 0 \mod \primeInt$, for $r \in \PositiveIntegers$, $1 \leq r \leq \primeInt-2$ and $\primeInt \geq 3$, which implies that for a nonzero polynomial $g(x) \in \singlevariablepolynomials{\finiteIntegerField}{x}$ of degree at most $\primeInt-2$, $\sum_{i = 0}^{\primeInt-1} g(i) \equiv 0 \mod \primeInt$. Now, $\primeInt \sum_{i = 0}^{\primeInt-1} i^{\primeInt-1} \equiv 0 \mod \primeInt$, and, for $l \in \PositiveIntegers$, such that $\primeInt+1 \le l \leq 2\primeInt-2$, $l\sum_{i = 0}^{\primeInt-1} i^{l-1} \equiv l\sum_{i = 0}^{\primeInt-1} i^{l-1-(\primeInt-1)} \equiv  l\sum_{i = 0}^{\primeInt-1} i^{l-\primeInt} \equiv 0 \mod \primeInt$, since $1 \leq l-\primeInt \leq \primeInt-2$. Thus, for a nonzero polynomial $h(x) \in \singlevariablepolynomials{\finiteIntegerField}{x}$ of degree at most $2\primeInt-2$, $\sum_{i = 0}^{\primeInt-1} h'(i) \equiv 0 \mod \primeInt$. The coefficients $\basel{c}{i}$, for $0 \leq i \leq \primeInt-1$, must be so chosen that the additional requirement that $f(x)+\sigma x$ is a polynomial of degree at most $2\primeInt-2$ can also be fulfilled. Now, let $\basel{\lambda}{i} \in \finiteIntegerField$, for $0 \leq i \leq \primeInt-1$, be chosen, such that the cardinality of the set $\Lambda = \{\basel{\lambda}{i}\,:\, 0 \leq i \leq \primeInt-1\}$ is at most $\primeInt-1$ and $\sum_{i = 0}^{\primeInt-1} \basel{\lambda}{i} = 0$. Then, $\basel{c}{j}-\basel{c}{0}$ are found from the condition $f'(j)+\sigma = \sum_{{\tiny{\begin{array}{c} i = 0\\ i \neq j \end{array}}}}^{\primeInt-1} \basel{a}{i} (j-i)^{-1}  - j^{-1} (\basel{c}{j}-\basel{c}{0}) = \basel{\lambda}{j}$, for $1 \leq j \leq \primeInt-1$, and hence, $ f'(0)+\sigma =  - \sum_{i = 1}^{\primeInt-1} i^{-1} \basel{b}{i}  = \basel{\lambda}{0}$, for all choices of $\basel{c}{0}$. Now, let $\sigma$ be chosen from $\finiteIntegerField \backslash \Lambda$, where the latter set is nonempty, since the cardinality of $\Lambda$ is at most $\primeInt-1$, by the choices of $\basel{\lambda}{i}$, for $0 \leq i \leq \primeInt-1$. Finally, $\basel{c}{0}$ is chosen, and $\basel{b}{j}$ and $\basel{c}{j}$, for $1 \leq j \leq \primeInt-1$, are determined by the aforementioned conditions.
 
 For a small prime number $\primeInt$, positive integers $l$ and $r$, such that $l \geq 2$ and $1 \leq r \leq l$, a bijective mapping $f(x) \in \singlevariablepolynomials{\finiteIntegerRing{\primeInt^{l}}}{x}$ and $y \in \finiteIntegerRing{\primeInt^{l}}$, the following procedure computes $\basel{x}{r} \in \finiteIntegerRing{\primeInt^{r}}$, such that $\basel{f}{r}(\basel{x}{r}) \equiv y \mod \primeInt^{r}$, assuming $\basel{x}{1} \in \finiteIntegerField$ is known, such that $\basel{f}{1}(\basel{x}{1}) \equiv y \mod \primeInt$, where $\basel{f}{r}(x) =  f(x) \mod \primeInt^{r}$, applying the $\mod \primeInt^{r} ~ $ operation only to the coefficients. Let $2 \leq r \leq l$, where $l \geq 2$, $s \in \PositiveIntegers$ be such that $\left \lceil \frac{r}{2} \right \rceil \leq s \leq r-1$ and $\basel{y}{r} = y \mod \primeInt^{r} \in \finiteIntegerRing{\primeInt^{r}}$, and $\basel{x}{s} = \basel{f^{-1}}{s}(\basel{y}{r} \mod \primeInt^{s}) \in \finiteIntegerRing{\primeInt^{s}}$ has been computed. Let $\basel{\hat{x}}{s} \in \finiteIntegerRing{\primeInt^{r}}$ be such that $\basel{\hat{x}}{s} \equiv \basel{x}{s} \mod \primeInt^{s}$. Since $\basel{f}{r}(\basel{\hat{x}}{s}) \equiv \basel{y}{r} \mod \primeInt^{s}$, it follows that $\basel{f}{r}(\basel{\hat{x}}{s}) = \basel{y}{r} + \primeInt^{s} \basel{g}{r,\,s }(\basel{\hat{x}}{s},\, \basel{y}{r})$, for some mapping $\basel{g}{r,\, s}(\basel{\hat{x}}{s},\, \basel{y}{r})$, and therefore, $\basel{f}{r}\bglb {\small{\basel{\hat{x}}{s} +  [ \basel{f'}{r}(\basel{\hat{x}}{s})]^{-1} \cdot [ \basel{y}{r} - \basel{f}{r}(\basel{\hat{x}}{s}) ]}} \bgrb \equiv \basel{f}{r}({\small{\basel{\hat{x}}{s}}}) +  \basel{f'}{r}({\small{\basel{\hat{x}}{s}}}) \cdot \bgls \basel{f'}{r}({\small{\basel{\hat{x}}{s}}})\bgrs^{-1} \cdot \bgls \basel{y}{r} - \basel{f}{r}({\small{\basel{\hat{x}}{s}}}) \bgrs \equiv  \basel{f}{r}({\small{\basel{\hat{x}}{s}}}) +  \bgls \basel{y}{r} - \basel{f}{r}({\small{\basel{\hat{x}}{s}}}) \bgrs  \equiv \basel{y}{r} \mod \primeInt^{r}$. Thus, $\basel{f^{-1}}{r}(\basel{y}{r}) = \basel{\hat{x}}{s} +  \bgls \basel{f'}{r}(\basel{\hat{x}}{s})\bgrs^{-1} \cdot \bgls \basel{y}{r} - \basel{f}{r}(\basel{\hat{x}}{s}) \bgrs \mod \primeInt^{r}$. If $r = l$, then the $f^{-1}(y)$ is just computed for $y \in \finiteIntegerRing{\primeInt^{l}}$, and the procedure can be stopped; otherwise, the previous steps are repeated, replacing the current value of $r$ by $\min\{2r,\, l\}$. 
\\
 
\noindent
{\underline{\small{\bf{Examples in} $\singlevariableexpressions{\scalars}{z}$}}}~~
 Let $\scalars$ be a finite field of $\primeInt^{n}$ elements, for some prime number $\primeInt$ and $n \in \PositiveIntegers$, such that $\primeInt^{n} \geq 3$, and let $\cnt = \primeInt^{n}-1$. Let $t \geq 2$ be a positive integer divisor of $\primeInt^{n}-1$, and let $\basel{H}{t} = \{x^{t} = 1 \, : \, x \in \nonzeroscalars\}$. Let $f(x)\in \singlevariablepolynomials{\Integers}{x}$ be such that $f(x) \mod t$ yields a polynomial mapping from $\finiteIntegerRing{t}$ onto itself. It may be recalled that, as a set, $\basel{\Integers}{t}$ is assumed to consist of integers $i$, where $0 \leq i \leq t-1$. Let $a$ be a primitive element in $\nonzeroscalars$. Now, for $x \in \basel{H}{t}$, since $x^{t} = 1$, applying $\basel{\log}{a}$ operation on both sides, $t \basel{\log}{a} x = 0 \mod \cnt$, which implies that $\basel{\log}{a} x$ is an integer multiple of $\frac{\cnt}{t} = \frac{\primeInt^{n}-1}{t}$, for every $x \in \basel{H}{t}$, and, since the cyclic subgroup generated by $a^{\frac{\cnt}{t}}$ is $\basel{H}{t}$, it follows that $\basel{\log}{a}$ is a bijective mapping of $\basel{H}{t}$ onto $\frac{\cnt}{t}\cdot \finiteIntegerRing{\cnt} = \{(\frac{i\cnt}{t}) \mod \cnt\,:\, 0 \leq i \leq t-1\}$. Now, $f(\basel{\log}{a}(x)) \mod \cnt$, for $x \in \basel{H}{t}$, is an injective mapping, when restricted to $\basel{H}{t}$, which can be modified appropriately, by changing its constant term, if necessary, to obtain a polynomial $g$, which results in a bijective mapping from $\frac{\cnt}{t}\cdot \finiteIntegerRing{\cnt}$ into itself, with respect to $\mod \cnt$ operation.  Then, the mapping $\eta(x) = a^{g(\basel{\log}{a} x)}$, for $x \in \nonzeroscalars$, is such that its restriction to $\basel{H}{t}$ is a bijective mapping from $\basel{H}{t}$ onto itself.

\subsubsection{Hybrid Single Variable Permutation Polynomials with Hashing}

\noindent
{\small{\fbox{\sf{Method 1}}}}~~
Let $\basel{\ell}{i}(x) \in \singlevariablepolynomials{\scalars}{x}$, $1 \leq i \leq k$, where $k \in \PositiveIntegers$, $k \geq 2$, be indicator functions of a partition $\{\basel{S}{i}\,:\, 1 \leq i \leq k\}$  of $\scalars$. Let $\sigma$ be a permutation on $\{1,\, \ldots,\, k\}$, such that the set cardinalities of $\basel{S}{i}$ and $\basel{S}{\sigma(i)}$ are equal, for $1 \leq i \leq k$. Let $\basel{g}{i}$ be a mapping from $\scalars$ into $\scalars$, such that $\basel{g}{i}\bglb \basel{S}{i} \bgrb = \basel{S}{\sigma(i)}$, for $1 \leq i \leq k$. Thus, $\basel{g}{i}$ is one-to-one when restricted to $\basel{S}{i}$, for $1 \leq i \leq k$. Let $\eta(x) \in \singlevariablepolynomials{\scalars}{x}$ be a permutation polynomial, and $\chi(x) = \sum_{i = 1}^{k} \basel{\ell}{i}(x) \eta({\small{\basel{g}{i}(x)}})$. Then, $\chi(\scalars) = \bigcup_{i = 1}^{k} \eta\bglb \basel{g}{i}(\basel{S}{i})\bgrb = \bigcup_{i = 1}^{k} \eta\bglb \basel{S}{\sigma(i)} \bgrb$, and  since $\{\basel{S}{\sigma(i)}\,:\, 1 \leq i \leq k\}$ is a partition of $\scalars$, $\chi(x)$ is a surjective (hence bijective) polynomial from $\scalars$ onto $\scalars$. For inverting $\chi(x) = y$, for fixed $y \in \scalars$, let $\xi = \eta^{-1}(y)$. Now, there exists exactly one index $i$, where $1 \leq i \leq k$, such that $\xi \in \basel{S}{\sigma(i)} = \basel{g}{i}\bglb\basel{S}{i}\bgrb$, and therefore, the unique element $x \in \basel{S}{i}$, such that $x  = \basel{g^{-1}}{i}(\xi)$, satisfies $\chi(x) = y$. If $\basel{f}{i}$, for $1 \leq i \leq k$, are mappings from $\scalars$ into $\scalars$, such that $\basel{f}{i}(\basel{g}{i}(x)) = x$, for $x \in \basel{S}{i}$, then $\chi^{-1}(y) = \sum_{i = 1}^{k} \basel{\ell}{\sigma(i)} \bglb \eta^{-1}(y)\bgrb \basel{f}{i}\bglb \eta^{-1}(y) \bgrb$, for $y \in \scalars$. The case of bijective mappings in $\singlevariableexpressions{\scalars}{x}$ can be similarly discussed. In the following examples, the corresponding examples in section \ref{Sec-partition-of-unity} are  revisited.

\vspace*{-0.2cm}

\begin{small}
\paragraph{\small{\bf{Examples.}}}   (A) ~ Let $T(x) = \sum_{i = 1}^{n} \basel{a}{i}x^{\primeInt^{i-1}}$, $\basel{a}{i} \in \scalars$, $1 \leq i \leq n$,  be of rank $t$, where  $t$ is a small positive integer, such as $t \in \{1,\, 2\}$,  as  described in the first example in section \ref{Sec-partition-of-unity} and let $V = \{x \in \scalars \, : \, T(x) = 0\}$. Then, there exist $k = \primeInt^{t}$ representative elements $\basel{b}{i} \in \scalars$, $1 \leq i \leq k$, such that $\{T(\basel{b}{i})\,:\,  1 \leq i \leq k\} = T(\scalars)$, and  $\basel{S}{i} =   V + \basel{b}{i} = \{x + \basel{b}{i} \, : \, x \in V\}$,  $1 \leq i \leq k$.  Let $\basel{f}{i}(x) = \basel{c}{i,\, 0}+ \sum_{i = 1}^{n} \basel{c}{i,\, j} x^{\primeInt^{j-1}}$, where $\basel{c}{i,\, j},\, x \in \scalars$, $0 \leq j \leq n$, be such that $V \subseteq \basel{f}{i} (V)$, for $1 \leq  i \leq k$.   Thus,  in the notation of the above discussion, the permutation polynomial $\basel{f}{i} (x)-\basel{b}{i}+\basel{b}{\sigma(i)}$ can be chosen to be $\basel{g}{i}(x)$, for $x \in \scalars$ and $1 \leq i \leq k$. ~~(B) ~ Let $f(z) = z^{t}$,  where $t$ is a large positive integer dividing $\primeInt^{n}-1$, as described in the second example of section \ref{Sec-partition-of-unity}.  Let $\basel{a}{1} = 0$ and $\basel{a}{i} \in \nonzeroscalars$, for $2 \leq i \leq k$, where $k = 1+\frac{(\primeInt^{n}-1)}{t}$, be such that $\{f(\basel{a}{i}) \,:\, 1 \leq i \leq k\}$ is the codomain of $f$. Let $\sigma$ be a permutation on $\{1,\, \ldots,\, k\}$, such that $\sigma(1) = 1$, and let $\basel{H}{t} = \{y \in \scalars \,:\, y^{t} = 1\}$. Then, $\basel{S}{i} = \basel{a}{i}\basel{H}{t} = \{\basel{a}{i}v \,:\, v \in \basel{H}{t}\}$, for $1 \leq i \leq k$. Let $\basel{h}{i}(x)$, $x \in \basel{H}{t}$, be a bijective mapping discussed in the previous section, for $2 \leq i \leq k$. Thus, representing elements $\basel{c}{i} \in \nonzeroscalars$ can be found easily, such that the mapping $\basel{g}{i}(x) = \basel{c}{i} \basel{h}{i}(\basel{a^{-1}}{i} x)$ satisfies $\basel{g}{i}\bglb \basel{S}{i} \bgrb = \basel{S}{\sigma(i)}$, for $x \in \basel{S}{i}$ and $2 \leq i \leq k$.
\\

\end{small}

\noindent
{\small{\fbox{\sf{Method 2}}}}~~
Let $\mygrp$ be $\nonzeroscalars$ or $\scalars$. Let $f$ and $h$ be mappings from $\mygrp$ into itself, such that $f$ is bijective and $h\bglb {\small{f(x)}} \bgrb = h(x)$, for $x \in \mygrp$. For instance, if (A) $f$ is such that the cyclic group generated by it, as a subgroup of bijective mappings from $\mygrp$ into $\mygrp$, with composition as the group operation, is of small order $\rho \geq 2$, (B) $g\, : \, \mygrp^{\rho} \rightarrow \scalars$ is a symmetric function, which can be  an expression in $\Expressions{\scalars}{z}{\rho}$, symmetric in all the $\rho$ variables, (C) $\basel{f}{0}(x) = x$ and  $\basel{f}{i}(x) = f\bglb\basel{f}{i-1}(x)\bgrb$, for $1 \leq i \leq \rho$,  and (D) $h(x) = g\bglb x,\, \basel{f}{1}(x),\, \ldots, \, \basel{f}{\rho-1}(x)\bgrb$, for $x \in \mygrp$, then $\basel{f}{\rho}(x) = x$ and  $h\bglb f(x)\bgrb = h(x)$, for $x \in \mygrp$.  Let $\sigma$ be a permutation on $\{1,\, \ldots, \, \rho\}$, and $\{\basel{S}{i}\, : \, 1 \leq i \leq k \}$, where $2 \leq k \leq \rho$, be a partition of $\scalars$, and let $\basel{\ell}{i}(x)$, $x \in \scalars$, be the indicator function of $\basel{S}{i}$, for $1 \leq i \leq k$. Let $\eta$ be a bijective mapping from $\mygrp$ into $\mygrp$, and $\zeta(x) = \sum_{i = 1}^{k} \basel{\ell}{i}\bglb h(x)\bgrb \eta\bglb\basel{f}{\sigma(i)}(x)\bgrb$, for $x \in \mygrp$. Let $x, \, y \in \mygrp$ be such that $\zeta(x) = \zeta(y)$, and let $i,\, j \in \{1,\, \ldots, \, k\}$ be such that $\basel{\ell}{i}(h(x)) = 1$ and $\basel{\ell}{j}(h(y)) = 1$. Then, $\eta\bglb\basel{f}{\sigma(i)}(x)\bgrb = \eta\bglb \basel{f}{\sigma(j)}(y)\bgrb$, and  since $\eta$ is bijective, it follows that $\basel{f}{\sigma(i)}(x) = \basel{f}{\sigma(j)}(y)$. If $\sigma(i) \leq \sigma(j)$, then $x = \basel{f}{\sigma(j)-\sigma(i)}(y)$, and since $h\bglb f(y)\bgrb = h(y)$, it follows that $h(x) = h(y)$, $\sigma(i) = \sigma(j)$ and $i = j$, and therefore, $x = y$. Thus, $\zeta^{-1}(y) = \sum_{i = 1}^{k} \basel{\ell}{i}\bglb h({\small{\eta^{-1}(y)}})\bgrb \basel{f^{-1}}{\sigma(i)}\bglb {\small{\eta^{-1}(y)}} \bgrb$, for $y \in \mygrp$.

\subsection{\label{Sec-nonparametric-multivariate-injective-mappings}Multivariate Injective Mappings without Parameters}
 
\subsubsection{Multivariate Injective Mappings from $\mygrp^{m}$ into $\eltSet^{m}$}
In this subsection, an iterative algorithm to construct a multivariate bijective mapping from $\mygrp^{m}$ into $\eltSet^{m}$, for $m\in \PositiveIntegers$, is described. The algorithm utilizes parametric univariate bijective mappings discussed in the previous sections. In later subsections, some variations involving hashing are described.

\begin{enumerate}
\item Let $\basel{f}{i}\,:\, \mygrp \rightarrow \mygrp$ and $\basel{g}{i}\,:\, \eltSet \rightarrow \eltSet$,  for $1 \leq i \leq m$, be bijective mappings. 

\item Let $\basel{h}{i}(\basel{z}{1},\, \ldots, \, \basel{z}{m-1};\, x)$ be parametric injective mappings from $\mygrp$ into $\eltSet$, for $1 \leq i \leq m$, $x \in \mygrp$ and $\basel{z}{1},\, \ldots, \, \basel{z}{m-1} \in \eltSet$ being parameters, constructed, for example, as described in section \ref{sec-parametrization-of-permutation-polynomials}.

\item Let  $\basel{\zeta}{i}(\xx) = \basel{h}{i}\bglb \basel{\zeta}{i+1}(\xx),\, \ldots,\,
 \basel{\zeta}{m}(\xx),\, \basel{x}{1},\, \ldots,\, \basel{x}{ i-1}\, ; ~
\basel{f}{i}(\basel{x}{i})\bgrb$ and  $\basel{\eta}{i}(\xx) = \basel{g}{i}\bglb \basel{\zeta}{i}(\xx) \bgrb$, for $\xx = (\basel{x}{1},\, \ldots,\, \basel{x}{m}) \in \mygrp^{m}$ and $1\leq i \leq m$. Let $\eta(\xx) = (\basel{\eta}{1}(\xx),\, \ldots,\, \basel{\eta}{m}(\xx))$.

\end{enumerate}
\noindent
For finding $\xx =(\basel{x}{1},\,\ldots,\, \basel{x}{m}) \in \mygrp^{m}$, such that $\eta(\xx) = \yy$, for any fixed $\yy = (\basel{y}{1},\,\ldots,\, \basel{y}{m}) \in \eltSet^{m}$,  let $\basel{\epsilon}{i} = \basel{g^{-1}}{i}(\basel{y}{i})$ and $\basel{\delta}{i} = \basel{h^{-1}}{i} (\basel{\epsilon}{i+1},\, \ldots, \, \basel{\epsilon}{m},\, \basel{x}{1},\, \ldots,\, \basel{x}{i-1};\, \basel{\epsilon}{i})$, for $1 \leq i \leq m$.  Then, $\basel{x}{i} = \basel{f^{-1}}{i} (\basel{\delta}{i})$, for $1 \leq i \leq m$.  Now, for $\eltSet = \scalars$ and $\mygrp = \nonzeroscalars$, if $\basel{g}{i}$ and $\basel{h}{i}$,  for $1 \leq i \leq m$, are bijective mappings and parametric bijective mappings, respectively, from $\nonzeroscalars$ into $\nonzeroscalars$, then the above procedure can be applied to obtain multivariate bijective mappings from $\mygrp^{m}$ into $\mygrp^{m}$. These mappings are required in appealing for a security that is immune to threats resulting from Gr\"{o}bner basis analysis. It can be observed that one level of exponentiation suffices for the purpose.

\subsubsection{Hybrid Multivariate Injective Mappings with Hashing}
For Method 1 of the previous subsection, in the first example, in place of $T(x)$, $x \in \scalars$,  $T\bglb\alpha(\xx)\bgrb$, $\xx \in \scalars^{m}$, and in the second example, in place of $f(z)$, $z \in \scalars$, $f\bglb \beta(\xx)\bgrb$, $\xx \in \scalars^{m}$, are chosen, where $\alpha \,: \, \scalars^{m} \rightarrow \scalars$ is a non constant affine mapping in the first example, and $\beta(\xx) = c \prod_{i = 1}^{m}\basel{x^{\basel{s}{i}}}{i}$, for some nonnegative integers  $\basel{s}{i}$, which, when positive, are relatively prime to $\primeInt^{n}-1$, and, when zero, for the corresponding subscript index $i$, the variable $\basel{x}{i}$ does not occur in the product, for $1 \leq i \leq m$, such that $\beta(\xx)$ is nonconstant, in the second example. Similarly, Method 2 hashing of the previous subsection can also be  extended to multivariate mappings, replacing $x$ with $\xx$. For instance, if $\mygrp = \nonzeroscalars$, $a$ is a primitive element in $\nonzeroscalars$ and $\cnt$ is the set cardinality of $\nonzeroscalars$, then $f(\xx)$ can be chosen to be $\bglb a^{\basel{\phi}{1}(\basel{\log}{a} \xx)}, \, \ldots, \,  a^{\basel{\phi}{m}(\basel{\log}{a} \xx)}\bgrb$, where $\basel{\log}{a} \xx = (\basel{\log}{a} \basel{x}{1},\, \ldots,\, \basel{\log}{a}\basel{x}{m})$ and $\Phi(\yy) = (\basel{\phi}{1}(\yy),\, \ldots,\, \basel{\phi}{m}(\yy)\bgrb$ is a bijective mapping from $\basel{\Integers^{m}}{\cnt}$ into itself, such that the cyclic subgroup generated by $\Phi$, with respect to function composition operation, has a group order $\rho$,  while $g$ can be chosen to be an expression from ${\mathcal {EXP}}\bglb \scalars \,; \, [\basel{z}{1,\,1}, \, \ldots, \, \basel{z}{m,\, 1},\, \ldots, \, \basel{z}{1,\,\rho}, \, \ldots, \, \basel{z}{m,\, \rho}]\bgrb$, which is symmetric in the $\rho$ vectors $(\basel{z}{1,\, i}, \, \ldots, \, \basel{z}{m,\, i}\bgrb$, for $1 \leq i \leq \rho$.  If $g = \pi^{t}$, for a symmetric mapping  $\pi$ obtained by taking product of terms as appropriate and a large positive integer divisor $t$ of $\cnt$, then since $t\Phi(\yy)$ is a bijective mapping from $t\basel{\Integers^{m}}{\cnt}$ into itself, the order of the cyclic subgroup generated by $t\Phi(\yy)$, as a subgroup of the group of bijective mappings from $t\basel{\Integers^{m}}{\cnt}$ into itself, can be ensured to be only a small divisor of $\rho$, resulting in a more efficient method of hashing, even for a very large and perhaps unknown $\rho$. It can be observed that $g$ can be chosen to depend only on a few scalar components from each vector, while maintaining symmetry in all its vector parameters, with each vector consisting of $m$ scalars components, and that the main objective in Method 2 hashing is to produce a hashing function $h$ that evaluates to the same same value, even if $f$ is applied on its arguments.

\section{\label{Sec-PKC-and-DS}Public Key Cryptography and Digital Signature}

  Let the number of elements in the plain message (or plain signature message) be $\mu$, and the number of elements in the encrypted message (or encrypted signature message) be $\nu$, where $\mu, \, \nu \in \PositiveIntegers$ and $\mu \leq \nu$.  Let $\eltSet$ be $\scalars$ or $\finiteIntegerRing{\cnt}$, and $\mygrp \subseteq \eltSet$ be the set from which plain message elements are sampled. If the number of plain and encrypted (or plain and signed) messages are the same, then a multivariate bijective mapping $P\,:\, \mygrp^{\mu} \rightarrow \mygrp^{\mu}$ is chosen and advertised in the public key lookup table $\LookupTable$, while $P^{-1}$ is saved in the back substitution table $\BackSubstitutionTable$. Let $\bglb\basel{\xi}{1},\, \ldots,\, \basel{\xi}{\mu}\bgrb \in \mygrp^{\mu}$ be plain message. For public key cryptography, the encrypted message is $\bglb\basel{\epsilon}{1},\, \ldots,\, \basel{\epsilon}{\mu}\bgrb  = P\bglb\basel{\xi}{1},\, \ldots,\, \basel{\xi}{\mu}\bgrb$, and the decryption is $P^{-1}\bglb\basel{\epsilon}{1},\, \ldots,\, \basel{\epsilon}{\mu}\bgrb$. For digital signature, the signed message is $\bglb\basel{\epsilon}{1},\, \ldots,\, \basel{\epsilon}{\mu}\bgrb  = P^{-1}\bglb\basel{\xi}{1},\, \ldots,\, \basel{\xi}{\mu}\bgrb$, and recovered message is $P\bglb\basel{\epsilon}{1},\, \ldots,\, \basel{\epsilon}{\mu}\bgrb$. In the remaining part of the section, it is assumed that $1 \leq \mu \leq \nu-1$. Let $\nu = \mu+\lambda$, for some positive integer $\lambda$. Let $\kappa$ be the number of padding message elements in the hashing keys.  Let $\xx = (\basel{x}{1},\, \ldots,\, \basel{x}{\mu}) \in \mygrp^{\mu}$ be the plain message,  $\yy = (\basel{y}{1},\, \ldots,\, \basel{y}{\nu})\in \eltSet^{\nu}$ be the encrypted or signed message, and  $\oomega= (\basel{\omega}{1},\, \ldots,\, \basel{\omega}{\kappa}) \in \mygrp^{\kappa}$ be a padding message. The multivariate mappings in the rest of this section are expressions from either $\multipolynomials{\eltSet}{t}{\sigma}{m}{n}$ or $\multiexpressions{\eltSet}{t}{\sigma}{m}{n}$, for some appropriate variable names $t$ and $\sigma$, and subscript numbers $m$ and $n$, depending on the context of occurrence and arity of the mappings.

\subsection{\label{Sec-HashKeys}Hashing Keys}
    
The following subroutine generates the hashing keys required
by the algorithms of sections \ref{Sec-PKC} and \ref{Sec-DS}.

   \paragraph{\underline{Subroutine for Generation of Hashing Keys}} The table generated is the private key hash table $\HashTable$,
containing the hashing keys.
\begin{small}
\begin{enumerate}

   \item The following inputs to the subroutine are taken: positive integers $\mu$, $\kappa$, $L$, $\lambda$, and a binary flag SIGN, where $L$ is the number of hashing keys, $L \leq \lambda$, and SIGN is set to the binary value $\true$, if this subroutine is called for digital signature, and set to $\false$ for public key cryptography.

  \item The private key hash table $\HashTable$ is initialized to empty set. The input parameters are saved in the private key  hash table $\HashTable$. Let $\nu = \mu+\lambda$. 
   
   \item Let $\basel{f}{l}(\xx,\, \oomega)$, for $1 \leq l \leq L$, be selected and saved in the private key hash table $\HashTable$. If $L < \lambda$, let $\basel{Q}{i}(\xx,\, \oomega)$, for $1 \leq i \leq \lambda-L$, be selected and saved in the private key hash table $\HashTable$. The chosen functions are required to evaluate to elements in $\mygrp$, for $(\xx,\, \oomega) \in \mygrp^{\mu+\kappa}$.  Let {\small{$F(\xx,\, \oomega) = \bglb \basel{f}{1}(\xx,\, \oomega),\, \ldots,\, \basel{f}{L}(\xx,\, \oomega)\bgrb$}.}
    
\item   Now, a parametric multivariate injective mapping
$\eta(\basel{y}{1}, \ldots, \basel{y}{\lambda-L}; \zz)$,
$\zz = (\basel{z}{1}, \ldots, \basel{z}{L})$, is selected
such that $\eta^{-1}(\basel{y}{1}, \ldots, \basel{y}{\lambda-L}; \zz)$
can be computed easily (discussed in section \ref{Sec-parametric-injective-mappings}). The multivariate mappings required to compute both
  $\eta(\basel{y}{1}, \ldots, \basel{y}{\lambda-L}; \zz)$ and
$\eta^{-1}(\basel{y}{1}, \ldots, \basel{y}{\lambda-L}; \zz)$
 are saved in the private key hash table $\HashTable$.
If SIGN is set to $\true$, then this procedure is called for generating  digital signature hashing keys, and hence, 
 let $\basel{g}{l}(\basel{y}{1}, \ldots, \basel{y}{\lambda})$
 $ =$
$\basel{\eta^{-1}}{l}(\basel{y}{1}, \ldots, \basel{y}{\lambda-L};\,
 (\basel{y}{\lambda-L+1}, \ldots, \basel{y}{\lambda}))$,  $1 \leq l \leq L$,
  which are also saved in the private key hash table $\HashTable$.
 
 \item Let {\small{$\basel{Q}{\lambda-L+i}(\xx,\, \oomega)$
$ =$
$\basel{\eta}{i}\bglb \basel{Q}{1}(\xx,\, \oomega),\, \ldots,\, \basel{Q}{\lambda-L}(\xx,\, \oomega);~\,
 F(\xx,\, \oomega) \bgrb$, $1 \leq i \leq L$},}
which are saved in the private key hash table $\HashTable$.
Thus, for {\small{$(\xx,\, \oomega) \in \mygrp^{\mu+\kappa}$,
  $F(\xx,\, \oomega)$
$=$
$\eta^{-1}\bglb \basel{Q}{1}(\xx,\, \oomega), \ldots, \basel{Q}{\lambda-L}(\xx,\,\oomega); ~
(\basel{Q}{\lambda-L+1}(\xx,\, \oomega), \ldots, \basel{Q}{\lambda}(\xx,\, \oomega)) \bgrb$}.} The parametric multivariate injective mapping $\eta(\basel{y}{1}, \ldots, \basel{y}{\lambda-L}; \zz)$ are required to be so chosen that ($i$) it is easily expressible as a multivariate mapping, for public key cryptography, and ($ii$) $\eta^{-1}(\basel{y}{1}, \ldots, \basel{y}{\lambda-L}; \zz)$ and $\basel{g}{l}(\basel{y}{1}, \ldots, \basel{y}{\lambda})$,  $1 \leq l \leq L$, are easily expressible as multivariate mappings, for digital signature, and for signature authentication, the multivariate mappings $\basel{Q}{i}(\xx,\, \oomega)$, $1 \leq i \leq \lambda$, must occur as public key mappings, which need to be easily expressible, as well.

\end{enumerate}
\end{small}

\subsection{\label{Sec-PKC}Public Key Cryptography (PKC)}

The input is the private key hash table $\HashTable$,
containing the hashing keys. 

\paragraph{\underline{Public Key Cryptography Key Generation Algorithm}}

The tables generated are as follows:  (1) the private key back substitution table $\BackSubstitutionTable$, containing information for decryption of
 public key encrypted message, and (2) the public key lookup table $\LookupTable$, containing the multivariate mappings for encrypting plain message.
\begin{small}
\begin{enumerate}

\item The subroutine for generation of hashing keys
(described in section \ref{Sec-HashKeys}) is called, which 
takes input parameters, {\em viz.}, positive integers
$\mu$, $\kappa$, $L$, $\lambda$, and a binary flag SIGN, 
which is set to $\false$ by the calling function,
generates the multivariate mappings
$\basel{f}{l}(\xx,\, \oomega)$, $1 \leq l \leq L$,
and $\basel{Q}{i}(\xx,\, \oomega)$, $1 \leq i \leq \lambda$,
sets $\nu = \mu+\lambda$, and saves them in the private key
hash table $\HashTable$. The private key back substitution table
$\BackSubstitutionTable$ is initialized to empty set.

 \item  A parametric multivariate injective mapping
$\zeta(\basel{z}{1},\,\ldots,\,\basel{z}{L}; \, \xx)$ is selected
such that the parametric inverse multivariate mapping
$\zeta^{-1}(\basel{z}{1},\,\ldots,\,\basel{z}{L}; \, \yy)$
 can be computed easily
 (discussed in section \ref{Sec-parametric-injective-mappings}).
 The information required to compute 
 $\zeta(\basel{z}{1},\,\ldots,\,\basel{z}{L}; \, \xx)$ and
$\zeta^{-1}(\basel{z}{1},\,\ldots,\,\basel{z}{L}; \, \yy)$ 
 is  saved in the private key back substitution table $\BackSubstitutionTable$.
 Let  $\basel{Q}{\lambda+i}(\xx,\, \oomega) =   
 \basel{\zeta}{i}(\basel{f}{1}(\xx,\,\oomega),\, \ldots,\, \basel{f}{L}(\xx,\, \oomega); \, \xx)$,
 $1 \leq i \leq \mu$.

\item An invertible affine linear transformation
 $T\,:\, \eltSet^{\nu} \to \eltSet^{\nu}$
 is selected, and its inverse transformation $T^{-1}$
 is saved in the back substitution table $\BackSubstitutionTable$.

 \item Let 
 $\bglb\basel{P}{1}(\xx,\, \oomega),\,\ldots,\,\basel{P}{\nu}(\xx,\, \oomega)\bgrb$
  $=$
  $T\bglb\basel{Q}{1}(\xx,\,\oomega),\,\ldots,\,\basel{Q}{\nu}(\xx,\, \oomega)\bgrb$  be the encryption multivariate mappings,
 which are advertised in the public key lookup table $\LookupTable$,  along with
 $\mu$, $\nu$, $\kappa$, $\eltSet$ and $\mygrp$.

\end{enumerate}
\end{small}

\paragraph{\underline{Encryption}} Let
$\bglb\basel{\xi}{1},\,\ldots,\,\basel{\xi}{\mu}\bgrb$ be the plain message.
The encryptor chooses padding message 
$ \basel{\omega}{1},\, \ldots,\, \basel{\omega}{\kappa} \in \scalars$,
computes $\basel{\epsilon}{i} = 
\basel{P}{i}\bglb\basel{\xi}{1},\,\ldots,\,\basel{\xi}{\mu},\, 
 \basel{\omega}{1},\, \ldots,\, \basel{\omega}{\kappa}\bgrb$,
 $1 \leq i \leq \nu$, and transmits  
$\bglb\basel{\epsilon}{1},\,\ldots,\,\basel{\epsilon}{\nu}\bgrb$
to the receiver.

\paragraph{\underline{Decryption}}  The input items required for decryption are read
from the private key hash table $\HashTable$ and the private key
back substitution table $\BackSubstitutionTable$. The decryption algorithm is as follows:

\begin{small}
\begin{enumerate}

\item  Let 
$\bglb\basel{\epsilon}{1},\ldots,\basel{\epsilon}{\nu}\bgrb \in \eltSet^{\nu}$
be the received encrypted message.

\item Let $\bglb \basel{v}{1},\, \ldots,\, \basel {v}{\nu}\bgrb = 
T^{-1}\bglb\basel{\epsilon}{1},\ldots,\basel{\epsilon}{\nu}\bgrb$.
Thus, $\basel{v}{i}$
$=$
$\basel{Q}{i}( \basel{\xi}{1},\, \ldots,\, \basel{\xi}{\mu},\, \basel{\omega}{1},\, \ldots,\, \basel{\omega}{\kappa})$, $1 \leq i \leq \nu$, where
$(\basel{\xi}{1},\, \ldots,\, \basel{\xi}{\mu})$ is the
plain message (to be decrypted in the subsequent steps),
and $(\basel{\omega}{1},\, \ldots,\, \basel{\omega}{\kappa})$
is the padding message, which will not be decrypted.  Let
$\basel{y}{l} = \basel{v}{l}$, $1 \leq l \leq \lambda$. 

\item Let $(\basel{z}{1},\, \ldots,\, \basel{z}{L}) = 
\eta^{-1}\bglb \basel{y}{1},\,\ldots,\,\basel{y}{\lambda-L};\,
(\basel{y}{\lambda-L+1},\, \ldots,\, \basel{y}{\lambda}) \bgrb$. 
It is clear that $\basel{z}{l}$
$=$
$\basel{f}{l}(\basel{\xi}{1},\,\ldots,\, \basel{\xi}{\mu},\, \basel{\omega}{1},\, \ldots,\, \basel{\omega}{\kappa})$, $1 \leq l \leq L$.

\item The plain message is 
$(\basel{\xi}{1},\, \ldots,\, \basel{\xi}{\mu}) ~ = ~
\zeta^{-1} \bglb \basel{z}{1},\, \ldots,\, \basel{z}{L};$
$\, (\basel{v}{\lambda+1},\ldots,\, \basel{v}{\nu}) \bgrb$.
\end{enumerate}
\end{small}

\subsection{\label{Sec-DS}Digital Signature (DS)}

The input is the private key hash table $\HashTable$,
containing the hashing keys. 

\paragraph{\underline{Digital Signature Key Generation Algorithm}}

The tables generated are as follows: (1) the private key digital signature table $\SignatureTable$, containing information for signing the plain message, (2) the public key signature verification table $\SignatureVerificationTable$, containing the multivariate mappings for recovery of plain message, and (3) the public key signature authentication table $\SignatureAuthenticationTable$, containing the multivariate mappings for verifying the authentication of the plain message.
\begin{small}
\begin{enumerate}

\item The subroutine for generation of hashing keys
(described in section \ref{Sec-HashKeys}) is called, which 
takes input parameters, {\em viz.}, positive integers
$\mu$, $\kappa$, $L$, $\lambda$, and a binary flag SIGN, 
which is set to $\true$ by the calling function now, 
generates the multivariate mappings $\basel{f}{l}(\xx,\, \oomega)$,
$\basel{g}{l}\bglb \basel{z}{1},\, \ldots,\basel{z}{\lambda}\bgrb$, $1 \leq l \leq L$, and $\basel{Q}{i}(\xx,\, \oomega)$, $1 \leq i \leq \lambda$,   such that $\basel{f}{l}(\xx,\, \oomega)$
$=$
$\basel{g}{l}\bglb\basel{Q}{1}(\xx,\, \oomega),\,\ldots,\,\basel{Q}{\lambda}(\xx,\, \oomega)\bgrb$,
for $(\xx,\, \oomega) \in \scalars^{\mu+\kappa}$, $1 \leq l \leq L$,  sets $\nu = \mu+\lambda$, and saves them in the private key hash table $\HashTable$.
The private key signature table $\SignatureTable$ is initialized to empty set.

  \item  A parametric multivariate bijective mapping
$\zeta(\basel{z}{1},\,\ldots,\,\basel{z}{L}; \, \xx)$
from $\mygrp^{\mu}$ into $\mygrp^{\mu}$ is selected
such that the parametric inverse
$\zeta^{-1}(\basel{z}{1},\,\ldots,\,\basel{z}{L}; \, \xx)$
 can be computed easily
 (discussed in section \ref{Sec-parametric-injective-mappings}).
 The information required to compute 
  $\zeta(\basel{z}{1},\,\ldots,\,\basel{z}{L}; \, \xx)$ and
 $\zeta^{-1}(\basel{z}{1},\,\ldots,\,\basel{z}{L}; \, \xx)$ 
 is saved in the private key signature table $\SignatureTable$.
 
  \item Let
  $\basel{P}{i}( \basel{y}{1},\, \ldots,\, \basel{y}{\nu}) =$
  $\basel{\zeta}{i} \bglb \basel{g}{1} (\basel{y}{1},\,\ldots,\, \basel{y}{\lambda}),$
   $\,\ldots,\,$
     $\basel{g}{L} (\basel{y}{1},\,\ldots,\, \basel{y}{\lambda})\, ;\, $
   $ (\basel{y}{\lambda+1},\, \ldots,\, \basel{y}{\nu}) \bgrb  $,  $1 \leq i \leq \mu$, be the signature verification multivariate mappings 
which are advertised in the public key signature verification table $\SignatureVerificationTable$, along with
 $\mu$, $\nu$, $\kappa$, $\eltSet$ and $\mygrp$, 
 and let the plain message authentication multivariate mappings be
 $\basel{S}{i}(\xx,\, \oomega)$
    $ = $
 $\basel{Q}{i}(\xx,\, \oomega)$,  $1 \leq i \leq \lambda$, which are advertised in the public key signature authentication table $\SignatureAuthenticationTable$, along with $\mu$, $\nu$, $\kappa$, $\eltSet$ and $\mygrp$. The signature verification table $\SignatureVerificationTable$ is advertised as a public key, with read  permissions for the intended receiver to access. There are two possibilities for signature authentication verification: ($i$) a public authority, that is responsible for providing signature authentication ascertainment and for possible issuance of a certification to that effect, is identified, which is referred to herein as a trusted authentication verifier (TAV), in which case, the padding message is transmitted to the intended receiver, possibly encrypting it by a public or shared key encryption algorithm, whereas the decryption key for the padding message and the signature authentication table $\SignatureAuthenticationTable$ are shared by the signer with only the TAV, or ($ii$) there is no TAV, in which case, the signature authentication table $\SignatureAuthenticationTable$ and the signature verification table $\SignatureVerificationTable$ are made available, with read access permissions, to the intended receiver as a public key. 
\end{enumerate}
\end{small}

\paragraph{\underline{Digital Signing Algorithm}}
Let {\small{$(\basel{\xi}{1},\, \ldots,\, \basel{\xi}{\mu})$}} be the
plain  message. The parameters required for digital signing are
read from the private key hash table $\HashTable$ and the private key
signature table $\SignatureTable$. The digital signing algorithm is as follows:
\begin{small}
\begin{enumerate}

\item  The signer chooses padding message
$(\basel{\omega}{1},\, \ldots,\, \basel{\omega}{\kappa}) \in \mygrp^{\kappa}$, either generating them randomly, or based on previous correspondences.

\item The signer computes the hash values $\basel{z}{l} = \basel{f}{l}(\basel{\xi}{1},\, \ldots,\, \basel{\xi}{\mu},\, 
  \basel{\omega}{1},\, \ldots,\, \basel{\omega}{\kappa})$,
$1 \leq l \leq L$, the authentication header entries
 $\basel{\epsilon}{i} = \basel{Q}{i}(\basel{\xi}{1},\, \ldots,\, \basel{\xi}{\mu},\,
  \basel{\omega}{1},\, \ldots,\, \basel{\omega}{\kappa})$, $1 \leq i \leq \lambda$, and the signed message entries
 $(\basel{\epsilon}{\lambda+1},\, \ldots,\, \basel{\epsilon}{\nu})$
 $ =$
$\zeta^{-1} (\basel{z}{1},\, \ldots,\, \basel{z}{L};\,(\basel{\xi}{1},\ldots,\, \basel{\xi}{\mu})) $.

 \item   The signature message is
 $\bglb \basel{\epsilon}{1},\,\ldots, \, \basel{\epsilon}{\nu}\bgrb$, 
 which is transmitted to the intended receiver,
 while the padding message $\bglb\basel{\omega}{1},\, \ldots,\, \basel{\omega}{\kappa}\bgrb$ is either transmitted to the intended receiver together with the signature, either on demand or for free, or communicated to a trusted authentication verifier (TAV), with which the signer registers the signature authentication table $\SignatureAuthenticationTable$. 

\end{enumerate}   
\end{small}

\paragraph{\underline{Digital Signature Verification Algorithm}}  The input items required for signature verification are public key signature verification table $\SignatureVerificationTable$, and the signature authentication table $\SignatureAuthenticationTable$ or a method for ascertaining by a trusted authentication verifier (TAV). The signature verification algorithm is as follows:
\begin{small}
\begin{enumerate}

\item Let $\bglb\basel{\epsilon}{1},\, \ldots,\, \basel{\epsilon}{\nu}\bgrb$
be the received signature message. The padding
$\bglb\basel{\omega}{1},\, \ldots,\, \basel{\omega}{\kappa}\bgrb$ may have also been optionally received.

\item Let 
{\small{$\basel{\xi}{i} = \basel{P}{i}(\basel{\epsilon}{1},\, \ldots,\, \basel{\epsilon}{\nu})$, $1 \leq i \leq \mu$}.} The plain signature message is {\small{ $(\basel{\xi}{1},\, \ldots,\, \basel{\xi}{\mu})$}.}
The public key signature verification table $\SignatureVerificationTable$ contains the information required in this step.

\item If the signature authentication table $\SignatureAuthenticationTable$ is available, then the authentication of the plain message can be verified by testing whether $\basel{S}{i}\bglb\basel{\xi}{1},\, \ldots,\, \basel{\xi}{\mu},\,
\basel{\omega}{1},\, \ldots,\, \basel{\omega}{\kappa} \bgrb = \basel{\epsilon}{i}$,
$1 \leq i \leq \lambda$;  otherwise, a public authority TAV, that is responsible for signature authentication ascertainment, may be approached.
\end{enumerate}   
\end{small}

\section{\label{Sec-complexity-of-computing-nonparametric-inverses-of-parametric-multivariate-polynomial-mappings}Complexity Analysis of Computing Left Inverse Mappings of Multivariate Injective Mappings and of Computing Right Inverse Mappings of Multivariate Surjective Mappings}
\vspace*{-0.2cm}
Model theory of fields and polynomial algebras
is extensively studied in mathematical logic 
\cite{vanDalen:1994,  vandenDries:2000,  Hodges:1993,  Marker:2000,  MMP:1996}.
Let $\scalars$ be a field, and let $\ArithmeticExpressions(\scalars)$
be the set of arithmetic expressions without quantifiers,
obtained by collecting the expressions involving any number
of finitely many variables, constructed using parentheses and
the binary or unary arithmetic operators of addition $+$,
subtraction $-$, multiplication $\cdot$, possibly division $/$,
exponentiation $^k$, where $k$ is a positive integer, and
binary valued relational operator $=$ (and possibly other
relational operators such as $<$, \, $>$, \, $\leq$ and $\geq$).
The relational operators allow construction of
assertions that evaluate to anyone of the special symbolic constants
$\false$ and $\true$, represented by $0$ and $1$, respectively.
In the sequel, the variables assume values from $\scalars$,
the arithmetic expressions evaluate to values in $\scalars$,
as defined by the arithmetic operations in $\scalars$,
and the assertions evaluate to values in $\{0,\, 1\}$.
A variable taking values in $\{0,\, 1\}$ is a boolean
variable. The arithmetic expressions in
$\ArithmeticExpressions(\basel{\Integers}{2})$ 
are boolean expressions. For any field $\scalars$,
a boolean variable $x$ can be obtained from the equation
$x^{2}-x = 0$.  For boolean variables $x$ and $y$,
~$\lnot x$ can be represented by $1-x$, ~$x \wedge y$
by $x\cdot y$, ~$x \vee y$ by $1-(1-x)\cdot(1-y)$,
~$x\oplus y$ by $(x-y)^{2}$, ~$x \rightarrow y$ by
$1-x \cdot(1-y)$,  ~and $x \leftrightarrow y$ by $1-(x-y)^{2}$,
where $\lnot$ denotes the logical ``negation'',
$\wedge$ the logical  ``and'',
$\vee$ the logical ``or'',
$\oplus$ the logical ``exclusive or'',
$\rightarrow$ the logical ``implies'',
and $\leftrightarrow$ the logical ``implies and
is implied by''. The inequality operator, denoted by $\not =$,
is a secondary binary operator defined as 
the logical negation of the equality operator.
Let $ \QuantifiedArithmeticExpressions(\scalars)$
be the set of arithmetic expressions in which
some (none, some or all) variables are constrained
by ``existential'' $\exists$ or ``universal'' $\forall$
quantifiers. A variable constrained by a quantifier is
called a {\em bound} variable. A variable that is not
bound is called a {\em free} variable. An arithmetic
expression in which all the variables are free is a
quantifier free arithmetic expression, {\em i.e.},
an expression in $\ArithmeticExpressions(\scalars)$. 
A quantified arithmetic expression is in prenex normal
form, if all the quantifiers occur before the otherwise
quantifier free arithmetic expression, {\em i.e},
a quantified arithmetic expression of the form
{\small{$\forall \basel{y}{1}\, \ldots\, \forall \basel{y}{\basel{k}{1}} 
\, \exists \basel{x}{1}\,\ldots\,
 \forall \basel{y}{\basel{k}{i-1}+1}\, \ldots\, \forall \basel{y}{\basel{k}{i}} 
\, \exists \basel{x}{i}\,\ldots\,
 \forall \basel{y}{\basel{k}{m-1}+1}\, \ldots\, \forall \basel{y}{\basel{k}{m}} 
\, \exists \basel{x}{m}$}}
{\small{$\, \forall \basel{y}{\basel{k}{m}+1}\, \ldots\, \forall \basel{y}{n}$}}
~~ {\small{$f(\basel{x}{1},\,\ldots,\, \basel{x}{m},\, 
\basel{y}{1},\, \ldots,\, \basel{y}{n})$}}, where $m$ and $n$
are positive integers, and $\basel{k}{i}$, for $1 \leq i \leq m$,
are nonnegative integers such that $\basel{k}{i} \leq \basel{k}{i+1}$,
for $1 \leq i \leq m-1$, and $\basel{k}{m} \leq n$.
The variables $\basel{y}{j}$, $1 \leq j \leq n$, are 
{\em independent} variables, as they are bound to universal
quantifiers. The variable $\basel{x}{i}$ depends on the variables
$\basel{y}{j}$, $1 \leq j \leq \basel{k}{i}$, $1 \leq i \leq m$,
and  is a {\em dependent} bound variable. 
 A tuple $\bglb \basel {a}{1},\, \ldots,\, \basel{a}{i},\,
 \basel{b}{1},\, \ldots,\, \basel{b}{\basel{k}{i}}\bgrb
\in \scalars^{i+\basel{k}{i}}$, $1 \leq i \leq m$,
 is {\em feasible} to a quantified arithmetic expression
 in prenex normal form with no free variables as described before,
 if either $i = m$ and  $f(\basel{a}{1},\,\ldots,\, \basel{a}{m},\, 
\basel{b}{1},\, \ldots,\, \basel{b}{\basel{k}{m}},\,
\basel{y}{\basel{k}{m}+1},\, \ldots,\,\basel{y}{n})$
evaluates to $1$, for $\basel{y}{\basel{k}{m}+1},\, \ldots,\,\basel{y}{n}$
$ \in $
$\scalars$,
or $1 \leq i \leq m-1$ and each tuple
{\small{$(\basel{a}{1},\, \ldots,\, \basel{a}{i},\, \basel{x}{i+1},\,
 \basel{b}{1},\, \ldots,\, \basel{b}{\basel{k}{i}},
 \basel{y}{\basel{k}{i}+1},\, \ldots,\, \basel{y}{\basel{k}{i+1}})$},}
for $\basel{y}{\basel{k}{i}+1},\, \ldots,\,\basel{y}{\basel{k}{i+1}} \in \scalars$,
and for some $\basel{x}{i+1} \in \scalars$, that may depend on $ \basel {a}{1},\, \ldots,\, \basel{a}{i},$
$\basel{b}{1},\, \ldots,\, \basel{b}{\basel{k}{i}},$
$\basel{y}{\basel{k}{i}+1},\, \ldots,\,\basel{y}{\basel{k}{i+1}}$
$ \in \scalars$ is feasible. If for every $\basel{b}{1},\, \ldots,\, \basel{b}{\basel{k}{1}} \in \scalars$, there exists $\basel{a}{1} \in \scalars$,
 such that the tuple $\bglb \basel {a}{1},\, 
  \basel{b}{1},\, \ldots,\, \basel{b}{\basel{k}{1}}\bgrb$
 is feasible, then the given instance of binary valued
 quantified arithmetic expression is {\em satisfiable}.
 The evaluation problem for quantified boolean expressions
 in prenex normal form with no free variables in 
 $\QuantifiedArithmeticExpressions(\scalars)$ is
 to find whether the given input instance is satisfiable.
 Let $\SatisfiableQuantifiedArithmeticExpressions(\scalars)
 \subseteq \QuantifiedArithmeticExpressions(\scalars)$ be the set
  of satisfiable binary valued quantified arithmetic expressions
 ({\em i.e.}, quantified arithmetic assertions) in prenex normal form
 with no free variables that evaluate to $\true$. Let
 $\QBF$ and $\QSAT$ be $\QuantifiedArithmeticExpressions(\basel{Z}{2})$
 and $\SatisfiableQuantifiedArithmeticExpressions(\basel{Z}{2})$,
 respectively. By the previous discussion, every boolean expression in $\QBF$,
 analogously in $\QSAT$, can be represented by some arithmetic expression
 in $\QuantifiedArithmeticExpressions(\scalars)$, analogously in
$\SatisfiableQuantifiedArithmeticExpressions(\scalars)$,
with equality binary relation, for any field $\scalars$.
The evaluation problem for quantified boolean expressions
in prenex normal form with no free variables in $\QBF$ is
$\PSPACE$-complete, where $\PSPACE$ is the set of formal
languages acceptable in polynomial space \cite{HMU:2007}.

\subsection{\label{Sec-CSP}Constraint Satisfaction Problem}
Let 
\hfill{\small{$\forall \basel{y}{1}\, \ldots\, \forall \basel{y}{\basel{k}{1}} 
\, \exists \basel{x}{1}\,\ldots\,
 \forall \basel{y}{\basel{k}{i-1}+1}\, \ldots\, \forall \basel{y}{\basel{k}{i}} 
\, \exists \basel{x}{i}\,\ldots\,
 \forall \basel{y}{\basel{k}{m-1}+1}\, \ldots\, \forall \basel{y}{\basel{k}{m}} 
\, \exists \basel{x}{m}$}}
{\small{$\, \forall \basel{y}{\basel{k}{m}+1}\, \ldots\, \forall \basel{y}{n}
~~f(\basel{x}{1},\,\ldots,\, \basel{x}{m},\, 
\basel{y}{1},\, \ldots,\, \basel{y}{n})$}} be an instance
in $\SatisfiableQuantifiedArithmeticExpressions(\scalars)$,
where $m$ and $n$ are positive integers, and $\basel{k}{i}$,
for $1 \leq i \leq m$, are nonnegative integers such that
$\basel{k}{i} \leq \basel{k}{i+1}$, for $1 \leq i \leq m-1$,
and $\basel{k}{m} \leq n$. A tuple
$(\basel {a}{1},\, \ldots,\, \basel{a}{r},\,
 \basel{b}{1},\, \ldots,\, \basel{b}{\basel{k}{r}})
\in \scalars^{r+\basel{k}{r}}$, $1 \leq r \leq m$, is 
{\em functionally feasible by quantifier free arithmetic expressions}
to the given constraint satisfaction problem,
if there exist quantifier free arithmetic expressions  
$\basel{g}{1}(\basel{y}{1},\,\ldots,\,\basel{y}{\basel{k}{1}})$
and $\basel{g}{i} (\basel{x}{1},\, \ldots, \basel{x}{i-1}, \,
\basel{y}{1},\,\ldots,\,\basel{y}{\basel{k}{i}})$, $2 \leq i \leq m$,
in $\ArithmeticExpressions \bglb \scalars \bgrb$, such that
the following holds:
$\forall \basel{y}{1}\, \ldots\, \forall \basel{y}{n}~
f(\basel{x}{1},\,\ldots,\, \basel{x}{m},\, 
\basel{y}{1},\, \ldots,\, \basel{y}{n})   ~ = ~ \true$, ~ where
{\small{$\basel{x}{1} = \basel{g}{1}(\basel{y}{1},\,\ldots,\,\basel{y}{\basel{k}{1}})$,
$ \basel{x}{i} = \basel{g}{i} (\basel{x}{1},\, \ldots, \basel{x}{i-1}, \,
 \basel{y}{1},\,\ldots,\,\basel{y}{\basel{k}{i}})$, $2 \leq i \leq m$,
 $\basel{g}{1}(\basel{b}{1},\,\ldots,\,\basel{b}{\basel{k}{1}})= \basel{a}{1}$
 and $\basel{g}{i} (\basel{a}{1},\, \ldots, \basel{a}{i-1}, \, 
\basel{b}{1},\,\ldots,\,\basel{b}{\basel{k}{i}})  = \basel{a}{i}$, 
$2 \leq i \leq r$}.} It can be observed that for a finite field $\scalars$,
a feasible tuple is also functionally feasible by quantifier free
arithmetic expressions. A solution to the constraint satisfaction
problem is to find quantifier free arithmetic expressions,
if and when they exist,
$\basel{g}{1}(\basel{y}{1},\,\ldots,\,\basel{y}{\basel{k}{1}})$
for $\basel{x}{1}$ and 
$\basel{g}{i} (\basel{x}{1},\, \ldots, \basel{x}{i-1}, \,
\basel{y}{1},\,\ldots,\,\basel{y}{\basel{k}{i}})$ for $\basel{x}{i}$,
$2 \leq i \leq m$, such that for all
$\basel{y}{1}\, \ldots\,  \basel{y}{n} \in  \scalars$,~
 $f(\basel{x}{1},\,\ldots,\, \basel{x}{m},\,
\basel{y}{1},\, \ldots,\, \basel{y}{n})$
$~ = ~ 1$,
where {\small{
$\basel{x}{1} = \basel{g}{1}(\basel{y}{1},\,\ldots,\,\basel{y}{\basel{k}{1}})$}}
and {\small{$\basel{x}{i} = \basel{g}{i}(\basel{x}{1},\, \ldots, \basel{x}{i-1},
\basel{y}{1},\,\ldots,\,\basel{y}{\basel{k}{i}})$},} for $2 \leq i \leq m$.
The constraint satisfaction problem is feasible, if it has a solution
in quantifier free arithmetic expressions.

\begin{theorem}
\label{polynomial-time-reducibility-of-EvaluationProblem-to-ConstraintSatisfactionProblem}
The constraint satisfaction problem for binary valued instances
in prenex normal form with no free variables in $\SatisfiableQuantifiedArithmeticExpressions (\scalars)$
is $\PSPACE$-hard.
\end{theorem}
\proof Let 
\begin{small}
\[
\forall \basel{y}{1}\, \ldots\, \forall \basel{y}{\basel{k}{1}} 
\, \exists \basel{x}{1}\,\ldots\,
 \forall \basel{y}{\basel{k}{m-1}+1}\, \ldots\, \forall \basel{y}{\basel{k}{m}} 
\, \exists \basel{x}{m}\, \forall \basel{y}{\basel{k}{m}+1}\, \ldots\, \forall \basel{y}{n}
~~f(\basel{x}{1},\,\ldots,\, \basel{x}{m},\, 
\basel{y}{1},\, \ldots,\, \basel{y}{n}) 
\]
\end{small}
\lspace where $m$ and $n$ are positive integers and
$\basel{k}{i}$, $1 \leq i \leq m$, are integers such that
$0 \leq \basel{k}{i} \leq \basel{k}{i+1} \leq n$, $1 \leq i \leq m-1$,
be a given  instance of binary valued quantified boolean expression
with no free variables in $\QBF$ for the evaluation problem. Let 
\begin{small}
\[
\begin{array}{l}
\ltab 
\exists \basel{w}{1} ~~~~ \forall \basel{y}{1}\, \ldots\, \forall \basel{y}{\basel{k}{1}} 
\, \exists \basel{x}{1}\,\ldots\, \forall \basel{y}{\basel{k}{m-1}+1}\,
 \ldots\, \forall \basel{y}{\basel{k}{m}} \, \exists \basel{x}{m} ~~
  \forall \basel{y}{\basel{k}{m}+1}\, \ldots\, \forall \basel{y}{n}  \\
  \ltab ~~~~ \exists \basel{t}{1}\, \ldots\, \exists \basel{t}{\basel{k}{1}} 
 \, \forall \basel{v}{1}\,\ldots\,  \exists \basel{t}{\basel{k}{m-1}+1}\,
  \ldots\, \exists \basel{t}{\basel{k}{m}} \, \forall \basel{v}{m} ~~
  \exists \basel{t}{\basel{k}{m}+1}\, \ldots\, \exists \basel{t}{n}\\
\ltab   [~ \basel{w}{1} \wedge
f(\basel{x}{1},\,\ldots,\, \basel{x}{m},\, 
\basel{y}{1},\, \ldots,\, \basel{y}{n}) ~] \vee ~
[~  (\lnot \basel{w}{1}) \wedge
(\lnot f(\basel{v}{1},\,\ldots,\, \basel{v}{m},\, 
\basel{t}{1},\, \ldots,\, \basel{t}{n})) ~]
\end{array}
\]
\end{small}
\lspace be an instance to the constraint satisfaction problem
with no free variables, which can be easily shown to be in $\QSAT$,
since feasibility coincides with functional feasibility by arithmetic
expressions for $\basel{\Integers}{2}$. The input binary valued quantified
boolean expression evaluates to $1$ if and only if $\basel{w}{1}$ is $1$
in any solution to the constructed instance of the constraint satisfaction
problem.  Now, as discussed at the beginning of the section, the field
$\basel{\Integers}{2}$, together with all its arithmetic and logical
operations, can be emulated by the arithmetic operations and equality
operator with any field $\scalars$. Thus, the constraint satisfaction
problem for $\SatisfiableQuantifiedArithmeticExpressions(\scalars)$,
which includes equivalent binary valued quantified arithmetic expressions
for those in $\QSAT$, is $\PSPACE$-hard. \qed

\subsection{\label{Sec-QEP}Quantifier Elimination Problem}

Let $\powerset(\scalars)$ be a set of parametric subsets of $\scalars$,
parametrized by variables assuming values in $\scalars$, such that the
binary valued characteristic functions of the sets are assertions in
$\ArithmeticExpressions(\scalars)$. For an instance in
$\QuantifiedArithmeticExpressions(\scalars)$,
the quantifier elimination problem for a given instance is to compute, 
for {\small{$\basel{x}{1}, \ldots,\, \basel{x}{i-1},\,
\basel{y}{1},\,\ldots,\, \basel{y}{\basel{k}{i}}$
$ \in $
$\scalars$}}, sets
{\small{$\basel{G}{i}(\basel{x}{1},\, \ldots, \, \basel{x}{i-1},\, 
 \basel{y}{1},\,\ldots,\, \basel{y}{\basel{k}{i}})$
in $\powerset(\scalars)$},}
$1 \leq i \leq m$, such that
{\small{ 
$\bglc \basel{x}{i} \in \scalars\,:\, $
$ (\basel{x}{1},\, \ldots, \, \basel{x}{i},\, $
$ \basel{y}{1},\,\ldots,\, \basel{y}{\basel{k}{i}}) ~$
$\textrm{is feasible to the given instance} \bgrc $
$=$
$\basel{G}{i}(\basel{x}{1},\, \ldots, \, \basel{x}{i-1},\, 
 \basel{y}{1},\,\ldots,\, \basel{y}{\basel{k}{i}})$
}}, $1 \leq i \leq m$.
If $\scalars$ is the field of real numbers, with the set of binary
relations $\{=,\, <,\, \leq,\, >,\, \geq\}$ and the set of constants $\{0,\, 1\}$,
then the emptiness testing of parametric subsets of $\scalars^{n}$, for an
arbitrary positive integer $n$, where the characteristic functions of the
parametric subsets are binary valued quantified arithmetic expressions,
is decidable (or computable), and quantifier elimination is possible,
{\em i.e.}, equivalent quantifier free arithmetic assertions can be computed
for the quantified arithmetic assertions as characteristic functions for the
parametric subsets of $\scalars^{n}$ \cite{vanDalen:1994,  vandenDries:2000, Marker:2000,  Tarski:1951}.
Thus, $\QuantifiedArithmeticExpressions (\scalars)$ admits quantifier elimination,
and the sets of feasibility tuples for instances in
 $\QuantifiedArithmeticExpressions(\scalars)$ have characteristic functions
in $\ArithmeticExpressions (\scalars)$, that can be computed by an algorithm.
Set solutions can be enumerated by backtracking method \cite{HSR:2007}.
By the same proof of Theorem \ref{polynomial-time-reducibility-of-EvaluationProblem-to-ConstraintSatisfactionProblem},
the quantifier elimination problem can be shown to be $\PSPACE$-hard.  

\begin{theorem}
\label{hardness-of-prenex-normal-form-constraint-satisfaction-problem-with-unique-solutions}
The constraint satisfaction problem for binary valued instances in prenex normal form
with no free variables in $\SatisfiableQuantifiedArithmeticExpressions (\scalars)$,
that have unique solutions, is $\PSPACE$-hard.
\end{theorem}
\proof  Let
\begin{small}
\[
\forall \basel{y}{1}\, \ldots\, \forall \basel{y}{\basel{k}{1}} 
\, \exists \basel{x}{1}\,\ldots\,
 \forall \basel{y}{\basel{k}{m-1}+1}\, \ldots\, \forall \basel{y}{\basel{k}{m}} 
\, \exists \basel{x}{m}\, \forall \basel{y}{\basel{k}{m}+1}\, \ldots\, \forall \basel{y}{n}
~~f(\basel{x}{1},\,\ldots,\, \basel{x}{m},\, 
\basel{y}{1},\, \ldots,\, \basel{y}{n})
\]
\end{small}
\lspace be an instance in $\QBF$ for the quantifier elimination problem 
in prenex normal form with no free variables. Let
{\small{$\basel{X}{i}(\basel{x}{1},\, \ldots,\, \basel{x}{i},\,
\basel{y}{1},\, \ldots,\, \basel{y}{\basel{k}{i}})$}}
be the characteristic function of the set
{\small{$\basel{G}{i}(\basel{x}{1},\, \ldots,\, \basel{x}{i-1},\,
\basel{y}{1},\, \ldots,\, \basel{y}{\basel{k}{i}})$},}
for {\small{$\basel{x}{1},\, \ldots,\, \basel{x}{i},\,
\basel{y}{1},\, \ldots,\, \basel{y}{\basel{k}{i}} $
$\in$ 
$\basel{\Integers}{2}$, $1 \leq i \leq m$}.}
Let {\small{$\basel{X}{i,\, b}(\basel{x}{1},\, \ldots,\, \basel{x}{i-1},\,
\basel{y}{1},\, \ldots,\, \basel{y}{\basel{k}{i}})$
$ =$
$\basel{X}{i}(\basel{x}{1},\, \ldots,\, \basel{x}{i-1},\, b,\,
\basel{y}{1},\, \ldots,\, \basel{y}{\basel{k}{i}})$},}
for {\small{$\basel{x}{1},\, \ldots,\, \basel{x}{i-1},\,
\basel{y}{1},\, \ldots,\, \basel{y}{\basel{k}{i}} \in 
\basel{\Integers}{2}$, $b \in \basel{\Integers}{2}$, $1 \leq i \leq m$}.}
In the remaining part of the proof, the sets
{\small{$\basel{G}{i}(\basel{x}{1},\, \ldots,\, \basel{x}{i-1},\,
\basel{y}{1},\, \ldots,\, \basel{y}{\basel{k}{i}})$}}
are represented by the pair of boolean functions
{\small{$\basel{X}{i,\, b}(\basel{x}{1},\, \ldots,\, \basel{x}{i-1},\,
\basel{y}{1},\, \ldots,\, \basel{y}{\basel{k}{i}})$},}
for {\small{$\basel{x}{1},\, \ldots,\, \basel{x}{i-1},\,
\basel{y}{1},\, \ldots,\, \basel{y}{\basel{k}{i}} \in 
\basel{\Integers}{2}$, $b \in \basel{\Integers}{2}$, $1 \leq i \leq m$}.}
Let the following instance to the constraint satisfaction problem be considered:
\begin{small}
\begin{eqnarray}
&& \ltab \ltab ~~ \forall \basel{y}{1}\, \ldots\, \forall \basel{y}{\basel{k}{1}} \,  ~~ \exists \basel{v}{1,\, 0}  ~~ \exists \basel{v}{1,\, 1} ~~ \forall \basel{x}{1}\,  \ldots\,
 \forall \basel{y}{\basel{k}{i-1}+1}\, \ldots\,
 \forall \basel{y}{\basel{k}{i}} 
 ~~ \exists \basel{v}{i,\, 0}  ~~ \exists \basel{v}{i,\, 1} ~~ \forall  \basel{x}{i}\,  \forall \basel{y}{\basel{k}{i}+1}
 \, \ldots\, \forall \basel{y}{\basel{k}{i+1}}  \nonumber \\
 &&    \ldots\, \forall \basel{y}{\basel{k}{m-1}+1}\, \ldots\,
 \forall \basel{y}{\basel{k}{m}} ~~~~
  \exists \basel{v}{m,\, 0} ~~~~   \exists \basel{v}{m,\, 1} ~~~~
 \forall  \basel{x}{m}\, \forall \basel{y}{\basel{k}{m}+1}\, \ldots\, \forall \basel{y}{n}  \nonumber \\
&& \ltab   \bigwedge_{i = 1}^{m-1} \bglc \tab \bgls~~ (\basel{x}{i} = 0) ~ \rightarrow ~  \bgls ~~ \basel{v}{i,\, 0}  ~~~~ \leftrightarrow ~~~~ (\basel{v}{i+1,\, 0} \vee \basel{v}{i+1,\, 1}) ~~\bgrs ~~ \bgrs \tab \tab  \wedge  \nonumber \\
&&  \tab \bgls~~ (\basel{x}{i} = 1) ~ \rightarrow ~  \bgls ~~ \basel{v}{i,\, 1}  ~~~~ \leftrightarrow ~~~~ (\basel{v}{i+1,\, 0} \vee \basel{v}{i+1,\, 1}) ~~\bgrs ~~ \bgrs \tab  \bgrc \tab \tab \tab \tab \bigwedge \nonumber \\
&& \ltab   \bglc \tab
\bgls~~ (\basel{x}{m} = 0) ~ \rightarrow ~  \bgls ~~ \basel{v}{m,\, 0}  ~~~~ \leftrightarrow ~~~~ 
f(\basel{x}{1},\,\ldots,\, \basel{x}{m},\, 
\basel{y}{1},\, \ldots,\, \basel{y}{n})  ~~\bgrs ~~ \bgrs  ~~~~   \wedge \nonumber \\
&&  \tab \bgls~~ (\basel{x}{m} = 1) ~ \rightarrow ~  \bgls ~~ \basel{v}{m,\, 1}  ~~~~ \leftrightarrow ~~~~ 
f(\basel{x}{1},\,\ldots,\, \basel{x}{m},\, 
\basel{y}{1},\, \ldots,\, \basel{y}{n}) ~~\bgrs ~~ \bgrs \tab   \bgrc \label{indicator-functions-of-complete-set-solutions-of-a-given-instance-of-CSP}
\end{eqnarray}
\end{small} 
\lspace The boolean functions
{\small{$\basel{X}{m,\, b}(\basel{x}{1},\, \ldots,\, \basel{x}{m-1},\,
\basel{y}{1},\, \ldots,\, \basel{y}{\basel{k}{m}})$}}, that encode the indicator function of the set
{\small{$\basel{G}{m}(\basel{x}{1},\, \ldots,\, \basel{x}{m-1},\,
\basel{y}{1},\, \ldots,\, \basel{y}{\basel{k}{m}})$}},
 are the solutions to the boolean variables $\basel{v}{m,\, b}$,
 $b \in \basel{\Integers}{2}$, respectively,
 for {\small{$\basel{x}{1},\, \ldots,\, \basel{x}{m-1},\,
\basel{y}{1},\, \ldots,\, \basel{y}{\basel{k}{m}} \in 
\basel{\Integers}{2}$},} for the following instance of
constraint satisfaction problem:
\begin{small}
\begin{eqnarray*}
&& \ltab 
\forall \basel{y}{1}\, \ldots\, \forall \basel{y}{\basel{k}{1}} \, 
\forall \basel{x}{1}\,  \ldots\,
 \forall \basel{y}{\basel{k}{m-1}+1}\, \ldots\,
 \forall \basel{y}{\basel{k}{m}} ~~~~
  \exists \basel{v}{m,\, 0} ~~~~   \exists \basel{v}{m,\, 1} ~~~~
 \forall  \basel{x}{m}\, \forall \basel{y}{\basel{k}{m}+1}\, \ldots\, \forall \basel{y}{n} \nonumber \\
&& \ltab   \bglc \tab
\bgls~~ (\basel{x}{m} = 0) ~ \rightarrow ~  \bgls ~~ \basel{v}{m,\, 0}  ~~~~ \leftrightarrow ~~~~ 
f(\basel{x}{1},\,\ldots,\, \basel{x}{m},\, 
\basel{y}{1},\, \ldots,\, \basel{y}{n})  ~~\bgrs ~~ \bgrs  ~~~~  \wedge \nonumber \\
&&  \tab \bgls~~ (\basel{x}{m} = 1) ~ \rightarrow ~  \bgls ~~ \basel{v}{m,\, 1}  ~~~~ \leftrightarrow ~~~~ 
f(\basel{x}{1},\,\ldots,\, \basel{x}{m},\, 
\basel{y}{1},\, \ldots,\, \basel{y}{n}) ~~\bgrs ~~ \bgrs \tab  \bgrc
\label{unique-set-solutions-of-a-given-instance-of-CSP-1}
\end{eqnarray*}
\end{small} 
\lspace 
 After obtaining the boolean functions
{\small{$\basel{X}{i+1,\, b} (\basel{x}{1},\, \ldots,\, \basel{x}{i},
~\basel{y}{1},\, \ldots,\, \basel{y}{\basel{k}{i+1}})$, $b \in \basel{\Integers}{2}$},} 
 of {\small{$\basel{G}{i+1}(\basel{x}{1},\, \ldots,\, \basel{x}{i},\,
\basel{y}{1},\, \ldots,\, \basel{y}{\basel{k}{i+1}})$}},
the boolean functions 
{\small{$\basel{X}{i,\, b}(\basel{x}{1},\, \ldots,\, \basel{x}{i-1},\,
\basel{y}{1},\, \ldots,\, \basel{y}{\basel{k}{i}})$, $b \in \basel{\Integers}{2}$},} 
of {\small{$\basel{G}{i}(\basel{x}{1},\, \ldots,\, \basel{x}{i-1},\,
\basel{y}{1},\, \ldots,\, \basel{y}{\basel{k}{i}})$},} for
{\small{$\basel{x}{1},\, \ldots,\, \basel{x}{i},\,
\basel{y}{1},\, \ldots,\, \basel{y}{\basel{k}{i+1}} \in \scalars$},}
 are the solutions to the boolean variables $\basel{v}{i,\, b}$,
 $1 \leq i \leq m-1$,  $b \in \basel{\Integers}{2}$,  respectively,  
 for the following instance of constraint satisfaction problem:
 \begin{small}
\begin{eqnarray*}
&& \ltab \ltab ~~
\forall \basel{y}{1}\, \ldots\, \forall \basel{y}{\basel{k}{1}} \, 
\forall \basel{x}{1}\,  \ldots\,
 \forall \basel{y}{\basel{k}{i-1}+1}\, \ldots\,
 \forall \basel{y}{\basel{k}{i}} 
 ~~ \exists \basel{v}{i,\, 0}  ~~ \exists \basel{v}{i,\, 1} ~~
 \forall  \basel{x}{i}\,  \forall \basel{y}{\basel{k}{i}+1}
 \, \ldots\, \forall \basel{y}{\basel{k}{i+1}} \nonumber \\
&& \ltab \ltab~~  \bglc \tab
  \bgls  ~~ (\basel{x}{i} = 0)  ~~ \rightarrow~~ \bgls ~~\basel{v}{i,\, 0} ~~~~ \leftrightarrow \\
&& \ltab  \bglb \basel{X}{i+1,\, 0} (\basel{x}{1},\, \ldots,\, \basel{x}{i},
~\basel{y}{1},\, \ldots,\, \basel{y}{\basel{k}{i+1}}) ~~ \vee ~~
\basel{X}{i+1,\, 1} (\basel{x}{1},\, \ldots,\, \basel{x}{i},
~\basel{y}{1},\, \ldots,\, \basel{y}{\basel{k}{i+1}})\bgrb ~~ \bgrs ~~  \bgrs \tab \wedge  \nonumber  \\
&& \ltab ~~  \bgls  ~~ (\basel{x}{i} = 1)  ~~ \rightarrow~~ \bgls ~~\basel{v}{i,\, 1} ~~~~ \leftrightarrow \\
&&  \bglb \basel{X}{i+1,\, 0} (\basel{x}{1},\, \ldots,\, \basel{x}{i},
~\basel{y}{1},\, \ldots,\, \basel{y}{\basel{k}{i+1}}) ~~ \vee ~~
\basel{X}{i+1,\, 1} (\basel{x}{1},\, \ldots,\, \basel{x}{i},
~\basel{y}{1},\, \ldots,\, \basel{y}{\basel{k}{i+1}})\bgrb ~~ \bgrs ~~  \bgrs \tab \bgrc 
\label{unique-set-solutions-of-a-given-instance-of-CSP-2}
\end{eqnarray*}
\end{small} 
\lspace In summary, the boolean functions
{\small{$\basel{X}{i,\, b}(\basel{x}{1},\, \ldots,\, \basel{x}{i-1},\,
\basel{y}{1},\, \ldots,\, \basel{y}{\basel{k}{i}})$}}, that encode the indicator function of the set
{\small{$\basel{G}{i}(\basel{x}{1},\, \ldots,\, \basel{x}{i-1},\,
\basel{y}{1},\, \ldots,\, \basel{y}{\basel{k}{i}})$}},
 are the solutions to the boolean variables $\basel{v}{i,\, b}$,
 $b \in \basel{\Integers}{2}$, respectively,
 for {\small{$\basel{x}{1},\, \ldots,\, \basel{x}{i-1},\,
\basel{y}{1},\, \ldots,\, \basel{y}{\basel{k}{i}} \in 
\basel{\Integers}{2}$}} and $1 \leq i \leq m$, in the proposed instance of constraint satisfaction problem. The actual solutions for $\basel{v}{i,\, b}$ can also depend on $\basel{v}{j,\, c}$, for $b,\, c \in \basel{\Integers}{2}$, $1 \leq j \leq i-1$ and $2 \leq i \leq m$, in the proposed instance of the constraint satisfaction problem. Nonetheless, the boolean formulas {\small{$\basel{X}{i,\, b}(\basel{x}{1},\, \ldots,\, \basel{x}{i-1},\, \basel{y}{1},\, \ldots,\, \basel{y}{\basel{k}{i}})$}} are assumed to be the main solutions for $\basel{v}{i,, b}$, for $b \in \basel{\Integers}{2}$ and $1 \leq i \leq m$, as these are the solutions of the instance for the quantifier elimination problem, which is reduced to the instance of the constraint satisfaction problem.  \qed
\\

 In the proof of
Theorem \ref{hardness-of-prenex-normal-form-constraint-satisfaction-problem-with-unique-solutions},
for avoiding the possibility of dependence of a solution
for $\basel{v}{i,\,1}$ on $\basel{v}{i,\,0}$, the formulas in the above
are encoded treating the variables in pairs, representing $\basel{v}{i,\,b}$,
$b \in \basel{\Integers}{2}$, by a single variable 
$\basel{v}{i} = (\basel{v}{i,\,0},\, \basel{v}{i,\,1})$, 
performing the required computations in $\basel{\Integers^{2}}{2}$,
for $1 \leq i \leq m$. Thus, the encoding contains only a single
dependent variable $\basel{v}{i}$, $1 \leq i \leq m$, and either
component of it depends only on the variables constrained
by quantifiers occurring before the lone existential quantifier.
The components $\basel{v}{i,\, \basel{x}{i}}$ are replaced by a projection $T(\basel{x}{i},\, \basel{v}{i})$, which can further be chosen to be linear in $\basel{v}{i}$ for each fixed $\basel{x}{i}$, for $1 \leq i \leq m$, to avoid duplication. There is a unique solution separately for each component $T(\basel{x}{i},\, \basel{v}{i})$ of the dependent variable $\basel{v}{i}$, for $1 \leq i \leq m$. Now, as discussed in the beginning of the section, the field $\basel{\Integers}{2}$, together with all its arithmetic and logical operations, can be emulated by the arithmetic operations and equality operator with any field $\scalars$.
The characteristic functions of set solutions to the quantifier elimination problem for binary valued instances in prenex normal form with no free variables in $\QuantifiedArithmeticExpressions(\scalars)$, which includes $\QBF$, are unique. Thus, the constraint satisfaction problem for instances in $\SatisfiableQuantifiedArithmeticExpressions (\scalars)$, that admit unique solutions, is $\PSPACE$-hard, since the stated set of instances also contains those instances encoding the characteristic functions for quantifier elimination problem for instances in $\QBF$.

\subsection{\label{Sec-simultaneous-multivariate-polynomial-equations}Simultaneous Multivariate Polynomial Equations over $\scalars$}
Let $l,\, m, \, m \in \PositiveIntegers$ and$ \basel{f}{i}\bglb \basel{x}{1},\,\ldots,\, \basel{x}{m}, \, \basel{y}{1},\, \ldots, \, \basel{y}{n}\bgrb  \in \ArithmeticExpressions(\scalars)$, for $1 \leq i \leq l$, be arithmetic expressions. A system of (multivariate) polynomial equations is the following:
\begin{equation}
\basel{f}{i}\bglb \basel{x}{1},\,\ldots,\, \basel{x}{m},\, \basel{y}{1},\, \ldots, \, \basel{y}{n}\bgrb  \tab = \tab 0\,,
  \tab\tab 1 \leq i \leq l\,, \label{System-of-polynomial-equations}
\end{equation}
where $\basel{y}{j}$, $1 \leq j \leq n$, are independent variables and $\basel{x}{i}$, $1 \leq i \leq m$, are dependent variables, assuming values from $\scalars$, both specified as part of an instance. A tuple $\bglb \basel{a}{1},\, \ldots, \, \basel{a}{i}, \,  \basel{b}{1},\, \ldots, \, \basel{b}{n}\bgrb$ is feasible to (\ref{System-of-polynomial-equations}), if either (1) $i = m$ and 
(\ref{System-of-polynomial-equations}) holds with $\basel{x}{r} = \basel{a}{r}$, for $1 \leq r \leq m$, and $\basel{y}{j} = \basel{b}{j}$, for $1 \leq j \leq n$, or (2) $1 \leq i \leq m-1$, and  $\bglb \basel{a}{1},\, \ldots, \, \basel{a}{i}, \, \basel{a}{i+1},\,  \basel{b}{1},\, \ldots, \, \basel{b}{n}\bgrb$ is feasible for some $\basel{a}{i+1}$ depending on $\bglb \basel{a}{1},\, \ldots, \, \basel{a}{i}, \, \ldots, \, \basel{b}{1},\, \ldots, \, \basel{b}{n}\bgrb$. Let $\powerset\bglb \scalars \bgrb$ be the collection of admissible subsets of $\scalars$, whose indicator functions are in $\ArithmeticExpressions \bglb \scalars \bgrb$. A complete solution to (\ref{System-of-polynomial-equations}) are parametric maximal sets
{\small{$\basel{G}{i}\bglb\basel{a}{1},\, \ldots,\, \basel{a}{i-1},\,
\basel{y}{1},\, \ldots,\, \basel{y}{n}\bgrb \in \powerset\bglb \scalars \bgrb$},} such that {\small{$\basel{G}{i}\bglb\basel{a}{1},\, \ldots,\, \basel{a}{i-1},\,
\basel{y}{1},\, \ldots,\, \basel{y}{n}\bgrb = \left \{ ~ \basel{a}{i} \in \scalars\, :\, \bglb\basel{a}{1},\, \ldots,\, \basel{a}{i-1},\, \basel{a}{i},\, 
\basel{y}{1},\, \ldots,\, \basel{y}{n}\bgrb ~~ \textrm{is feasible}~ \right \}$},} for  $1 \leq i \leq m$.

In the above system, the ordering of the variables $\basel{x}{1},\ldots, \basel{x}{m}$ appears specified. However, this ordering can be made innocuous by additional constraints as follows:

\begin{small}
\begin{eqnarray*}
&& \tab \tab \basel{w^{2}}{i,\, j} ~~ = ~~ \basel{w}{i,\,j}\,, \tab 1 \leq i,\, j \leq m\\
&& \sum_{j = 1}^{m} \basel{w}{i,\,j} ~~ = ~~ 1 \tab \textrm{and} \tab
\sum_{j = 1}^{m} \prod_{
\scriptsize{\begin{array}{c}
 k = 1 \\
k \neq j \end{array} } }^{m} (1-\basel{w}{i,\,k}) ~~ = ~~ 1\,, \tab 1 \leq i \leq m \\
&& \sum_{i = 1}^{m} \basel{w}{i,\,j} ~~ = ~~ 1 \tab \textrm{and} \tab
\sum_{i = 1}^{m} \prod_{
\scriptsize{\begin{array}{c}
 k = 1\\
k \neq i \end{array} } }^{m} (1-\basel{w}{k,\,j}) ~~ = ~~ 1\,, \tab 1 \leq j \leq m\\
&&\\
&&\left[ \begin{array}{cccc}
\basel{w}{1,\,1}& \basel{w}{1,\,2}& \ldots & \basel{w}{1,\,m}\\
\basel{w}{2,\,1}& \basel{w}{2,\,2}& \ldots & \basel{w}{2,\,m}\\
\vdots & \vdots & \vdots  & \vdots \\
\basel{w}{m,\,1}& \basel{w}{m,\,2}& \ldots & \basel{w}{m,\,m}\\
\end{array} \right] 
\left[
\begin{array}{c}
\basel{x}{1}\\
\basel{x}{2}\\
\vdots\\
\basel{x}{m}\\
\end{array}\right] - 
\left[
\begin{array}{c}
\basel{x}{m+1}\\
\basel{x}{m+2}\\
\vdots\\
\basel{x}{2m}\\
\end{array}\right]  ~~ = ~~ 
\left[
\begin{array}{c}
0\\
0\\
\vdots\\
0\\
\end{array}\right] ~~, \tab \mathrm{and}\\
&&\\
&&
\basel{f}{i}\bglb \basel{x}{m+1},\,\ldots,\, \basel{x}{2m}, \, \basel{y}{1},\, \ldots, \, \basel{y}{n}\bgrb  \tab = \tab 0\,,
  \tab\tab 1 \leq i \leq l
\end{eqnarray*}
\end{small}
\lspace where $\basel{y}{j}$, $1 \leq j \leq n$, are independent variables, and all the remaining variables are dependent variables. The ordering is concealed by allowing the system to choose an appropriate ordering of the variables $\basel{x}{m+1},\, \ldots, \,\basel{x}{2m}$, while allowing $\basel{x}{1}, \ldots,\, \basel{x}{m}$ to appear in the specified order. In the above set of constraints, for  each row of the matrix 
$\basel{\left[\basel{w}{i,\,j}\right]}{1 \leq i,\, j \leq m}$, for the constraints on $i$, $1 \leq i \leq m$, and for each column of the matrix $\basel{\left[\basel{w}{i,\,j}\right]}{1 \leq i,\, j \leq m}$, for the constraints on $j$, $1 \leq j \leq m$, the first
constraint requires at least one entry of $1$, and the second constraint requires $(m-1)$ entries of $0$, in the respective row or column, and the matrix $\basel{\left[\basel{w}{i,\,j}\right]}{1 \leq i,\, j \leq m}$ is a permutation matrix.

\begin{theorem}
\label{reducibility-of-QSAT-to-system-of-polynomial-equations}
The quantifier elimination problem for instances in $\QuantifiedArithmeticExpressions (\basel{\Integers}{2})$ is polynomial time subroutine equivalent to the problem of solving systems of multivariate polynomial equations, for the field $\basel{\Integers}{2}$.
\end{theorem}
\proof 
Let
\begin{small}
\[
\forall \basel{y}{1}\, \ldots\, \forall \basel{y}{\basel{k}{1}} 
\, \exists \basel{x}{1}\,\ldots\,
 \forall \basel{y}{\basel{k}{m-1}+1}\, \ldots\, \forall \basel{y}{\basel{k}{m}} 
\, \exists \basel{x}{m}\, \forall \basel{y}{\basel{k}{m}+1}\, \ldots\, \forall \basel{y}{n}
~~f(\basel{x}{1},\,\ldots,\, \basel{x}{m},\, 
\basel{y}{1},\, \ldots,\, \basel{y}{n})
\]
\end{small}
\lspace be an instance in $\QBF$ in prenex normal form with no free variables, for some positive integers $m$ and $n$, and nonnegative integers $\basel{k}{i}$, such that $\basel{k}{i-1} \leq \basel{k}{i} \leq n$, for $1 \leq i \leq n$, where $\basel{k}{0} = 0$. Let $\basel{\chi}{m-i+1,\, a}(\basel{x}{1},\, \ldots, \basel{x}{m-i},\, \basel{y}{1},\, \ldots, \, \basel{y}{\basel{k}{m-i+1}}) \in \ArithmeticExpressions(\basel{\Integers}{2})$ be the solution for $\basel{v}{m-i+1,\, a}$, for $a \in \basel{\Integers}{2}$ and $1 \leq i \leq m$, such that the following holds:
\begin{small}
\begin{equation}
\ltab  \left.
\begin{array}{r}
 \basel{v}{m,\, a} ~~ \leftrightarrow ~~  \forall \basel{y}{\basel{k}{m}+1}\, \ldots\, \forall \basel{y}{n}
~~f(\basel{x}{1},\,\ldots,\, \basel{x}{m-1},\, a,\,  
\basel{y}{1},\, \ldots,\, \basel{y}{n})\,, ~~ \textrm{and} \\
 \basel{v}{m-i+1,\, a} ~~ \leftrightarrow ~~  \forall \basel{y}{\basel{k}{m-i+1}+1}\, \ldots\, \forall \basel{y}{\basel{k}{m-i+2}} ~~ \bgls \shiftright \\
 \tab  \basel{\chi}{m-i+2,\, 0}(\basel{x}{1},\, \ldots, \basel{x}{m-i},\, a,\, \basel{y}{1},\, \ldots, \, \basel{y}{\basel{k}{m-i+2}}) ~~ \vee \\
\tab \tab   \basel{\chi}{m-i+2,\, 1}(\basel{x}{1},\, \ldots, \basel{x}{m-i},\, a,\,  \basel{y}{1},\, \ldots, \, \basel{y}{\basel{k}{m-i+2}}) ~~ \bgrs \\
 \shiftright \textrm{for} ~~a \in \basel{\Integers}{2} ~~ \textrm{and} ~~  2 \leq i \leq m
 \end{array} \right\} \label{main-equation}
\end{equation}
\end{small}
\lspace The above equations can also be expressed as follows:
\begin{small}
\begin{displaymath}
 \ltab 
\begin{array}{l}
 \lnot \basel{v}{m,\, a} ~~ \leftrightarrow ~~  \exists \basel{u}{\basel{k}{m}+1}\, \ldots\, \exists \basel{u}{n}
~~\lnot f(\basel{x}{1},\,\ldots,\, \basel{x}{m-1},\, a,\,  
\basel{y}{1},\, \ldots,\, \basel{y}{\basel{k}{m}},\, \basel{u}{\basel{k}{m}+1},\, \ldots,\, \basel{u}{n})\,, ~~ \textrm{and} \\
 \lnot \basel{v}{m-i+1,\, a} ~~ \leftrightarrow ~~  \exists \basel{u}{\basel{k}{m-i+1}+1}\, \ldots\, \exists \basel{u}{\basel{k}{m-i+2}} ~~ \bgls \\
 \tab  \lnot \basel{\chi}{m-i+2,\, 0}(\basel{x}{1},\, \ldots, \basel{x}{m-i},\, a,\, \basel{y}{1},\, \ldots, \, \basel{y}{\basel{k}{m-i+1}},\, \basel{u}{\basel{k}{m-i+1}+1},\, \ldots,\, \basel{u}{\basel{k}{m-i+2}}) ~~ \wedge \\
\tab \tab \lnot  \basel{\chi}{m-i+2,\, 1}(\basel{x}{1},\, \ldots, \basel{x}{m-i},\, a,\, \basel{y}{1},\, \ldots, \, \basel{y}{\basel{k}{m-i+1}},\, \basel{u}{\basel{k}{m-i+1}+1},\, \ldots,\, \basel{u}{\basel{k}{m-i+2}}) ~~ \bgrs \\
\shiftright \shiftright  \textrm{for} ~~a \in \basel{\Integers}{2} ~~ \textrm{and} ~~  2 \leq i \leq m 
 \end{array}
\end{displaymath}
\end{small}
\lspace Let $\basel{G}{m,\, a,\, j,\, b}(\basel{x}{1},\,\ldots,\, \basel{x}{m-1},\,  
\basel{y}{1},\, \ldots,\, \basel{y}{\basel{k}{m}},\, \basel{u}{\basel{k}{m}+1},\, \ldots,\, \basel{u}{j-1})$,  for the ground case $\basel{x}{m} = a$ and $\basel{u}{j} = b$, for $a,\, b \in \basel{\Integers}{2}$ and $\basel{k}{m}+1 \leq j \leq n$, be indicator functions of complete solutions in the following system of simultaneous multivariate equations:
\begin{small}
\[
  f(\basel{x}{1},\,\ldots,\, \basel{x}{m-1},\, a,\,  
\basel{y}{1},\, \ldots,\, \basel{y}{\basel{k}{m}},\, \basel{u}{\basel{k}{m}+1},\, \ldots,\, \basel{u}{n})
~~ = ~~ 0
\]
\end{small}
\lspace Then, the indicator function,  $\basel{\chi}{m,\, a}(\basel{x}{1},\, \ldots, \basel{x}{m-1},\, \basel{y}{1},\, \ldots, \, \basel{y}{\basel{k}{m}})$, which is the solution for $\basel{v}{m,\, a}$, for $a \in \basel{\Integers}{2}$, is given by
\begin{small}
\begin{eqnarray*}
&& \ltab \ltab  \lnot \bgls  \basel{G}{m,\, a,\, \basel{k}{m}+1,\, 0}(\basel{x}{1},\,\ldots,\, \basel{x}{m-1},\,  
\basel{y}{1},\, \ldots,\, \basel{y}{\basel{k}{m}}) ~ \vee ~ \basel{G}{m,\, a,\, \basel{k}{m}+1,\, 1}(\basel{x}{1},\,\ldots,\, \basel{x}{m-1},\,  
\basel{y}{1},\, \ldots,\, \basel{y}{\basel{k}{m}}) \bgrs
\end{eqnarray*}
\end{small}
\lspace Now, for $1 \leq i \leq m-1$, after obtaining $\basel{\chi}{m-i+1,\, a}(\basel{x}{1},\, \ldots, \basel{x}{m-i},\, \basel{y}{1},\, \ldots, \, \basel{y}{\basel{k}{m-i+1}})$, let $\basel{G}{m-i,\, a,\, j,\, b}(\basel{x}{1},\,\ldots,\, \basel{x}{m-i-1},\,  
\basel{y}{1},\, \ldots,\, \basel{y}{\basel{k}{m-i}},\, \basel{u}{\basel{k}{m-i}+1},\, \ldots,\, \basel{u}{j-1})$, for the ground case $\basel{x}{m-i} = a$ and $\basel{u}{j} = b$, for $a,\, b \in \basel{\Integers}{2}$ and $\basel{k}{m-i}+1 \leq j \leq \basel{k}{m-i+1}$, be indicator functions of complete solutions in the following system of simultaneous multivariate equations:
\begin{small}
\begin{eqnarray*}
&& \basel{\chi}{m-i+1,\, 0}(\basel{x}{1},\, \ldots, \basel{x}{m-i},\, \basel{y}{1},\, \ldots, \, \basel{y}{\basel{k}{m-i}},\, \basel{u}{\basel{k}{m-i}+1},\, \ldots,\, \basel{u}{\basel{k}{m-i+1}}) ~~ = ~~ 0\,, \tab \textrm{and}\\
&& \basel{\chi}{m-i+1,\, 1}(\basel{x}{1},\, \ldots, \basel{x}{m-i},\, \basel{y}{1},\, \ldots, \, \basel{y}{\basel{k}{m-i}},\, \basel{u}{\basel{k}{m-i}+1},\, \ldots,\, \basel{u}{\basel{k}{m-i+1}}) ~~ = ~~ 0\\
&& \textrm{where}~~ \basel{x}{m-i} ~~ \textrm{is set to the ground value}~~ a\,.
\end{eqnarray*}
\end{small}
\lspace Then, the indicator function,  $\basel{\chi}{m-i,\, a}(\basel{x}{1},\, \ldots, \basel{x}{m-i-1},\, \basel{y}{1},\, \ldots, \, \basel{y}{\basel{k}{m-i}})$, which is the solution for $\basel{v}{m-i,\, a}$, for the ground instance $\basel{x}{m-i} = a$ and $a \in \basel{\Integers}{2}$, is given by
\begin{small}
\begin{eqnarray*}
&& \ltab \bgls \lnot \basel{G}{m-i,\, a,\, \basel{k}{m-i}+1,\, 0}(\basel{x}{1},\,\ldots,\, \basel{x}{m-i-1},\,  
\basel{y}{1},\, \ldots,\, \basel{y}{\basel{k}{m-i}}) \bgrs  ~~ \wedge \\
&& \bgls \lnot \basel{G}{m-i,\, a,\, \basel{k}{m-i}+1,\, 1}(\basel{x}{1},\,\ldots,\, \basel{x}{m-i-1},\,  
\basel{y}{1},\, \ldots,\, \basel{y}{\basel{k}{m-i}}) \bgrs
\end{eqnarray*}
\end{small}
\lspace for $1 \leq i \leq m-1$. Thus, finding complete solutions for systems of simultaneous multivariate equations over $\basel{\Integers}{2}$ is $\PSPACE$-hard, as it is logically equivalent to the quantifier elimination problem.  \qed
\\

For a positive integer $k$ and bit sequences $(\basel{u}{1},\, \ldots,\, \basel{u}{k}),\, $
$(\basel{t}{1},\, \ldots,\, \basel{t}{k})$
$\in$
$\basel{\Integers^{k}}{2}$ let ``$\succeq$'' be the ``successor or equal to'' relation
with respect to the dictionary ordering of finite binary sequences, such that the comparison
of corresponding bits is performed starting from least subscript index and up towards higher
subscript indexes, as follows:
\begin{small}
\begin{eqnarray*}
&& (\basel{u}{1},\, \ldots,\, \basel{u}{k}) \succeq (\basel{t}{1},\, \ldots,\, \basel{t}{k}) ~~~~ \textrm{exactly when the following holds :}\\
&& \ltab \ltab \bglb ~ (\basel{u}{1} = 1)  \wedge  (\basel{t}{1} = 0) ~\bgrb ~ \vee ~
 \bigvee_{i = 2}^{k} \bglb ~ \bigwedge_{j = 1}^{i-1} (\basel{u}{j} = \basel{t}{j})  \wedge  (\basel{u}{i} = 1)  \wedge  (\basel{t}{i} = 0) ~ \bgrb ~ \vee ~
 \bigwedge_{j = 1}^{k} (\basel{u}{j} = \basel{t}{j}) 
\end{eqnarray*}
\end{small}
\lspace Now, the prenex normal form equivalent formula for (\ref{main-equation}) is the following:
\begin{small}
\begin{eqnarray} 
&& 
\forall \basel{y}{1}\, \ldots\, \forall \basel{y}{\basel{k}{m}}~~
\exists \basel{v}{m,\, a} \exists \basel{u}{\basel{k}{m}+1,\, a}\, \ldots\, \exists \basel{u}{n,\, a}
~~ \forall \basel{z}{\basel{k}{m}+1}\, \ldots\, \forall \basel{z}{n} \tab \underbrace{\bgls}_{0} \nonumber \\
&&  \ltab \underbrace{\bglb}_{1}  [ ~ \basel{v}{m,\, a} \wedge \bigwedge_{j = \basel{k}{m}+1}^{n} (\basel{u}{j,\, a} = 0) ~] ~~~~ \wedge \nonumber \\
&&  \tab \tab  [~ \basel{G}{m}(\basel{x}{1},\,\ldots,\, \basel{x}{m-1},\, a,\,  
\basel{y}{1},\, \ldots,\, \basel{y}{\basel{k}{m}},\, \basel{z}{\basel{k}{m}+1},\, \ldots,\, \basel{z}{n})~]  \underbrace{\bgrb}_{1} \nonumber \\
&& \ltab \bigvee  
\underbrace{\bglb}_{2}~~   [ ~\lnot \basel{v}{m,\, a}~] ~ \wedge ~ [~\lnot \basel{G}{m}(\basel{x}{1},\,\ldots,\, \basel{x}{m-1},\, a,\,  
\basel{y}{1},\, \ldots,\, \basel{y}{\basel{k}{m}},\, \basel{u}{\basel{k}{m}+1,\, a},\, \ldots,\, \basel{u}{n,\, a})~]  ~~~~  \wedge \nonumber \\
&&  \shiftright  \bgls ~~~~(n = \basel{k}{m}) ~~ \vee ~~ [ ~(\basel{z}{\basel{k}{m}+1},\, \ldots,\, \basel{z}{n}) ~ \succeq ~
(\basel{u}{\basel{k}{m}+1,\, a},\, \ldots,\, \basel{u}{n,\, a}) ~ ] ~~ \vee \nonumber \\
&& \tab \tab [~ \basel{G}{m}(\basel{x}{1},\,\ldots,\, \basel{x}{m-1},\, a,\,  
\basel{y}{1},\, \ldots,\, \basel{y}{\basel{k}{m}},\, \basel{z}{\basel{k}{m}+1},\, \ldots,\, \basel{z}{n})~]
~~~~ \bgrs ~~ \underbrace{\bgrb}_{2} \tab \underbrace{\bgrs}_{0}
\label{unique-set-solutions-of-a-given-instance-of-CSP-1}
\end{eqnarray}
\end{small}
\lspace where
\begin{small}
\begin{eqnarray*}
&&   \basel{G}{m}(\basel{x}{1},\,\ldots,\, \basel{x}{m-1},\, \basel{x}{m},\,  
\basel{y}{1},\, \ldots,\, \basel{y}{\basel{k}{m}},\, \basel{y}{\basel{k}{m}+1},\, \ldots,\, \basel{y}{n})  ~~ = \\
&& \tab \tab  f(\basel{x}{1},\,\ldots,\, \basel{x}{m-1},\, \basel{x}{m},\,  
\basel{y}{1},\, \ldots,\, \basel{y}{\basel{k}{m}},\, \basel{y}{\basel{k}{m}+1},\, \ldots,\, \basel{y}{n})
\end{eqnarray*}
\end{small}
\lspace The solutions for the dependent variables bound by the existential quantifiers are unique. Let $\basel{\chi}{m-i+2,\, a}(\basel{x}{1},\, \ldots, \basel{x}{m-i+1},\, \basel{y}{1},\, \ldots, \, \basel{y}{\basel{k}{m-i+1}},\, \basel{y}{\basel{k}{m-i+1}+1},\, \ldots,\, \basel{y}{\basel{k}{m-i+2}})$ be the solution for the variable $\basel{v}{m-i+2,\, a}$, for $2 \leq i \leq m$ and $a \in \basel{\Integers}{2}$, and let 
\begin{small}
\begin{eqnarray*}
&& \ltab  \basel{G}{m-i+1}(\basel{x}{1},\,\ldots,\, \basel{x}{m-i},\, \basel{x}{m-i+1},\,  
\basel{y}{1},\, \ldots,\, \basel{y}{\basel{k}{m-i+1}},\, \basel{y}{\basel{k}{m-i+1}+1},\, \ldots,\, \basel{y}{\basel{k}{m-i+2}})  ~~ = \\
&& \tab   \bglb \tab  \basel{\chi}{m-i+2,\, 0}(\basel{x}{1},\, \ldots, \basel{x}{m-i+1},\, \basel{y}{1},\, \ldots, \, \basel{y}{\basel{k}{m-i+1}},\, \basel{y}{\basel{k}{m-i+1}+1},\, \ldots,\, \basel{y}{\basel{k}{m-i+2}}) ~~
 \vee \\
&& \tab \tab \tab  \basel{\chi}{m-i+2,\, 1}(\basel{x}{1},\, \ldots, \basel{x}{m-i+1},\, \basel{y}{1},\, \ldots, \, \basel{y}{\basel{k}{m-i+1}},\, \basel{y}{\basel{k}{m-i+1}+1},\, \ldots,\, \basel{y}{\basel{k}{m-i+2}}) \tab  \bgrb 
\end{eqnarray*}
\end{small}
\lspace and 
\begin{small}
\begin{eqnarray} 
&& \forall \basel{y}{1}\, \ldots\, \forall \basel{y}{\basel{k}{m-i+1}}~~
\exists \basel{v}{m-i+1,\, a} \exists \basel{u}{\basel{k}{m-i+1}+1,\, a}\, \ldots\, \exists \basel{u}{\basel{k}{m-i+2},\, a}
~~ \forall \basel{z}{\basel{k}{m-i+1}+1}\, \ldots\, \forall \basel{z}{\basel{k}{m-i+2}} \nonumber  \\
&&  \ltab \underbrace{\bgls}_{0} \tab  \underbrace{\bglb}_{1} ~~ [ ~ \basel{v}{m-i+1,\, a} \wedge \bigwedge_{j = \basel{k}{m-i+1}+1}^{\basel{k}{m-i+2}} (\basel{u}{j,\, a} = 0) ~] ~~~~ \wedge \nonumber \\
&&  \tab \tab \tab   [~ \basel{G}{m-i+1}(\basel{x}{1},\,\ldots,\, \basel{x}{m-i},\, a,\,  
\basel{y}{1},\, \ldots,\, \basel{y}{\basel{k}{m-i+1}},\, \basel{z}{\basel{k}{m-i+1}+1},\, \ldots,\, \basel{z}{\basel{k}{m-i+2}})~] ~~ \underbrace{\bgrb}_{1} \nonumber \\
&& \ltab \bigvee ~~ 
\underbrace{\bglb}_{2} \tab  [ ~\lnot \basel{v}{m-i+1,\, a}~] ~ \wedge \nonumber  \\
&&\tab \tab  [~\lnot \basel{G}{m-i+1}(\basel{x}{1},\,\ldots,\, \basel{x}{m-i},\, a,\,  
\basel{y}{1},\, \ldots,\, \basel{y}{\basel{k}{m-i+1}},\, \basel{u}{\basel{k}{m-i+1}+1,\, a},\, \ldots,\, \basel{u}{\basel{k}{m-i+2},\, a})~]  ~  \wedge \nonumber \\
&& \ltab  \bgls ~~~~ (\basel{k}{m-i+2} = \basel{k}{m-i+1}) ~~ \vee ~~ [ ~(\basel{z}{\basel{k}{m-i+1}+1},\, \ldots,\, \basel{z}{\basel{k}{m-i+2}}) ~ \succeq ~
(\basel{u}{\basel{k}{m-i+1}+1,\, a},\, \ldots,\, \basel{u}{\basel{k}{m-i+2},\, a}) ~ ] \nonumber \\
&&  \tab \vee ~~ [ ~~ \basel{G}{m-i+1}(\basel{x}{1},\,\ldots,\, \basel{x}{m-i},\, a,\,  
\basel{y}{1},\, \ldots,\, \basel{y}{\basel{k}{m-i+1}},\, \basel{z}{\basel{k}{m-i+1}+1},\, \ldots,\, \basel{z}{\basel{k}{m-i+2}}) ~]
\tab  \bgrs \nonumber \\
&& \shiftright
  \underbrace{\bgrb}_{2}  \tab  \underbrace{ \bgrs}_{0}
  \label{unique-set-solutions-of-a-given-instance-of-CSP-2}
\end{eqnarray}
\end{small}
\lspace  for $2 \leq i \leq m$ and $a \in \basel{\Integers}{2}$. Again, the solutions for the dependent variables bound by the existential quantifiers are unique. This discussion is summarized in the following:

\begin{corollary}
\label{reducibility-of-CSP-to-system-of-instances-in-Sigma-3-with-unique-solutions}
The constraint satisfaction problem for instances of the form 
\begin{small}
\[
\forall \basel{y}{1}\, \ldots\, \forall \basel{y}{k} 
~~ \exists \basel{x}{1}\,\ldots\, \exists\basel{x}{m}
~~ \forall \basel{y}{k+1}\, \ldots\, \forall \basel{y}{n} 
~~f(\basel{x}{1},\,\ldots,\, \basel{x}{m},\, 
\basel{y}{1},\, \ldots,\, \basel{y}{n})
\]
\end{small}
\lspace that are in $\SatisfiableQuantifiedArithmeticExpressions (\basel{\Integers}{2})$, 
 where $k$, $m$ and $n$ are positive integers, such that $1 \leq k \leq n$
and $\basel{x}{i}$, for $1 \leq i \leq m$, admit unique solutions, is $\PSPACE$-hard.
\end{corollary}
\proof Follows from the discussion preceding the statement. \qed
\\

\subsection{\label{Sec-parametric-polynomial-mappings}Parametric Multivariate Polynomial Mappings and their Nonparametric Inverses}
In this subsection, let $\scalars$ be a finite field. For integers $l \geq 0$,
$m \geq 1$ and $n \geq 1$, a parametric multivariate polynomial mapping, with
$\basel{z}{1}, \ldots,\, \basel{z}{l}$ as parameters, is
$\eta(\zz;\, \xx) = \bglb\basel{\eta}{1}(\zz;\, \xx),\, \ldots,\, \basel{\eta}{n}(\zz;\, \xx)\bgrb$,
where $\zz = (\basel{z}{1},\, \ldots,\, \basel{z}{l})$,  $\xx = (\basel{x}{1},\, \ldots, \, \basel{x}{m})$,
and $\basel{\eta}{i}(\zz;\, \xx)$
 $ \in $ 
$\singlevariablepolynomials{\scalars}{\basel{z}{\mathrm{1}},\, \ldots,\,
\basel{z}{\mathit{l}},\, \basel{x}{\mathrm{1}},\,\ldots,\, \basel{x}{\mathit{m}}}$,
 for $1 \leq i \leq n$.
A parametric left inverse $\eta^{\leftinverse}(\zz;\, \yy)$,
$\yy = (\basel{y}{1},\, \ldots,\, \basel{y}{n})$, of a parametric multivariate
polynomial mapping $\eta(\zz;\, \xx)$ on $X \subseteq \scalars^{l} \times \scalars^{m}$ is as follows: for every $\zz \in \scalars^{l}$,
$\xx \in \scalars^{m}$ and $\yy \in \scalars^{n}$, such that
$(\basel{z}{1},\, \ldots,\, \basel{z}{l},\, \basel{x}{1},\, \ldots,\, \basel{x}{m})$
$ \in $
$ X$, if  $\eta(\zz;\, \xx) = \yy$, then $\eta^{\leftinverse}(\zz;\, \yy) = \xx$.
A parametric right inverse $\eta^{\rightinverse}(\zz;\, \yy)$ on $Y \subseteq \scalars^{l} \times \scalars^{n}$  of a parametric multivariate polynomial mapping $\eta(\zz;\, \xx)$ is as follows: for every $\zz \in \scalars^{l}$,
$\xx \in \scalars^{m}$ and $\yy \in \scalars^{n}$, such that
$(\basel{z}{1},\, \ldots,\, \basel{z}{l},\, \basel{y}{1},\, \ldots,\, \basel{y}{n})$
$ \in $
$Y$, if  $\eta^{\rightinverse}(\zz;\, \yy) = \xx$, then $\eta(\zz;\, \xx) = \yy$.
For $\zz \in \scalars^{l}$, let $\basel{S^{\leftinverse}}{\eta} (\zz;\, \xx)$,
 $\xx \in \scalars^{m}$, and $\basel{S^{\rightinverse}}{\eta} (\zz;\, \yy)$,
 $\yy \in \scalars^{n}$,  be as follows:
$\basel{S^{\leftinverse}}{\eta} (\zz;\, \xx) = \bglc (\basel{z}{1},\, \ldots,\, \basel{z}{l},\, \basel{y}{1},\, \ldots,\, \basel{y}{n})$
$ \in $
$\scalars^{l} \times \scalars^{n} \, : \, \eta(\zz;\, \xx) = \yy \bgrc$,
and $\basel{S^{\rightinverse}}{\eta}(\zz;\, \yy)
  = \bglc (\basel{z}{1},\, \ldots,\, \basel{z}{l},\, \basel{x}{1},\, \ldots,\, \basel{x}{m})  \in \scalars^{l}\times \scalars^{m}\,:\,
  \eta(\zz;\, \xx) = \yy \bgrc$. Now, the following statements hold:
  ~ (1)~  for $\zz \in \scalars^{l}$ and $\xx \in \scalars^{m}$, the set
   $\basel{S^{\leftinverse}}{\eta} (\zz;\, \xx)$ contains exactly one element;
   ~(2)~  for $\zz \in \scalars^{l}$ and $\yy \in \scalars^{n}$,
 the set $\basel{S^{\rightinverse}}{\eta} (\zz;\, \yy)$ may be empty or nonempty; ~(3)~  a parametric left inverse
$\eta^{\leftinverse}(\zz;\, \yy)$ can be defined on the set
 $\bigcup_{(\basel{z}{1},\, \ldots,\, \basel{z}{l},\, \basel{x}{1},\, \ldots,\, \basel{x}{m})\in X} \basel{S^{\leftinverse}}{\eta}(\zz;\, \xx)$  if and only if
$\basel{S^{\leftinverse}}{\eta}(\zz;\, \xx)$
$ \cap $
$\basel{S^{\leftinverse}}{\eta}(\zz;\, \xx')  = \emptyset$,
 for {\small{$(\basel{z}{1},\, \ldots,\, \basel{z}{l},\, \basel{x}{1},\, \ldots,\, \basel{x}{m})$
 $ \in X$}}
 and 
{\small{$(\basel{z}{1},\, \ldots,\, \basel{z}{l},\, \basel{x'}{1},\, \ldots,\, \basel{x'}{m})$
$ \in X$}}, whenever $\xx \neq \xx'$;
  ~ and ~(4)~  a parametric right inverse
 $\eta^{\rightinverse}(\zz;\, \yy)$ can be defined on the set $Y$
 if and only if $\basel{S^{\rightinverse}}{\eta}(\zz;\, \yy) \neq \emptyset$,
 for every  $(\basel{z}{1},\, \ldots,\, \basel{z}{l},\, \basel{y}{1},\, \ldots,\, \basel{y}{n}) \in Y$. 
 If  a parametric left inverse (similarly, a parametric right inverse) of
 a parametric multivariate polynomial mapping does not depend on the parameters,
 then it is nonparametric. Let 
\begin{small}
\begin{eqnarray*}
\basel{T^{\leftinverse}}{\eta}(\xx) \tab =  \bigcup_{
(\basel{z}{1},\, \ldots,\, \basel{z}{l},~ \basel{x}{1},\, \ldots,\, \basel{x}{m}) \in X}  \bglc \yy \in \scalars^{n}\,:\,
\eta(\zz;\, \xx) = \yy\bgrc\\
\shiftright \textrm{for fixed} ~~\xx = (\basel{x}{1},\, \ldots,\, \basel{x}{m}) \in \scalars^{m}\,, \tab \textrm{and} \\
\basel{T^{\rightinverse}}{\eta}(\yy) \tab  =  \bigcap_{
(\basel{z}{1},\, \ldots,\, \basel{z}{l}, ~  \basel{y}{1},\, \ldots,\, \basel{y}{n}) \in Y} \bglc \xx \in \scalars^{m}\,:\,
\eta(\zz;\, \xx) = \yy \bgrc \\
\shiftright \textrm{for fixed} ~~ \yy = (\basel{y}{1},\, \ldots,\, \basel{y}{n}) \in \scalars^{n}
\end{eqnarray*}
\end{small}
\lspace  Then, on the set 
  $\bigcup_{(\basel{z}{1},\, \ldots,\, \basel{z}{l},~ \basel{x}{1},\, \ldots,\, \basel{x}{m}) \in X} \basel{S^{\leftinverse}}{\eta}(\zz;\, \xx)$,
a nonparametric left inverse $\eta^{\leftinverse}(\yy)$ can be defined 
 if and only if 
$\basel{T^{\leftinverse}}{\eta}(\xx) \cap \basel{T^{\leftinverse}}{\eta}(\xx') = \emptyset$,
for $\xx, \, \xx' \in \scalars^{m}$, $\xx \neq \xx'$, and on the set $Y$,
a nonparametric right inverse $\eta^{\rightinverse}(\yy)$ can be defined if and
only if $\basel{T^{\rightinverse}}{\eta}(\yy) \neq \emptyset$, for 
$(\basel{z}{1},\, \ldots,\, \basel{z}{l},\, \basel{y}{1},\, \ldots,\, \basel{y}{n}) \in Y$
and $\yy \in \scalars^{n}$. A parametric inverse is simultaneously a parametric
left inverse and a parametric right inverse. If a parametric inverse does not 
depend on the parameters, then it is nonparametric.

\begin{theorem}
\label{hardness-of-computing-nonparametric-inverse-of-parametric-multivariate-polynomial-mappings}
The computational problems of  $(1)$ finding nonparametric left inverses as quantifier free
arithmetic expressions  of parametric multivariate polynomial mappings,
  and $(2)$ nonparametric right inverses as quantifier free
arithmetic expressions of parametric multivariate polynomial mappings,
 with specified conditions on the domains of validity,
for the instances for which the stated inverses exist,
are both $\PSPACE$-hard.
\end{theorem}
\proof  Let {\small{$\forall \basel{t}{1}\, \ldots\, \forall \basel{t}{k}$
$~~ \exists \basel{w}{1} \, \ldots\, \exists \basel{w}{m} ~~$
$ \forall \basel{t}{k+1}\, \ldots\, \forall \basel{t}{n} ~~$
$f(\basel{w}{1},\, \ldots,\, \basel{w}{m},\,\basel{t}{1},\, \ldots,\, \basel{t}{n})$
$ \in $
$\QSAT$},}
for some  positive integers $k$, $m$ and $n$, such that $k \leq n$, be an instance 
for the constraint satisfaction problem, admitting unique solutions for each of the dependent
variables $\basel{w}{i}$ separately as quantifier free boolean expressions
$\basel{g}{i}(\basel{t}{1},\, \ldots, \basel{t}{k}) \in \ArithmeticExpressions(\basel{\Integers}{2})$,
for $1 \leq i \leq m$.

\paragraph{Part (1)} The proof is given by subroutine reduction taking one variable at a time,
starting from $m$ down to $1$.
For $\basel{w}{i},\, \basel{t}{j} \in \basel{\Integers}{2}$, $1 \leq i \leq m$
and $1 \leq j \leq n$, let 
\begin{small}
\begin{eqnarray*}
  \basel{\phi}{m}(\basel{w}{1},\, \ldots,\, \basel{w}{m},\, \basel{t}{1},\, \ldots,\, \basel{t}{n})~~
& = &
  f(\basel{w}{1},\, \ldots \,  \basel{w}{m},\, \basel{t}{1},\, \ldots,\, \basel{t}{n}) \,, \tab \textrm{and} \\
  \basel{h}{m}(\basel{w}{1},\, \ldots,\, \basel{w}{m-1},\, \basel{t}{1},\, \ldots,\, \basel{t}{n})~~
& = & 
  f(\basel{w}{1},\, \ldots \, \basel{w}{m-1},\, 0,\, \basel{t}{1},\, \ldots,\, \basel{t}{n}) ~~ \oplus \\
&& \tab   f(\basel{w}{1},\, \ldots \, \basel{w}{m-1},\, 1,\, \basel{t}{1},\, \ldots,\, \basel{t}{n}) 
\end{eqnarray*}
\end{small}
\lspace Let $\basel{\zeta}{(m)}(\zz;\,\xx)$, $\zz = (\basel{z}{1},\,\ldots,\, \basel{z}{n-k})$
and $\xx = (\basel{x}{1},\, \ldots,\, \basel{x}{m+k})$,
be a parametric multivariate polynomial mapping from
$\basel{\Integers^{m+k}}{2}$ into $\basel{\Integers^{m+k+1}}{2}$, with
parameters $\basel{z}{1},\,\ldots,\, \basel{z}{l}$,  where $l = n-k$, as follows:
\begin{small}
\[
\basel{\zeta}{(m,\,j)}(\zz;~\xx) ~~ = 
\left \{
\begin{array}{l}
 \basel{x}{j}\,,\tab  \textrm{for}~~ 1 \leq j \leq m-1\, ,\\
 \basel{x}{j+1}\,,\tab  \textrm{for}~~ m \leq j \leq m+k-1\, ,\\
 \basel{h}{m}(\basel{x}{1},\, \ldots,\, \basel{x}{m-1},\, \basel{x}{m+1},\, \ldots,\, \basel{x}{m+k},\,\basel{z}{1},\,\ldots,\, \basel{z}{l} )\,, ~~\textrm{for}~~  j = m+k \,,\\ 
 \bglb ~ \basel{\zeta}{(m,\, m+k)}(\zz;~\xx)   ~ \wedge ~ \basel{\phi}{m}(\basel{x}{1},\, \ldots,\,\basel{x}{m},\,  \basel{x}{m+1},\, \ldots,\, \basel{x}{m+k},\,\basel{z}{1},\,\ldots,\, \basel{z}{l} ) ~\bgrb \\
 \vee ~~
\bglb ~ \bglb ~ \lnot  ~ \basel{\zeta}{(m,\, m+k)}(\zz;~\xx) ~\bgrb  ~ \wedge ~ \basel{x}{j-k-1} ~\bgrb\,,
\tab    \textrm{for}~~     j = m+k+1\,,\\
\end{array} 
\right.
\]
\end{small}
\lspace The variables occurring in the above, in comparison with
the given instance of constraint satisfaction problem, are as follows:
$\basel{x}{j} = \basel{w}{j}$, for $1 \leq j \leq m$,
$\basel{t}{j} = \basel{x}{m+j}$, for $1 \leq j \leq k$, and
$\basel{t}{j} = \basel{z}{j-k}$, for $k+1 \leq j \leq n$. 
Let $\xx,\, \xx'  \in \basel{\Integers^{m+k}}{2}$ and $\zz \in \basel{\Integers^{n-k}}{2}$
be such that  $\basel{\zeta}{(m)}(\zz; \, \xx) = \basel{\zeta}{(m)}(\zz; \, \xx')$.
If $\xx \neq \xx'$, then it can only be the case that $\basel{x}{m} \neq \basel{x'}{m}$.
Since $\basel{\zeta}{(m,\, m+k)}(\zz; \, \xx) = \basel{\zeta}{(m,\,m+k)}(\zz; \, \xx')$,
it follows that
$ \basel{h}{m}(\basel{x}{1},\, \ldots,\, \basel{x}{m-1},\, \basel{x}{m+1},\, \ldots,\, \basel{x}{m+k},\,\basel{z}{1},\,\ldots,\, \basel{z}{l} )$
$ = $
$ \basel{h}{m}(\basel{x'}{1},\, \ldots,\, \basel{x'}{m-1},\, \basel{x'}{m+1},\, \ldots,\, \basel{x'}{m+k},\,\basel{z}{1},\,\ldots,\, \basel{z}{l} )$,
and since $\basel{\zeta}{(m,\, m+k+1)}(\zz; \, \xx)$
$ = $
$\basel{\zeta}{(m,\, m+k+1)}(\zz; \, \xx')$,
it follows that $\basel{x}{m} = \basel{x'}{m}$.
Now, a nonparametric left inverse of $\basel{\zeta}{(m)}$ is sought, 
which is valid on a maximal domain $\basel{X}{m} \subseteq \basel{\Integers^{m+n}}{2}$,
subject to the following conditions :
\begin{enumerate}
\item if $(\basel{x}{1},\, \ldots,\, \basel{x}{m+k},\,\basel{z}{1},\,\ldots,\, \basel{z}{l} )$
$\in$
$\basel{X}{m}$, then 
$(\basel{x}{1},\, \ldots,\, \basel{x}{m+k},\,\basel{z'}{1},\,\ldots,\, \basel{z'}{l} )$
$\in$
$\basel{X}{m}$, for every 
$(\basel{z'}{1},\,\ldots,\, \basel{z'}{l})$
$\in$
$\basel{\Integers^{l}}{2}$; 

\item for each fixed $(\basel{x}{m+1},\, \ldots,\, \basel{x}{m+k}) \in \basel{\Integers^{k}}{2}$,
there exists $(\basel{x}{1},\, \ldots,\, \basel{x}{m}) \in \basel{\Integers^{m}}{2}$,
such that $(\basel{x}{1},\, \ldots,\, \basel{x}{m+k},\,\basel{z}{1},\,\ldots,\, \basel{z}{l} )$
$\in$
$\basel{X}{m}$, for every 
$(\basel{z}{1},\,\ldots,\, \basel{z}{l})$
$\in$
$\basel{\Integers^{l}}{2}$, where $l = n-k$; and

\item a nonparametric left inverse of  $\basel{\zeta}{(m)}$ 
 can be defined on $\basel{\zeta}{(m)}(\basel{X}{m})$.
\end{enumerate}
The left inverse formula is as follows:
let $\basel{\zeta}{(m)}(\zz;~\xx) = \yy$, 
for some fixed
 $\yy = (\basel{y}{1},\, \ldots,\, \basel{y}{m+k+1})$
 $ \in $
 $\basel{\Integers^{m+k+1}}{2}$;
then {\small{$\basel{x}{j} = \basel{y}{j}$},} for {\small{$1 \leq j \leq m-1$},} 
{\small{$\basel{x}{j+1} = \basel{y}{j}$},} for {\small{$m \leq j \leq m+k-1$},} 
{\small{$\basel{x}{m} $
$ =$
$( \basel{y}{m+k} \wedge \basel{\rho}{m}(\yy) ) \vee 
( (\lnot \basel{y}{m+k} ) \wedge  \basel{y}{m+k+1} ) $,
for some function $\basel{\rho}{m}$ from $\basel{\Integers^{m+k+1}}{2}$ 
into $\basel{\Integers}{2}$. The domain $\basel{X}{m}$ of validity of
the left inverse satisfies the following inclusion:
\begin{eqnarray*}
&&\shiftleft \ltab \ltab  \basel{X}{m} \tab  \supseteq \tab  \{~ (\basel{x}{1},\, \ldots,\, \basel{x}{m+k},\, \basel{z}{1},\, \ldots,\, \basel{z}{n-k}) \in \basel{\Integers^{m+n}}{2}\,:\, \\
&& \basel{x}{j} = \basel{g}{j}(\basel{x}{m+1},\, \ldots,\, \basel{x}{m+k})\,, ~~\textrm{for}~ 1 \leq j \leq m ~\}
\end{eqnarray*}
and the function $\basel{\rho}{m}$ satisfies the following:
\begin{eqnarray*}
&& \ltab \ltab \basel{\rho}{m}\bglb \basel{g}{1}(\basel{y}{m},\, \ldots,\, \basel{y}{m+k-1}), \, \ldots,\, \basel{g}{m-1}(\basel{y}{m},\, \ldots,\, \basel{y}{m+k-1}), \, \basel{y}{m},\, \ldots,\, \basel{y}{m+k-1},\, 1,\, 0\bgrb ~~ = \\
&& \shiftright \shiftright  
\lnot \basel{g}{m}(\basel{y}{m},\, \ldots,\, \basel{y}{m+k-1})\,, ~~ \textrm{and} \\
&& \ltab \ltab  \basel{\rho}{m}\bglb \basel{g}{1}(\basel{y}{m},\, \ldots,\, \basel{y}{m+k-1}), \, \ldots,\, \basel{g}{m-1}(\basel{y}{m},\, \ldots,\, \basel{y}{m+k-1}), \, \basel{y}{m},\, \ldots,\, \basel{y}{m+k-1},\, 1,\, 1\bgrb ~~ = \\
&& \shiftright \shiftright  \tab 
 \basel{g}{m}(\basel{y}{m},\, \ldots,\, \basel{y}{m+k-1})
\end{eqnarray*}
Specification of the conditions on the domain of validity is part of the left inverse function computational problem, as required by the proof.

Now, after obtaining left inverses of $\basel{\zeta}{(i+1)},\, \ldots,\, \basel{\zeta}{(m)}$,
together with the hidden functions $\basel{\rho}{i+1},\,\ldots, \, \basel{\rho}{m}$, 
for some index $i$, where $1 \leq i \leq  m-1$,  the above procedure is repeated with the following
\begin{small}
\begin{eqnarray*}
&& \ltab \lspace \lspace \basel{\phi}{i}(\basel{w}{1},\, \ldots,\, \basel{w}{i},\, \basel{t}{1},\, \ldots,\, \basel{t}{n}) ~ = ~    \basel{\phi}{i+1} \bglb \basel{w}{1},\, \ldots \, \basel{w}{i}\, ,\,\,
  \basel{\rho}{i+1}(\basel{w}{1},\, \ldots \, \basel{w}{i},\, \basel{t}{1},\, \ldots,\, \basel{t}{k},\, 1,\, 1)\,,\\
 && \shiftright \shiftright \shiftright   \basel{t}{1},\, \ldots,\, \basel{t}{k},\, \basel{t}{k+1},\, \ldots,\, \basel{t}{n}\bgrb \,,  \tab \textrm{and} \\
&& \ltab  \basel{h}{i}(\basel{w}{1},\, \ldots,\, \basel{w}{i-1},\,  \basel{t}{1},\, \ldots,\, \basel{t}{n})
~~  = ~~  \basel{\phi}{i}(\basel{w}{1},\, \ldots \, \basel{w}{i-1},\, 0,\, \basel{t}{1},\, \ldots,\, \basel{t}{n}) ~~ \oplus \\
&& \shiftright \shiftright   \shiftright \basel{\phi}{i}(\basel{w}{1},\, \ldots \, \basel{w}{i-1},\, 1,\, \basel{t}{1},\, \ldots,\, \basel{t}{n}) 
\end{eqnarray*}
\end{small}
\lspace Thus, in the following instance of the constraint satisfaction problem :
\[
\forall \basel{t}{1}\, \ldots\, \forall \basel{t}{k}
~~ \exists \basel{w}{1} \, \ldots\, \exists \basel{w}{i} ~~
 \forall \basel{t}{k+1}\, \ldots\, \forall \basel{t}{n}~~
  \basel{\phi}{i}(\basel{w}{1},\, \ldots,\, \basel{w}{i},\, \basel{t}{1},\, \ldots,\, \basel{t}{n})
\]
the function $\basel{g}{j}(\basel{y}{1},\, \ldots,\, \basel{y}{k})$
is the unique solution for the variable $\basel{w}{j}$,  for $1 \leq j \leq i$.
Let $\basel{\zeta}{(i)}(\zz;\,\xx)$, $\zz = (\basel{z}{1},\,\ldots,\, \basel{z}{n-k})$
and $\xx = (\basel{x}{1},\, \ldots,\, \basel{x}{i+k})$,
be a parametric multivariate polynomial mapping from
$\basel{\Integers^{i+k}}{2}$ into $\basel{\Integers^{i+k+1}}{2}$, with
parameters $\basel{z}{1},\,\ldots,\, \basel{z}{l}$,  where $l = n-k$, as follows:
\begin{small}
\[
\basel{\zeta}{(i,\, j)}(\zz;~\xx) ~~ = 
\left \{
\begin{array}{l}
 \basel{x}{j}\,,\tab  \textrm{for}~~ 1 \leq j \leq i-1\, ,\\
 \basel{x}{j+1}\,,\tab  \textrm{for}~~ i \leq j \leq i+k-1\, ,\\
 \basel{h}{i}(\basel{x}{1},\, \ldots,\, \basel{x}{i-1},\, \basel{x}{i+1},\, \ldots,\,  \basel{x}{i+k},\,\basel{z}{1},\,\ldots,\, \basel{z}{l} )\,, ~~\textrm{for}~~  j = i+k \,,\\ 
 \bglb ~ \basel{\zeta}{(i, i+k)}(\zz;~\xx)   ~ \wedge ~ \basel{\phi}{i}(\basel{x}{1},\, \ldots,\,\,\basel{x}{i},\,  \basel{x}{i+1},\, \ldots,\, \basel{x}{i+k},\,\basel{z}{1},\,\ldots,\, \basel{z}{l} ) ~\bgrb \\
 \vee ~~
\bglb ~ \bglb ~ \lnot  ~ \basel{\zeta}{(i,\, i+k)}(\zz;~\xx) ~\bgrb  ~ \wedge ~ \basel{x}{j-k-1} ~\bgrb\,,~~~~   \textrm{for}~~     j = i+k+1\,,\\
\end{array} 
\right.
\]
\end{small}
\lspace It may be observed that 
\begin{eqnarray*}
&& \ltab \ltab \basel{\rho}{1}\bglb \basel{y}{1},\, \ldots,\, \basel{y}{k},\, 1,\, 0\bgrb ~~ = ~~
\lnot \basel{g}{1}(\basel{y}{1},\, \ldots,\, \basel{y}{k})\,, ~~ \textrm{and} \\
&& \ltab \ltab  \basel{\rho}{1}\bglb \basel{y}{1},\, \ldots,\, \basel{y}{k},\, 1,\, 1\bgrb ~~ = ~~
 \basel{g}{1}(\basel{y}{1},\, \ldots,\, \basel{y}{k})
\end{eqnarray*}
Thus, computing nonparametric left inverses of parametric multivariate
  polynomial mappings is $\PSPACE$-hard.

\paragraph{Part (2)}
Let $\eta(\zz;~\xx)$, $\zz = (\basel{z}{1},\,\ldots,\, \basel{z}{n-k})$
and $\xx = (\basel{x}{1},\, \ldots,\, \basel{x}{m+k})$, be a parametric
multivariate polynomial mapping from $\basel{\Integers^{m+k}}{2}$
into $\basel{\Integers^{k+1}}{2}$, with parameters
$\basel{z}{1},\,\ldots,\, \basel{z}{n-k}$ as follows: 
{\small{$\basel{\eta}{i}(\zz;~\xx)  = 
\basel{x}{m+i}$},} for {\small{$1 \leq i \leq k$},} and
{\small{$\basel{\eta}{k+1}(\zz;~\xx) =
f(\basel{x}{1},\, \ldots,\, \basel{x}{m+k},\, \basel{z}{1},\, \ldots,\, \basel{z}{n-k})$}.}
Now, if $\eta(\zz;~\xx) = \yy$, where
$\yy = (\basel{y}{1},\, \ldots,\, \basel{y}{k},\, 1)$
$ \in $
$ \basel{\Integers^{k}}{2} \times \{1\}$,
with $\basel{y}{k+1} = 1$, then $\basel{x}{i} = \basel{y}{i}$, for $1 \leq i \leq k$,
$\basel{x}{k+i} = \basel{g}{i}(\basel{y}{1},\, \ldots, \basel{y}{k})$, 
for $1 \leq i \leq m$, by the uniqueness of the solution for the given
instance of constraint satisfaction problem, and hence,
 computing nonparametric right inverses of parametric multivariate
  polynomial mappings is $\PSPACE$-hard. \qed 
\\

The construction of parametric injective mappings described in Theorem
\ref{hardness-of-computing-nonparametric-inverse-of-parametric-multivariate-polynomial-mappings}
shows how a general one-to-one mapping from $\mygrp^{m}$ into $\mygrp^{n}$,
where $m$ and $n$ are positive integers, with $m \leq n$, and $\mygrp$ is
a nonempty subset of a finite field $\scalars$, can be obtained: for a carefully
chosen bijective mapping $P(\yy)$ from $\mygrp^{n}$ into itself and hashing keys
$\basel{f}{i}(\xx)$, for $\xx = (\basel{x}{1},\, \ldots,\, \basel{x}{m}) \in \mygrp^{m}$ and $1 \leq i \leq n-m$,
the argument vector $(\basel{f}{1}(\xx),\, \ldots,\, \basel{f}{n-m}(\xx),\,  \basel{x}{1},\, \ldots,\, \basel{x}{m})$ 
is substituted for $\yy \in \mygrp^{n}$ in $P(\yy)$. Thus, $Q(\xx) = P(\basel{f}{1}(\xx),\, \ldots,\, \basel{f}{n-m}(\xx),\,  \basel{x}{1},\, \ldots,\, \basel{x}{m})$ is a generic multivariate one-to-one mapping from $\mygrp^{m}$ into $\mygrp^{n}$.

\vspace*{-0.2cm}

\section{Conclusions}

\subsection{\label{Sec-Security-Analysis}Security Analysis}

   The classical analysis of multivariate simultaneous equations can be applied only to polynomial equations \cite{vanDalen:1994, vandenDries:2000, Manin:2010, Marker:2000, MMP:1996, Tarski:1951}, and the Gr\"{o}bner basis analysis \cite{Buchberger:1965, Faugere:1999, Faugere:2002} cannot be extended to mappings involving functions as exponents. For a security that is immune to threats from Gr\"{o}bner basis analysis, parametric injective mappings from $\mygrp^{\mu}$ into $\eltSet^{\nu}$, with $\kappa$ parameters, for $\mygrp = \nonzeroscalars$, $\eltSet = \scalars$ and $\mu,\, \nu,\, \kappa \in \PositiveIntegers$, where $1 \leq \mu \leq \nu$ and $\scalars$ is a finite field, with component mappings taken as expressions from $\multiexpressions{\scalars}{x}{\omega}{\mu}{\kappa}$, restricting values of $\basel{x}{i}$ and $\basel{\omega}{j}$ to $\nonzeroscalars$, for $1 \leq i \leq \mu$ and $1 \leq j \leq \kappa$, with one level of exponentiation as described in section \ref{Sec-modular-exponentiation-over-Finite-Fields}, are adequate. 

The relevance of the complexity analysis described in section \ref{Sec-parametric-polynomial-mappings} is as follows: let $\xx = (\basel{x}{1},\, \ldots,\, \basel{x}{\mu})$,
 $\oomega= (\basel{\omega}{1},\, \ldots,\, \basel{\omega}{\kappa})$
 and $\yy = (\basel{y}{1},\, \ldots,\, \basel{y}{\nu})$.
For public key cryptography, the problem of computing 
$\xx$ from the equations $\basel{P}{i}(\xx,\,\oomega) = \basel{y}{i}$,
$1 \leq i \leq \nu$, with parameters
$\basel{\omega}{1},\, \ldots,\, \basel{\omega}{\kappa}$,
requires computation of nonparametric left inverse for
{\small{$(\basel{x}{1},\, \ldots, \basel{x}{\mu},\,
\basel{\omega}{1},\, \ldots,\, \basel{\omega}{\kappa})$
$\in$
$ \scalars^{\mu+\kappa}$}}.
For digital signature, the problem of computing $\yy$ from the equations 
$\basel{P}{i}(\yy) = \basel {x}{i}$, $1 \leq i \leq \mu$,
$\basel{y}{i} = \basel{x}{\mu+i}$, where
$\basel{S}{i}(\xx,\, \oomega) = \basel{x}{\mu+i}$,
$1 \leq i \leq \lambda$, with parameters
$\basel{\omega}{1},\, \ldots,\, \basel{\omega}{\kappa}$,
requires computation of nonparametric right inverse for
$(\basel{x}{1},\, \ldots, \basel{x}{\mu},\,
\basel{x}{\mu+1},\, \ldots,\, \basel{x}{\nu},\,
\basel{\omega}{1},\, \ldots,\, \basel{\omega}{\kappa}) \in \scalars^{\nu+\kappa}$,
constrained by $\basel{S}{i}(\xx,\, \oomega) = \basel{x}{\mu+i}$,
$1 \leq i \leq \lambda$. It can be observed that, for digital signature,
$\oomega$ could be under control of a trusted authentication verifier (TAV),
which ensures existential unforgeability.

\vspace*{-0.2cm}

\subsection{Summary}

In this paper, a new public key data encryption method is proposed, where the plain and encrypted messages are arrays. The method can also be used for digital certificate or digital signature applications. The key generation algorithm is particularly simple, easy and fast, facilitating changes of keys as frequently as required, and fast algorithms for polynomial multiplication and modular arithmetic \cite{BZ:2010, Pan:1992}, whenever appropriate, can be adapted in the encryption and decryption algorithms.

\end{document}